***The Nature of Low-Albedo Small Bodies from 3-µm Spectroscopy: One Group that Formed Within the Ammonia Snow Line and One that Formed Beyond It***

Andrew S. Rivkin[*1], Joshua P. Emery[*2], Ellen S. Howell[*3], Theodore Kareta[4], John W. Noonan[5], Matthew Richardson[6], Benjamin N. L. Sharkey[3], Amanda A. Sickafoose[6], Laura M. Woodney[7], Richard J. Cartwright[*8], Sean Lindsay[9], Lucas T. Mcclure[2]

1. Johns Hopkins University Applied Physics Laboratory
2. Northern Arizona University
3. University of Arizona
4. Lowell Observatory
5. Auburn University
6. Planetary Science Institute
7. California State University, San Bernardino
8. SETI Institute
9. University of Tennessee

---

[*] *Visiting Astronomer at the Infrared Telescope Facility, which is operated by the University of Hawaii under contract 80HQTR19D0030 with the National Aeronautics and Space Administration.*

# 1. Abstract

We present evidence, via a large survey of 191 new spectra along with previously-published spectra, of a divide in the 3-µm spectral properties of the low-albedo asteroid population. One group ("Sharp-types" or ST, with band centers < 3 µm) has a spectral shape consistent with carbonaceous chondrite meteorites, while the other group ("not-Sharp-types" or NST, with bands centered > 3 µm) is not represented in the meteorite literature but is as abundant as the STs among large objects. Both groups are present in most low-albedo asteroid taxonomic classes, and except in limited cases taxonomic classifications based on 0.5–2.5-µm data alone cannot predict whether an asteroid is ST or NST.

Statistical tests show the STs and NSTs differ in average band depth, semi-major axis, and perihelion at confidence levels ≥98%, while not showing significant differences in albedo.

We also show that many NSTs have a 3-µm absorption band shape like Comet 67P, and likely represent an important small-body composition throughout the solar system. A simple explanation for the origin of these groups is formation on opposite sides of the ammonia snow line, with the NST group accreting $H_2O$ and $NH_3$ and the ST group only accreting $H_2O$, with subsequent thermal and chemical evolution resulting in the minerals seen today. Such an explanation is consistent with recent dynamical modeling of planetesimal formation and delivery, and suggests that much more outer solar system material was delivered to the main asteroid belt than would be thought based on the number of D-class asteroids found today.

# 2. Introduction and Background

Over most of the last half-century, the fundamental spectral divide in the asteroid belt has been seen as that between the S-complex asteroids and the C-complex asteroids. These groups were distinguished by different albedos (the S asteroids have average albedos of roughly 23%, the average C asteroid albedos are typically 6%: DeMeo and Carry 2013) and absorption features (S asteroids are seen to have absorptions due to silicates, while some C asteroids have absorptions due to phyllosilicates near 0.7 µm, but are typically otherwise featureless in the 0.5–2.5-µm region). The recognition of this spectral divide was facilitated by large photometric and spectral surveys of asteroids beginning in the 1970s (Chapman et al. 1973), and the spectral taxonomies that were constructed were inspired by the existing stony/carbonaceous/iron meteorite groupings used by meteoriticists. Other groups of asteroids outside the S-C dichotomy were seen as much more rare, and thus relatively unimportant compared to those two large complexes.

Measurements of asteroids beyond 2.5 µm have been made for decades (Larson et al. 1979, Lebofsky et al. 1981, Feierberg et al. 1985, Jones et al. 1990) but have been slower to accumulate, given the difficulties of observing at those wavelengths. However, these measurements have shown that there is important variety in the C complex, including absorption bands that allow compositions to be determined for these objects in the same way that silicate absorption bands at shorter wavelengths are used to demonstrate that both chondritic and achondritic objects are present in the S-complex population. The discovery of objects exhibiting cometary activity on typical main-belt orbits (the "main-belt comets" or "activated asteroids": Hsieh and Jewitt 2006) pointed toward the existence of icy bodies in the Themis family and elsewhere within the orbit of Jupiter (Jewitt 2012), underlined by the discovery of absorption bands interpreted as due to surficial ice frost on (24) Themis, (90) Antiope, (65) Cybele, among other asteroids (Rivkin and Emery 2010, Campins et al. 2010b, Licandro et al. 2011, Takir et al. 2012, Hargrove et al. 2015, Rivkin et al. 2019). Other large asteroids, including Ceres, are additionally suspected of having an icy nature based on shape measurements (McCord and Sotin 2005, Hanus et al. 2020, Yang et al. 2020, Vernazza et al. 2020, Vernazza et al. 2021) or orbital hydrogen measurements (Prettyman et al., 2017). The ways in which the physical properties of icy asteroids might affect their representation in the meteorite and interplanetary dust collections were considered by Rivkin et al. (2014) and Rivkin et al. (2015a), with Vernazza et al. (2015), Vernazza et al. (2017), and Vernazza et al. (2021) addressing the problem via adaptive optics shape modeling.

Over the same time period, our understanding of comets has deepened immensely thanks to missions like Stardust, Deep Impact, and Rosetta. Analysis of Stardust samples shows intriguing signs that aqueous alteration occurred on comet 81P/Wild 2, though direct evidence remains elusive (Berger et al 2011, Hicks et al. 2017, Rietmeijer 2019). Nuclear spectra of comet 67P/Churyumov–Gerasimenko show absorptions attributed to submicron ice, ammoniated minerals, and organic materials, but no evidence of phyllosilicates (Raponi et al. 2020), and a qualitative similarity between the spectrum of 67P and some large asteroids has been noted (Rivkin et al. 2019, Poch et al. 2020).

The WISE survey provided a large number of albedos and diameters for main-belt asteroids and showed that low-albedo asteroids represent a large fraction of the population in all parts of the asteroid belt (Masiero et al. 2011). Dynamical models of early solar system history suggest that the low-albedo asteroids may have been delivered to their current orbits from distant formation locations (Walsh et al. 2012, Raymond and Izodoro 2017). Given the recent but numerous lines of evidence that there is a great deal of overlap in properties between the cometary and low-albedo asteroidal populations, we use spectroscopy in the 3-µm region to study the nature of low-albedo asteroids in terms of their hydrated minerals and their similarity to other low-albedo populations like comets.

### 2.1 Visible and Near-IR Taxonomy

Before going further, we must discuss how we will use existing asteroid taxonomy in this work. To quickly recap the situation, asteroids are classified using data in the 0.4—1.0-µm region in the Tholen (1984) and Bus (Bus and Binzel 2002) taxonomies, with the Bus-DeMeo (DeMeo et al. 2009) taxonomy extending the Bus taxonomy to 2.5 µm. While classifications are consistent from one taxonomic system to another in terms of major groupings, the details can differ due to the different datasets used as input, and it is not uncommon for objects to belong to different (but closely related) classes according to the different systems. For instance, (1) Ceres and (13) Egeria are both classified as G-class asteroids in the Tholen taxonomy, but are in the C and Ch classes, respectively, in the Bus taxonomy (which lacks a G class). (24) Themis is classified as a C asteroid in the Tholen taxonomy, a B asteroid in the Bus taxonomy, and again as a C asteroid in the Bus-Demeo taxonomy, while (150) Nuwa is classified as a CX, Cb, and C in the Tholen, Bus, and Bus-DeMeo taxonomies, respectively. These differences do not invalidate the taxonomies, of course, but do serve as a reminder that they are not compositional tools per se, and that particular classes may not be easily compared between systems even if they are given the same name in those systems.

Many low-albedo objects are distinguished from one another by spectral slope. Marsset et al. (2020) found that the intrinsic 1-σ uncertainty on spectral slope in measurements using the SpeX instrument (Rayner et al. 2003) is 4.7% per μm in the 0.8–2.4-μm region, which should generally allow discrimination between the major low-albedo asteroid spectral classes, though they note that effects such as object placement within the slit can lead to much larger effects on spectral slope if not caught. Because spectral slope can be a function of non-compositional factors like phase angle, a single object can in principle be classified differently depending on the observation used, if these effects are not accounted for. Finally, some of the objects in our sample have only been classified in the Tholen system, some only in the Bus system, and some in the Bus-DeMeo system, leading to the possibility of biasing results for classes that are only present in one of those taxonomies vs. those that are present (if not necessarily with the same definition) in all. For instance, the Ch class is defined by an absorption near 0.7 μm and is closely tied to a specific hydrated mineral composition (Rivkin et al. 2015), but this class does not exist in the Tholen taxonomy. In the past 20 years large asteroid surveys have generally used the 0.8—2.5 μm region due to the widespread use of the SpeX instrument (Rayner et al. 2003) for asteroid observations, with recent work done in the 0.7-μm region focusing instead on specific low-albedo families (for instance, de Leon et al. 2016, and Morate et al. 2016, 2018, 2019) discussed further in Section 6.3. Furthermore, there is potential for confusion because the currently-used Bus-DeMeo taxonomic classifications can be done using the full 0.45—2.45-μm spectral range or with the more limited 0.85—2.45-μm range, though the latter range misses some important spectral indicators (as just noted).

Given that the thrust of this work involves wavelengths beyond 2.5 μm where existing taxonomies do not extend, and also involves comparison of asteroid populations that are grouped via the taxonomies just discussed, there is the danger of inconsistencies. When referring to taxonomic classes, we will adopt the Bus classification from Bus and Binzel (2002) because it includes the important Ch class and has the largest number of classified objects, unless a Tholen or Bus-DeMeo classification is specifically called out. Because there are few members of the D and T classes in our sample (5 of each including objects for which D or T are the most likely classification according to Tholen 1984), and because these classes neighbor each other in principal component space in the Tholen, Bus, and Bus-DeMeo taxonomies, we will generally group the D and T classes together when discussed and tabulated below.

### 2.2 3-μm Taxonomy

While the existence of different 3-μm band shapes has been known for decades (Larson et al. 1979), there is no standard classification scheme for them. Previous studies (Rivkin et al. 2015, 2019, Takir and Emery 2012, Takir et al. et al. 2015) identified 3-4 groups, but give them different names and define some of them differently. Rivkin et al. (2019) focused on three groups named after type asteroids (Pallas-type, Ceres-type, and Themis-type: Figure 1), while recognizing both that other groups might be meaningfully split from the Themis types or conversely that the Ceres- and Themis-types might represent the same or very similar compositions. They also found no consistent and quantitative way to separate the Ceres types from the Themis types using simple criteria based on band centers and band depths, suggesting more sophisticated techniques (principal component analysis, machine learning algorithms, etc.) might be necessary to make those distinctions (Richardson et al. 2020). Takir and Emery (2012) defined 4 groups: “Sharp” (equivalent to the Pallas types), “Ceres-like” (having a band minimum near 3.05 µm within a broader band: effectively equivalent to the Rivkin et al. (2019) Ceres types), “Rounded” (effectively equivalent to the Themis types), and “Europa-like” (with a band center near 3.15 µm and a less-rounded appearance that the previous group).

Rivkin et al. (2019) noted that there was a qualitative similarity between some spectra of (324) Bamberga and comet 67P/Churyumov–Gerasimenko, despite nearly two orders of magnitude difference in their sizes. Other objects with close spectral similarity to 67P are included in this work. It is not yet clear whether this similarity represents similar origins and evolution despite their size differences, similar minerals formed despite different histories, or different compositions with similar-seeming spectra due to specific common molecular absorptions (Section 4.4). Which explanation would be preferred by William of Ockham as the simplest one could be a matter of what scientific company he was keeping, but in this work we will be assuming the first one while keeping the others in mind as reasonable alternatives.

In this work, we discuss objects that are smaller in size than the objects in the Rivkin et al. (2019) sample, and in most cases were observed when fainter and with lower data quality than that sample as well. We accordingly will use a very simple division of objects into “Sharp Type” and “not Sharp Type” (equivalent to “Pallas type” (PT) and “not Pallas type” in the Rivkin et al. 2019 scheme) through most of this work. As in Takir and Emery (2012), Sharp-type objects have a broad absorption band stretching to roughly 3.2–3.3 µm or longward before reaching the continuum level again, with a band minimum shortward of 3.0 µm. Based on laboratory spectra of analog meteorites and appropriate minerals, this band minimum is expected to fall in the 2.7—2.8-µm region that is unobservable in typical ground-based data (and omitted from the groundbased spectra included in this work). NST objects comprise everything else with a detectable absorption band. As will be discussed, few objects have no detectable absorption band at all in the 3-µm region, but they will be called out as appropriate.

We note that in a technical sense, we can classify spectra as ST or NST, but must use caution applying a classification to an entire object or to unobserved parts of a body. For instance, we might imagine referring to the interior of a body as ST or NST depending on the typical reflectance spectrum of minerals we expect to find there. While we may appear to throw caution to the wind in the remainder of the paper, readers should keep in mind that the ST/NST classification (like all taxonomic classification) is for convenience in discussing groups of objects with similar spectra rather than a replacement for actual compositional studies.

### 2.3 The Relationship Between Comets and Asteroids

Our understanding of the relationship between comets and asteroids has evolved significantly since the turn of the century, as dynamical models and solar-system formation models have been developed in conjunction with sample and astronomical measurements from the ground and space. A similarity between the visible-near IR spectra of some outer-belt asteroids and cometary nuclei was predicted by Gradie and Veverka (1980) and found shortly thereafter by Hartmann et al. (1982) in a study of asteroidal and cometary colors. Given the paradigms of small-body orbit evolution available at the time, it was not imagined that the P- and D-class asteroids literally migrated from a group of as-yet undiscovered transneptunian objects.

With the development of the Nice Model and the Grand Tack (Gomes et al. 2005, Walsh et al. 2012), it has been realized that comets and at least some asteroids may have a common origin (Levison et al. 2009). Sample data shows evidence of two broad classes of meteorites, a "carbonaceous" group and a "non carbonaceous" group, thought to have accreted either within (non-carbonaceous) or beyond (carbonaceous) the orbit of Jupiter, with little mixing (Warren 2011, Kleine et al. 2020). As discussed further in Section 6, there is also evidence that the ammonia ice line may also have been an important site for planetesimal formation (Dodson-Robinson 2009) and that all the low-albedo main-belt asteroids were delivered from beyond the orbit of Jupiter (Raymond and Izodoro 2017).

Spectral measurements have also further illuminated possible connections between comets and asteroids. The discovery of cometary activity in Themis-family asteroids (Hsieh and Jewitt 2006 ) was reinforced by the discovery of a 3-µm feature on Themis interpreted as due to ice frost (Rivkin and Emery 2010, Campins et al. 2010b) to show that icy bodies still existed in the main belt.

While low-albedo asteroids are abundant in the asteroid belt they are thought to be rarely represented in the meteorite collection. This apparent mismatch was considered by Rivkin et al. (2014) in the context of Ceres, who concluded that icy bodies might be less able to create family members with the coherence needed to survive as NEOs or meteorites. Work by Rivkin and DeMeo (2019) showed that the fraction of C-complex NEOs was roughly one-third of what might be expected from simple models of meteorite delivery and the population of the main belt, again suggesting that low-albedo asteroids may have physical properties that lead to fewer NEOs and meteorites than high-albedo asteroids. It has also been suggested that large low-albedo asteroids are represented among the interplanetary dust particles (IDPs) rather than the larger meteorites. Bradley et al. (1996) noted the similarity in 0.45—0.8-µm spectral slope between the chondritic smooth IDPs and C-type asteroids on the one hand and between chondritic porous IDPs and P/D asteroids (in the Tholen taxonomy) on the other. Vernazza et al. (2015) modeled a compiled set of mid-infrared measurements of asteroids, IDPs, and comets and argued that the non-Ch/Cgh C-complex asteroids are represented by pyroxene-rich IDPs while the P/D asteroids (again in the Tholen taxonomy) are represented by a mixture of pyroxene-rich and olivine-rich IDPs. Follow-up papers by Vernazza et al. (2017) and Vernazza et al. (2021) argue that the non-Ch/Cgh C-complex asteroids and P/D asteroids may be related, with the former group being aqueously altered versions of the latter one.

In the sections below, we present a large survey of low-albedo asteroids in the 3-µm spectral region, and interpret the results of that survey and the literature in a comprehensive, quantitative, and statistically-grounded manner. We argue that most of the spectra fall into one of two groups in terms of spectral shape and band center (for the wavelengths covered by the data), and that these two groups differ in some orbital and physical properties. One of the two groups bears spectral resemblance to the spectrum of comet 67P. Neither of the two groups is easily mapped onto shorter-wavelength taxonomic classes in a one-to-one manner save the Ch/Cgh class. We follow this with a discussion of the implications of the work, and finally include future directions for follow-up study.

## 3. Data Collection and Reduction

The data presented here were collected in the L-band Main-belt and NEO Observing Program (LMNOP) since 2002, as well as separate efforts led by Ellen Howell and Eric Volquardsen, all using the SpeX instrument (Rayner at al 2003) at the NASA Infrared Telescope Facility (IRTF). The Volquardsen data was obtained via the IRTF on-line archive, and included in this work with his encouragement. Several previous publications focused on particular objects or subsets of asteroids have also used observations taken during the course of the LMNOP (Rivkin et al. 2006, Rivkin et al. 2015, Rivkin et al. 2019, among others), with observing and reduction strategies in common with the observations presented for the first time in this work. Table 1 shows the data sources for this paper.

All of the data were obtained using SpeX in its long-wave cross-dispersed ("LXD") mode in the shortest wavelength setting. A 15" slit is used, with a beam switch of 7" between pairs of images. The specific exposure time and number of co-adds varies depending on the specific observing conditions, but the time between beam switches is kept below 120 seconds in order to allow subtraction of A-B pairs to correct for the effects of changes in atmospheric conditions that occur on the timescale of minutes. The limiting factor for exposure time of an image (or coadd) is typically thermal emission from the atmosphere. Several solar-type standard stars are observed in a typical night, with airmasses matched as closely as possible to the asteroid observations. The reduction pipeline minimizes the effect of airmass mismatches via an additional reduction step, as discussed below. SpeX was upgraded in 2014, and data in this work include observations obtained both before and after the upgrade, which have slightly different wavelength ranges for LXD mode. For consistency, we focus on the 2.2—4.0-µm wavelength range for both "New SpeX" and "Classic SpeX". Table 2 shows the distribution of SNR in the newly-presented measurements in the 2.9—3.4-µm region.

Table A1 in Appendix A shows the observing circumstances for the objects discussed in following sections, including V magnitude, distances from the Sun and Earth, and phase angle. Table A2 compiles physical and orbital properties for all the objects.

Reduction of LMNOP SpeX data has several steps. Extraction of spectra was done with Spextool (Cushing et al. 2003), a set of IDL routines designed for SpeX reduction developed and provided by the IRTF. After extraction, every combination of asteroid spectrum and star spectrum is run through an IDL-based set of routines developed by Bobby Bus and Eric Volquardsen and further developed by this team that correct for sub-pixel shifts between asteroid and star observations. It also uses an ATRAN model of the atmosphere (Lord 1992) to estimate the amount of precipitable water at the time of observation for each asteroid and star combination and remove it. This process has been used in several projects using SpeX data in the 0.8–2.5 and 2–4-µm regions (Clark et al. 2004, Rivkin et al. 2006, Rivkin et al. 2015, Binzel et al. 2019, among others). Following this step, a weighted average of the spectrum for each corrected asteroid-star pair was created for each asteroid, leading to a final asteroid spectrum. Bad pixels are flagged and omitted from the averaging process.

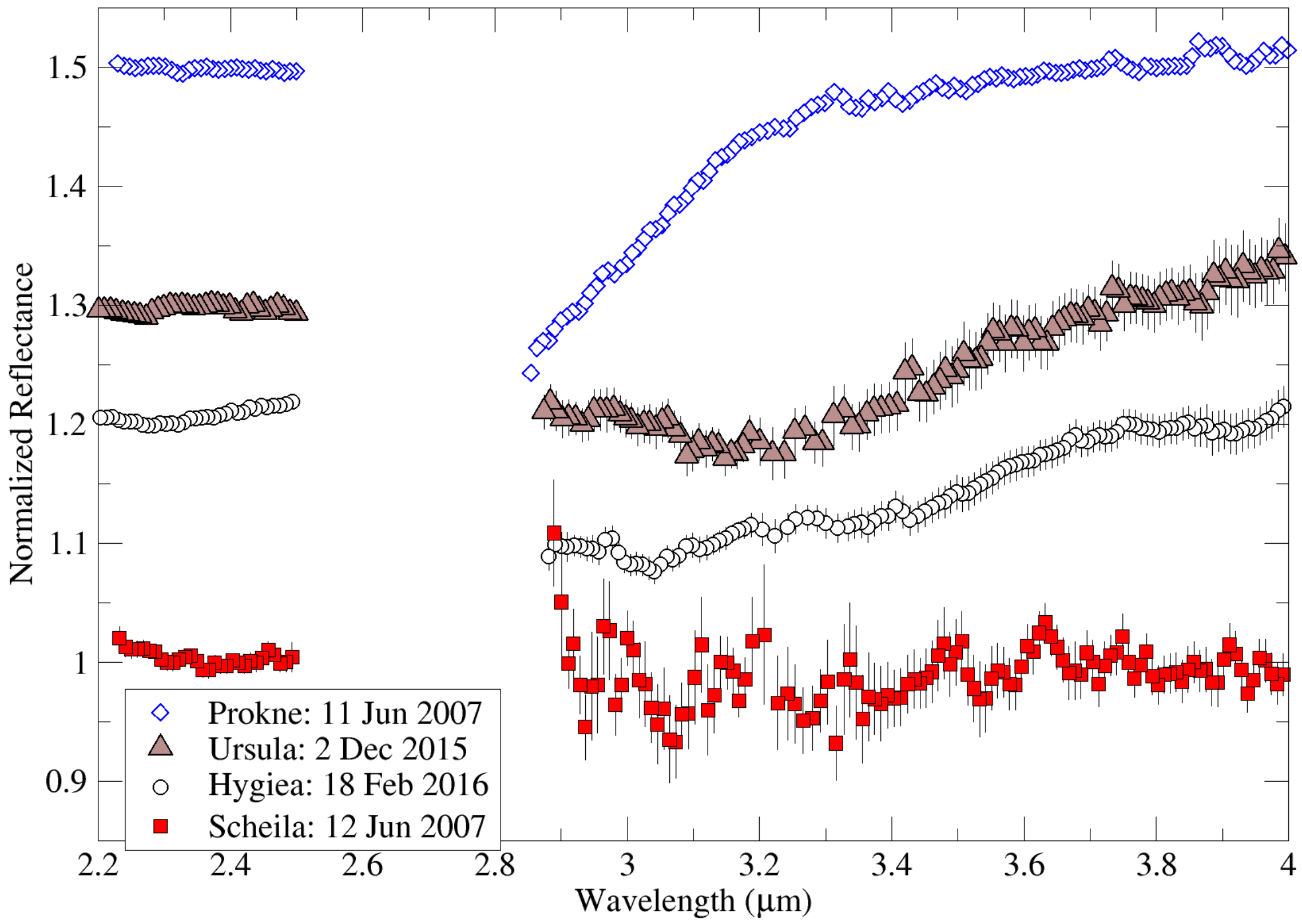

*Figure 1: 3-μm band shapes as seen in the new observations: (194) Prokne is an example of the Sharp-type (or "Pallas-type") shape, (375) Ursula the Themis-type shape, (10) Hygiea the Ceres-type shape, and (596) Scheila is interpreted to have no band. Objects with band centers at wavelengths > 3.0 μm are grouped as "Non-Sharp-types" or NSTs in this work. These spectra have had a linear continuum removed and are offset from one another for clarity*

Given the nature of the observing program, several asteroids were observed multiple times over the years, sometimes with an eye to testing whether surface variations were present as discussed in Rivkin et al. (2019) and in Section 5.1. Typically, all observations of an object within a single night were averaged into one spectrum, and observations on separate nights were never averaged with one another. The only exceptions to this practice were 2 separate visits to (2) Pallas on 23 June 2011 and 3 separate visits to Pallas on 6 December 2013, which were all maintained as separate spectra (Table A1). Ten objects were observed more than three times over the course of the LMNOP: Pallas and (10) Hygiea were each observed 10 times (including the separate Pallas visits just mentioned), Bamberga was observed 9 times, (65) Cybele 7 times, (704) Interamnia 6 times, (87) Sylvia, (52) Europa, and (476) Hedwig 5 times, and (335) Roberta and (31) Euphrosyne 4 times each. Most of the Hygiea, Bamberga, Interamnia, Europa, and Euphrosyne measurements were published in Rivkin et al. (2019), but additional data is reported here as well.

*Table 1: Data used in this work*

| Description | Number of spectra | Number of individual objects |
|---|---|---|
| New measurements (LMNOP) | 148 | 108 |
| New measurements (Howell PI) | 30 | 19 |
| New measurements (Volquardsen PI) | 13 | 12 |
| Ch asteroids from Rivkin et al. (2015b) | 42 | 36 |
| Large asteroids from Rivkin and Emery (2010) and Rivkin et al. (2019) | 34 | 9 |
| Total, removing duplicates | 267 | 159 |

*Table 2: SNR distribution of new measurements*

| Average SNR (2.9—3.4 µm, R~350) | Number of spectra |
|---|---|
| >5 | 191 (100%) |
| >10 | 187 (98%) |
| >25 | 136 (71%) |
| >50 | 58 (30%) |
| >100 | 14 (7%) |

Rivkin et al. (2019) noted the presence of some features interpreted as artifacts due to the extraction switching from one order to another. Similar artifacts were occasionally found during data reduction in this work, and are attributed to diminished sensitivity near the ends of orders. Combining images before extraction diminishes or removes artifacts in some cases, and this approach was taken in most cases, but artifacts remain in some spectra. We have addressed the issue by omitting the spectra that are most affected, and by inspecting the reduced data to ensure that artifacts are not mistaken for real absorptions. As part of this process, we also examined some older data, and re-reduced several spectra with possible artifacts after downloading them from the IRTF Legacy Archive. The spectrum of Interamnia from 12 September 2007 was noted by Rivkin et al. (2019) as having the longer-wavelength artifact, and after re-reduction as described above, a ST spectrum resulted rather than the NST spectrum originally published. Tables and figures reflect the values for the newly-reduced spectrum. Other re-reductions result in spectra that were only trivially different from the Rivkin et al. (2019) versions, and we use their published values in tables below.

## 3.1. Thermal Corrections and Continuum Selection

The temperatures of main-belt asteroids are sufficiently high to show detectable thermal emission in LXD spectra, particularly at their long-wavelength end near 4 µm. As a result, a correction to remove thermal emission is also made, which has a side benefit of providing some information about target thermal properties (Section 4.1).

As in previous work, we use a modified version of the Standard Thermal Model (STM: Lebofsky et al. 1986) in order to calculate and remove the thermal flux from the asteroids in this sample, allowing a reflectance-only spectrum to be analyzed. The inputs to the STM include both physical properties like radius, albedo, and emissivity, and observational circumstances like distance to the Sun and Earth and phase angle. All of the inputs are well known for the objects in the sample except for the "beaming parameter" (η), which is a free parameter abstractly representing a variety of factors that can change asteroidal temperatures like shape, surface roughness, obliquity, and thermal inertia.

The choice of η affects the amount of thermal flux removed, with an implied continuum associated with each choice. In this work, we assume linear continuum behavior, and in most cases (112 out of 191 spectra, or 59%) calculate an expected reflectance ratio at 3.75 µm based on data extrapolated from wavelengths unaffected by thermal flux. Where 2MASS data (Sykes et al. 2010) or MITHNEOS (Binzel et al. 2019) data were available, they were used for the extrapolation. If multiple datasets were available, the average extrapolation was used. If necessary, 52-color data (Bell et al. 1988) was used. If no appropriate datasets were available, a representative value was chosen for 3.75-µm continuum reflectance based on taxonomic class. The thermal correction routine iterated over a range of η values until one was found that resulted in the appropriate 3.75-µm reflectance value. In the remaining cases, poor-quality data near 3.75 µm or inconsistencies between the extrapolated value and the real spectral slope as seen in the LXD data required alternate approaches for determining the appropriate continuum. In these cases, one of three alternates were used: 1) requiring the spectral slope of a selected wavelength region > 2.5 µm to equal that of a different wavelength region < 2.5 µm (14 cases, or 7%); 2) requiring the spectral slope of a selected wavelength region to equal zero (30 cases, or 16%); or 3) requiring the reflectance at some wavelength to equal a value linearly extrapolated from a set of other wavelengths in the data (35 cases, or 18%).

# 4. Results

## 4.1. Thermal Properties of the Sample

The thermal properties of the asteroid sample, or subsamples, can be constrained by the results of the thermal corrections (Section 3.1). The η values from Rivkin et al. (2015b) and Rivkin et al. (2019) were included in the analysis. We note that we used the Gaussian fits discussed in Section 5 to determine whether an object fell into the NST or ST category. We also note that some spectra have no discernable 3-µm band and fall into neither ST nor NST categories (Section 4.7), so the sum of NST and ST groups in Table 3 does not match the total of the number of objects as divided by size. We find that the average value for η is close to 1 (and is equal to 1 within the uncertainties for most groups) for not only the entire sample, but for most subsets of objects grouped by spectral properties (Table 3). The relationship between η and phase angle is uncertain due to the abstract nature of η itself and its ambiguous physical meaning. Population studies suggest that η tends to increase over large phase angle ranges at an average rate of roughly 0.01 per degree (Alí-Lagoa et al. 2018). The range of average phase angle (Table 3) in these groups is limited (10°–15°) and can only account for a small fraction of the largest η differences seen in Table 3. Furthermore, Harris and Drube (2016) argue there is no phase angle dependence of η at phase angles < 20°, which includes all of the groups in Table 3.

The distinctions between the beaming parameters for the NST vs. ST asteroids are statistically significant at the 99% confidence level, as are those between the C-complex NSTs and either of the C-complex NST groups. However, the physical consequences of these differences in η are modest: because subsolar temperature is proportional to $\eta^{1/4}$, the difference in subsolar temperature caused by a change in η from 0.9 to 1.1 is only 5%, roughly 10-12 K at the subsolar point at typical temperatures in the middle of the asteroid belt, with the temperature change falling off with distance from the subsolar point. Following Harris and Drube (2016), objects with the average NST and ST values of η would have estimated thermal inertias of roughly 55 and 35 in SI units, respectively, given typical conditions and assumptions (8-hour rotation period, observed 2.8 AU from the Sun while at 45° subsolar latitude). Different assumptions change the specific thermal inertia estimates, but reasonable values for average solar distance, rotation period, and subsolar latitude all result in thermal inertias ≤ 100.

*Table 3: Beaming parameter for sample subgroups*

| Group | N of spectra | Average η | Average phase angle (degrees) |
|---|---|---|---|
| NST | 125 | 1.045 ± 0.330 | 11.5 ± 6.8 |
| ST | 128 | 0.938 ± 0.202 | 13.8 ± 7.4 |
| ST, Ch/Cgh | 45 | 0.899 ± 0.087 | 14.1 ± 7.2 |
| ST, C complex, not Ch/Cgh | 48 | 0.911 ± 0.090 | 15.0 ± 7.2 |
| NST, C Complex | 83 | 1.025 ± 0.258 | 12.5 ± 6.9 |
| Diameter > 120 km | 179 | 0.995 ± 0.327 | 12.8 ± 7.3 |
| Diameter < 120 km | 88 | 0.996 ± 0.243 | 11.7 ± 7.3 |

## 4.2. Spectral Classification by Visual Inspection

In keeping with previous work (Rivkin et al. 2015, 2019), we performed an initial classification via inspection prior to, and as a comparison for, a more quantitative study (Section 5) and as a means of checking data quality. Of the 191 new observations, 100 were visually classified as ST, 69 as NST, and 14 with no band (NB). A total of eight spectra were ambiguous as to which classification was most appropriate, typically due to lower-quality data near 2.9–3.0 µm, and were left unclassified. Figure A1 shows continuum-removed versions of all 191 spectra.

The sample discussed here includes members of the main low-albedo spectral classes as well as some rarer ones. Asteroids in the Ch or Cgh class have Sharp-type band shapes in the 3-µm region when that determination can be made (Rivkin et al. 2015). The other classes (whether in Tholen, Bus, or Bus-DeMeo schemes) all contain at least one ST and one NST object. This suggests, at least at face value, that the taxonomic classes defined from 0.5—2.5 µm, other than a Ch/Cgh classification, are of limited use in predicting the hydrated compositions of specific asteroids. As discussed in Section 2, the Ch and Cgh classes are defined by the presence of an absorption band in both the Bus and Binzel (2002) and the related DeMeo et al. (2009) taxonomies, while the definitions of other low-albedo classes are generally derived from spectral slopes and ranges of values of principal components in one or both of those taxonomies.

### 4.3. Band Shapes from Selected Band Depths

Figure 2 shows the 2.9-µm and 3.2-µm band depths, used as a simple proxy for band shape (Sato et al. 1997, Rivkin et al. 2015, 2019). The plots include four objects with distinctive 3-µm band shapes: (51) Nemausa, Ceres, Themis, and comet 67P, along with lines connecting them to the origin. Nemausa rather than Pallas is included as the example of the Sharp-types because its band depth is much deeper than Pallas' and it appears to be a more appropriate representative endmember as a result. Also included is a dashed line showing where the 2.9- and 3.2-µm band depths are equal. Rivkin et al. (2015, 2019) showed that CM chondrites from Takir et al. (2013) and Ch chondrites occupy the same area of the plot, generally straddling the Nemausa—origin line, while most large (diameter > 200 km) C-complex asteroids have spectral shapes that placed them in the Themis-67P area of the plot.

Figure 2 shows the distribution of ST and NST objects, plotting the 252 spectra in this work, Rivkin et al. (2015b), and Rivkin et al. (2019) for which band centers can be quantitatively fit (Section 5) and that can be classified as either ST or NST. Laboratory measurements of carbonaceous chondrites from Takir et al. (2013) and Bates et al. (2021) are also included, and it is evident that the Bates et al. measurements are consistent with the Takir et al. ones in this representation. The NST and ST groups dominate different parts of the plot, with a transition from one group to the other near but not at the 1:1 line, as might be expected for groups defined using a different but nearby wavelength (Table 4). To confirm, we fit the data using a Gaussian Mixture Model (GMM), implemented through Python's sklearn.mixture (Pedregosa et al. 2011). The optimal number of clusters, $k$=2, was determined by the minimum BIC (Bayesian Information Criterion). The same results were achieved when using all of the data in Fig. 2 as well as removing the high-error points (band-depth error > 0.05, to prevent biased results). We note that the AIC (Akaike Information Criterion) had a minimum at $k$=4 or 5. Therefore, a more complex model might eventually provide a better fit but simply dividing the data into two groups is reasonable.

The carbonaceous chondrites are found among the ST asteroids or in areas where both ST and NST asteroids are found, while no meteorites are found in exclusively NST regions. Two ST spectra are found relatively far into the NST region–these measurements, of (336) Lacadeira and (444) Gyptis, both have artifacts near 3.2 µm that artificially increase the 3.2-µm band depth. Inspection of these spectra confirms the Gaussian band-center determination that these are ST spectra.

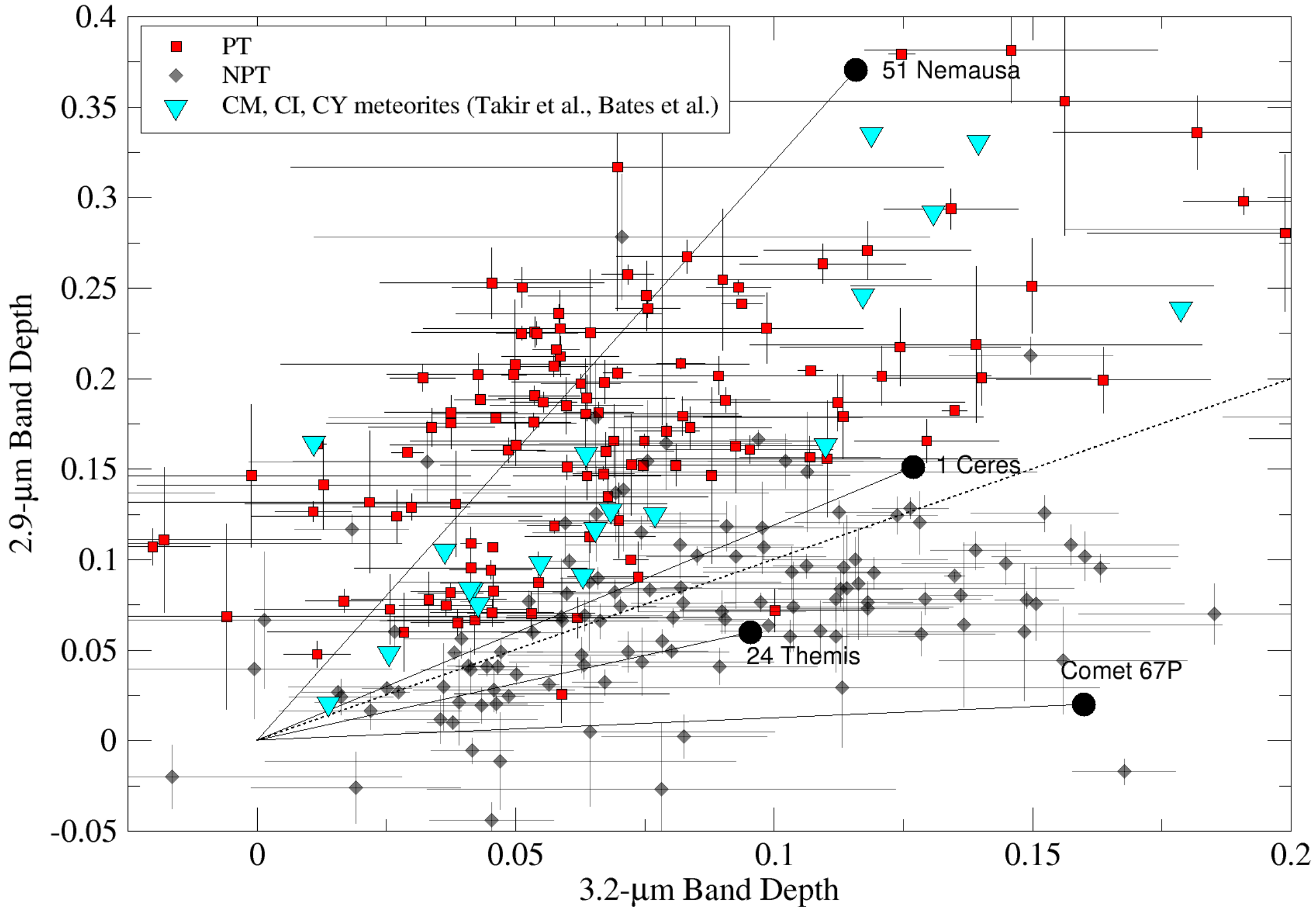

*Figure 2: Band Depth-Band Depth Plot for ST and NST objects. Also shown are laboratory measurements of hydrated carbonaceous chondrites from Takir et al. (2013) and Bates et al (2020). The NST and ST asteroids are largely separated from one another when plotted in this space, and the meteorites are found in areas where ST asteroids are more numerous. Two ST observations with artifacts near 3.2 µm are found in the NST area, but inspection and fits to band center confirm them as ST.*

*Table 4: 2.9- vs 3.2-µm Band Depth (BD) s for ST and NST objects*

| NB excluded | BD_2.9 > BD_3.2 | BD_2.9 < BD_3.2 |
|---|---|---|
| ST | **79** | **3** |
| NST | **28** | **35** |

Another way to investigate differences in distributions among one or more parameters is Kernel Density Estimation (KDE), a methodology which takes discrete data points and displays them as a distribution according to some width or weighting scheme. While typically used to re-engineer probability density distributions or other continuous functions from multiple individual measurements, it can be used generically on any dataset for which trends larger than individual datapoints are to be investigated. We recast the data in Figure 2 as three KDE plots for all the asteroids considered as well as the STs and NSTs separately. In Figure 3, each asteroid is plotted as a 2-dimensional gaussian with widths along each axis corresponding to its 1-sigma error in those parameters, and the plots are shown with logarithmically scaled color bars and logarithmically spaced contours.

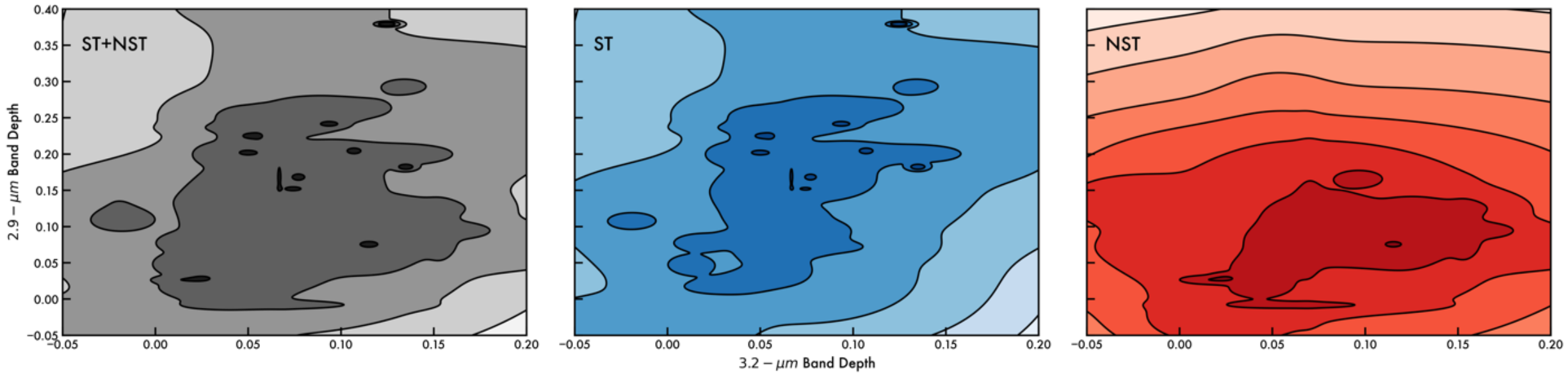

*Figure 3. Kernel Density Distributions (KDEs) for all the asteroids in our sample (greys, left) as well as the ST (blues, center) and NST (reds, right) groups separately. The two groupings of objects are clearly separated using this technique even when accounting for the errors on individual spectra.*

The ST and NST populations are shown to be distributed differently in the parameter space defined by the 2.9- and 3.2-micron band depths, with some overlap in the region where neither band is very deep for reasons described earlier in this subsection. None of the objects defined as NST have significant 2.9-micron band depths, while some ST objects typically have stronger 2.9 absorptions compared to 3.2. We note that KDE analyses traditionally use identical widths for each of the datapoint distributions, but we wanted to include our uncertainties to guard against overinterpretation of the data at hand. In other words, uniform widths for these KDE plots would result in even sharper contrast between the observed populations than is shown here. When combined with the other differences between the classes described elsewhere in this paper, both quantitative and qualitative, we view this as strong evidence that these groups represent distinct spectral classes.

### 4.4. Similarity to cometary spectra:

Figure 4 compares the spectra of Cybele, (431) Nephele, and (139) Juewa to 67P (Raponi et al. 2020), offset from one another for clarity. There are several additional objects in the sample (including objects from Rivkin et al. 2019) that have spectra consistent with 67P, including some with larger observational uncertainties and thus less robust matches. In order to quantitatively measure the similarity of asteroid spectra to the spectrum of 67P, we first model the continuum-removed 67P spectrum from 2.8–3.8 µm as the sum of 3 Gaussians with their amplitude, center wavelengths, and widths as free parameters. This allows the cometary spectrum to be calculated at specific wavelengths of interest in that interval. We then used two different measures of spectral similarity: 1) calculating the difference between that spectrum and the 67P spectrum at each wavelength, squaring it, and summing that squared difference over the 2.9–3.3-µm range, and 2) calculating the median absolute value of the difference at each wavelength from 2.9–3.3 µm for that spectrum. For these calculations, linear continua spanning 2.9—3.6 µm are removed from the asteroid spectra. Table 5 and Table 6 show the 10 asteroid spectra with the smallest differences from 67P using these measures, and Figure 5 shows some of these closest matches along with the spectrum of 67P calculated using the 3-Gaussian fit mentioned above. We note that in a few cases, a low-quality data point at 2.9 μm can result in a continuum that is unrepresentative, and Figure 5 additionally compares the spectrum of 90 Antiope from 28 August 2011 to 67P, with the asteroid continuum adjusted to provide a better match.  In order to minimize arbitrary data processing and because the intention in this work is to highlight the existence and implications of these similarities rather than exhaustively document them, we did not attempt to construct best-fit continua in this way for other asteroids.

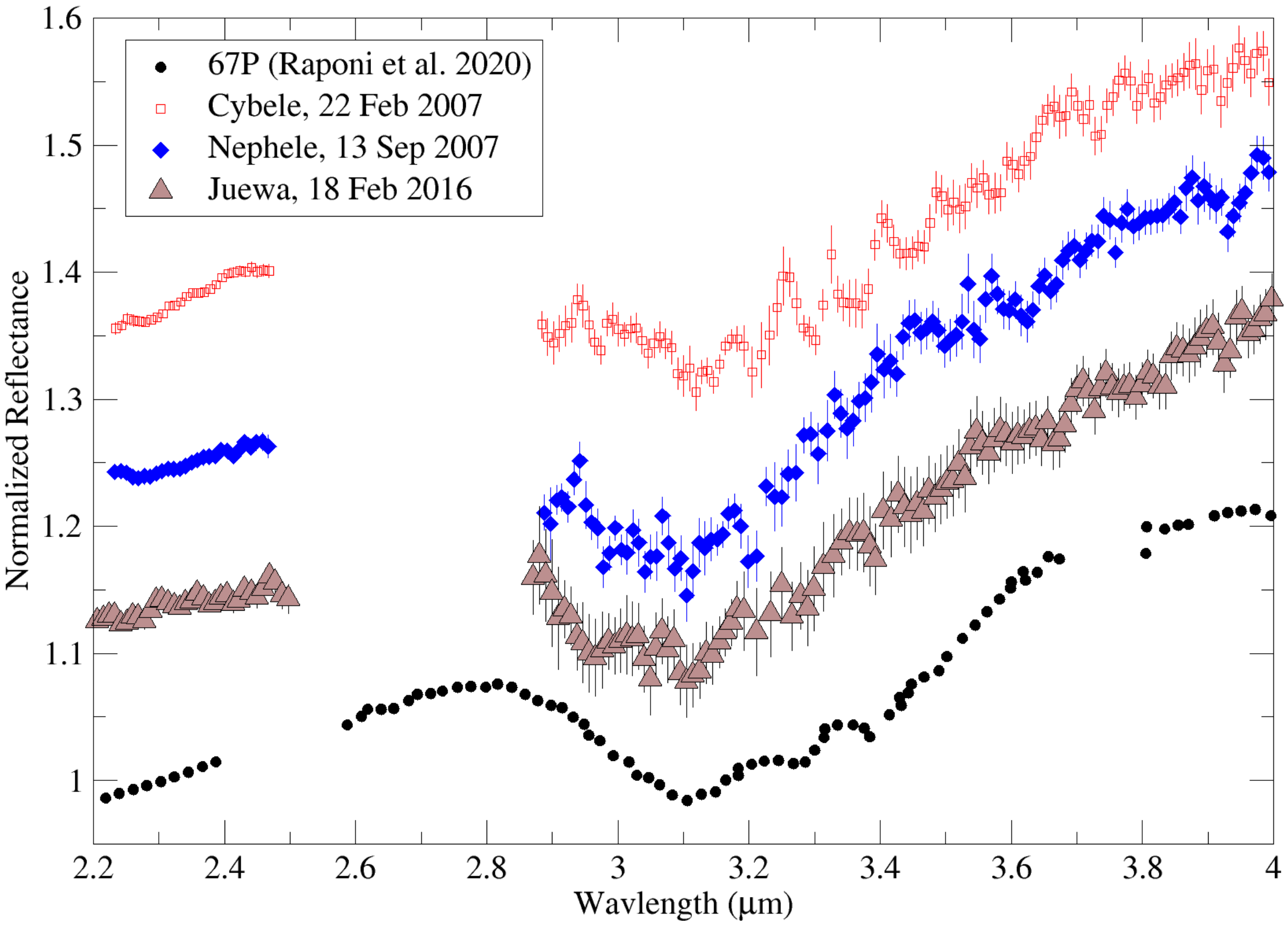

*Figure 4: Three NST asteroids compared to the Raponi et al. (2020) global average spectrum of comet 67P. The asteroids show an absorption band of similar depth and shape as the cometary spectrum, suggestive of similar compositions. The spectra are offset from one another for clarity.*

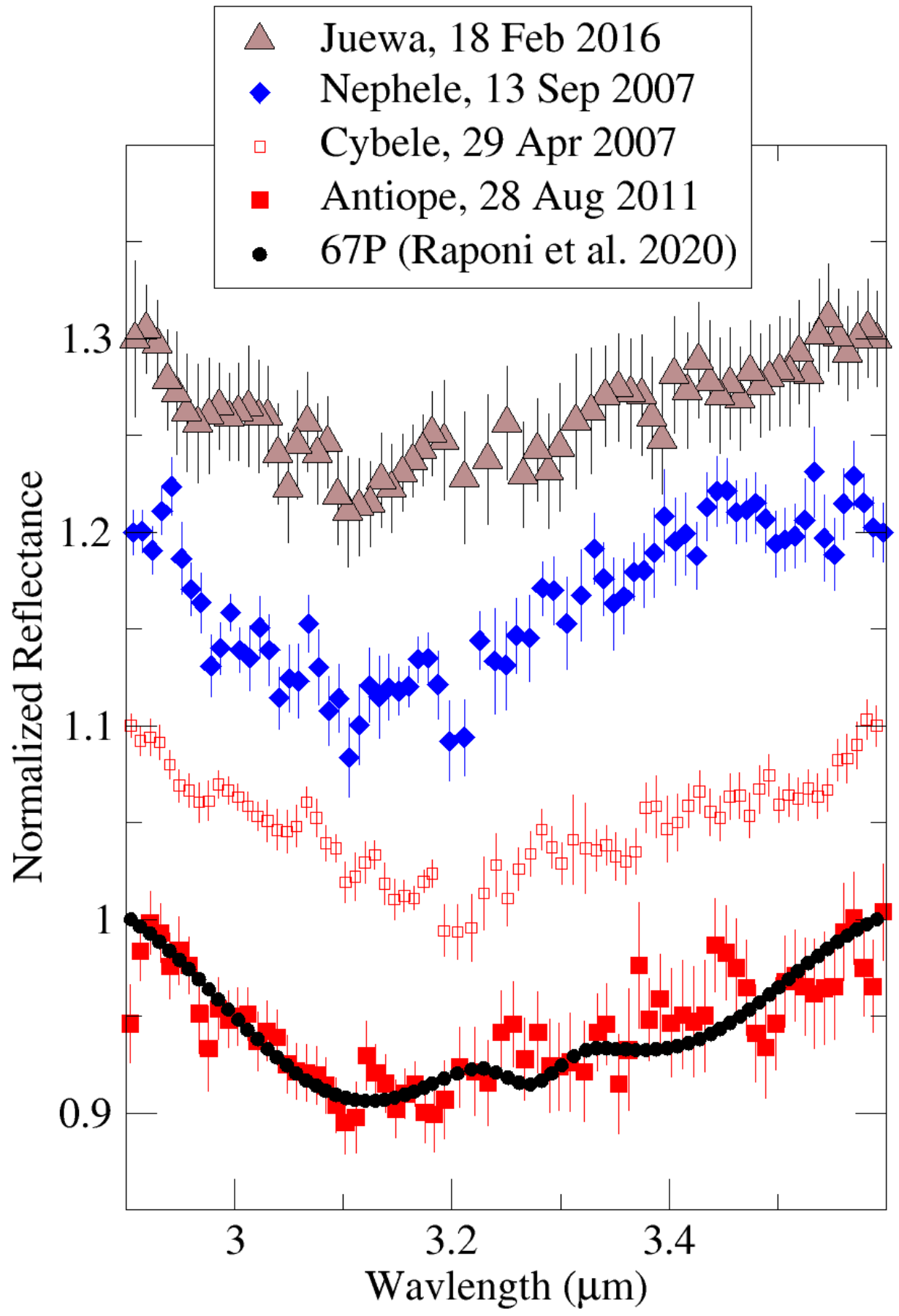

*Figure 5: Same as Figure 4, with a linear 2.9–3.6-μm continuum removed, and focusing on that wavelength region. Also added is Antiope, with its continuum adjusted to better fit 67P. Most of the NST spectra are qualitatively similar to 67P, though spectra with larger observational uncertainties*

We note that all of the asteroids in Tables 5 and 6 and Figures 3-4 are larger than 100 km, from different spectral complexes (for instance, Nephele is in the C complex, Juewa and Cybele in the X complex) and regions (Juewa in the middle asteroid belt, Nephele in the outer asteroid belt, and Cybele orbits beyond the asteroid belt). The similarity between the spectra and absorption features is evident by inspection as well as the calculations shown in Tables 5 and 6. Given this spectral similarity, we can posit that the hydrated compositions of these objects are also similar despite the different spectral slopes shortward of 2.5 μm. Raponi et al. (2020) attributes the 67P spectrum to a mixture of very fine grained ( < 1 μm) water-ice, ammoniated minerals, and organic materials, all of which have been detected or proposed on asteroid surfaces (Rivkin and Emery 2010, De Santis et al. 2012, among others). Nevertheless, the similarity is striking given the large size difference between 67P and these asteroids as well as the different lengths of time they are thought to have spent in the inner solar system. We do note that the semi-major axis of 67P is similar to that of the Cybele asteroids, though its eccentricity leads to an aphelion beyond Jupiter and a perihelion that technically classifies it as a Near-Earth Object. We also note that whatever is causing the spectral similarity can persist at relatively high temperatures and/or is replenished through an object's orbit: The volatility of the material creating NST spectra on asteroids and 67P is an important question, which we cannot yet answer but return to several times in the remainder of the paper.

*Table 5: Spectra that match average 67P spectrum most closely in median value of point-to-point differences.*

| Rank | Med val diff | spectrum |
|---|---|---|
| 1 | 0.012609 | Nephele, 13 Sep 2007 |
| 2 | 0.013081 | Themis, 7 Jan 2008 |
| 3 | 0.013576 | Palma, 22 Sep 2005 |
| 4 | 0.013802 | Bertha, 28 Feb 2003 |
| 5 | 0.0140985 | Cybele, 29 Apr 2007 |
| 6 | 0.0144035 | Juewa, 18 Feb 2016 |
| 7 | 0.01502 | Cybele, 20 Mar 2007 |
| 8 | 0.016828 | Erminia, 11 Oct 2005 |
| 9 | 0.017835 | Euphrosyne, 21 Sep 2005 |
| 10 | 0.020416 | Bamberga, 3 Jul 2012 |

The relationship between the subset of NSTs that are closer matches to 67P spectrally and the rest of the NSTs is not obvious. In some cases, poorer matches may be a function of lower-quality data masking the similarity. Interestingly, there are many more asteroids that appear to provide a good match to 67P near 3.1 µm, but a poorer match near 3.3 µm where carbon-bearing minerals have absorptions. While it is tempting to wonder if there is a connection to cometary nuclei that are depleted vs. non-depleted in particular carbon species (Cochran et al. 2012, 2020), there is far too little data to support anything beyond speculation. We discuss further implications of asteroids and 67P having similar spectra in Sections 6-7.

*Table 6: Spectra that match average 67P spectrum most closely in the point-to-point differences, squared and then summed.*

| Rank | Sum of sq diff | spectrum |
|---|---|---|
| 1 | 0.0138044 | Cybele, 29 Apr 2007 |
| 2 | 0.0144632 | Juewa, 18 Feb 2016 |
| 3 | 0.0176547 | Nephele, 13 Sep 2007 |
| 4 | 0.0195879 | Themis, 7 Jan 2008 |
| 5 | 0.0229714 | Bertha, 28 Feb 2003 |
| 6 | 0.0253854 | Cybele, 20 Mar 2007 |
| 7 | 0.0263807 | Bamberga, 3 Jul 2012 |
| 8 | 0.0288186 | Euphrosyne, 21 Sep 2005 |
| 9 | 0.0306076 | Palma, 22 Sep 2005 |
| 10 | 0.0332505 | Europa, 21 Jul 2013 |

## 4.5. Ryugu and Bennu

There are relatively few Near-Earth Objects (NEOs) that have been observed in the 3-µm region. Their very different thermal histories compared to typical main-belt asteroids and the very large thermal fluxes to be removed from their spectra makes a thorough comparison beyond the scope of this paper. The object (3200) Phaethon, for instance, has been reported to have no 3-µm band (Takir et al. 2020), but this may be a function of its very low perihelion and high surface temperatures rather than a primordial trait.

We will note that there is an intriguing similarity between the average NIRS3 spectrum of (162713) Ryugu (Kitazato et al. 2019, Yada et al. 2021) and NST asteroids. Figure 6a shows this similarity, with data in the 2.5—2.85-µm region removed for Ryugu to simulate how it might appear to Earth-based telescopes. Pilorget et al. (2021) presented reflectance spectra of the samples that Hayabusa2 returned from Ryugu, which show similar but muted absorptions at wavelengths > 2.8 µm compared to the Kitazato et al. spectrum. There is at least one inner-belt family associated with an NST parent (Svea) from which NEOs with NST spectra could originate.

The most prominent inner-belt, low-albedo family is the Polana family, which is often identified as a possible source for (101955) Bennu and/or Ryugu among other low-albedo NEOs (Campins et al. 2010a, 2013; Bottke et al. 2015, de León et al. 2018). The parent body (142) Polana was observed in this program but it has an uncertain classification from its 3-µm spectrum: its visual classification is ambiguous. The Gaussian fits described in Section 5 result in a band center of 2.98 ± 0.04 µm but its SNR < 20 in the 2.9–3.4-µm spectral range. Figure 6b compares the average OVIRS spectrum of Bennu (Zou et al. 2021) to Polana, with a linear continuum removed from the Polana spectrum. While improved spectra of Polana should determine whether it is ST, NST, or NB, the data currently available suggests that Bennu and Polana are consistent with one another in this wavelength region.

On its face, the spectral evidence that Bennu and Ryugu are spectrally most consistent with ST and NST objects, respectively, can be used to make predictions for the sample investigations. We certainly might expect geochemical evidence that Bennu and Ryugu came from different parent bodies. The fact that both have qualitatively different 3-µm spectra despite being similar in size and being delivered from roughly the same part of the asteroid belt is consistent with exogenic material being a minor component in global-scale spectra. If conversely, Bennu and Ryugu are thought likely to have originated as part of the same object after sample analysis, many of the scenarios considered in Section 6 will need to be revised or discarded. We note that Tatsumi et al. (2021) cannot rule out from geophysical data that Bennu and Ryugu could represent the same parent body, but based on space weathering trends as well as difference in 2.7-µm band depth they suggest that is a “lower possibility”.

## 4.6. Similarity of NST spectra to irradiated ice residues:

The absorptions seen on objects in the sample are attributed to phyllosilicates, ammoniated minerals, carbonates, organic materials, and water-ice. The phyllosilicates seen in ST (and some NST) spectra are thought to have been formed through parent-body aqueous alteration of anhydrous silicates (Krot et al. 2015). Recent work by Urso et al. (2020) suggests that the materials found on NSTs can be generated through irradiation of frozen water, methanol, and ammonia mixtures. Figure 7 compares the spectra of 67P from Raponi et al. (2020), Cybele, and the digitized spectra of two residues from Urso et al. formed by such irradiation and subsequent heating to 300 K. The residue spectra were originally presented in optical depth, and are converted to transmission and increased in contrast to more easily show their shapes. The matches between the residues, Cybele, and 67P are shown in Figure 7, with residue #2 ($H_2O$:$CH_3OH$:$NH_3$ at 3:1:1) matching the band centers near 3.12 µm and residue #3 ($H_2O$:$CH_3OH$:$NH_3$ at 1:1:1) better matching the shape at longer wavelengths.

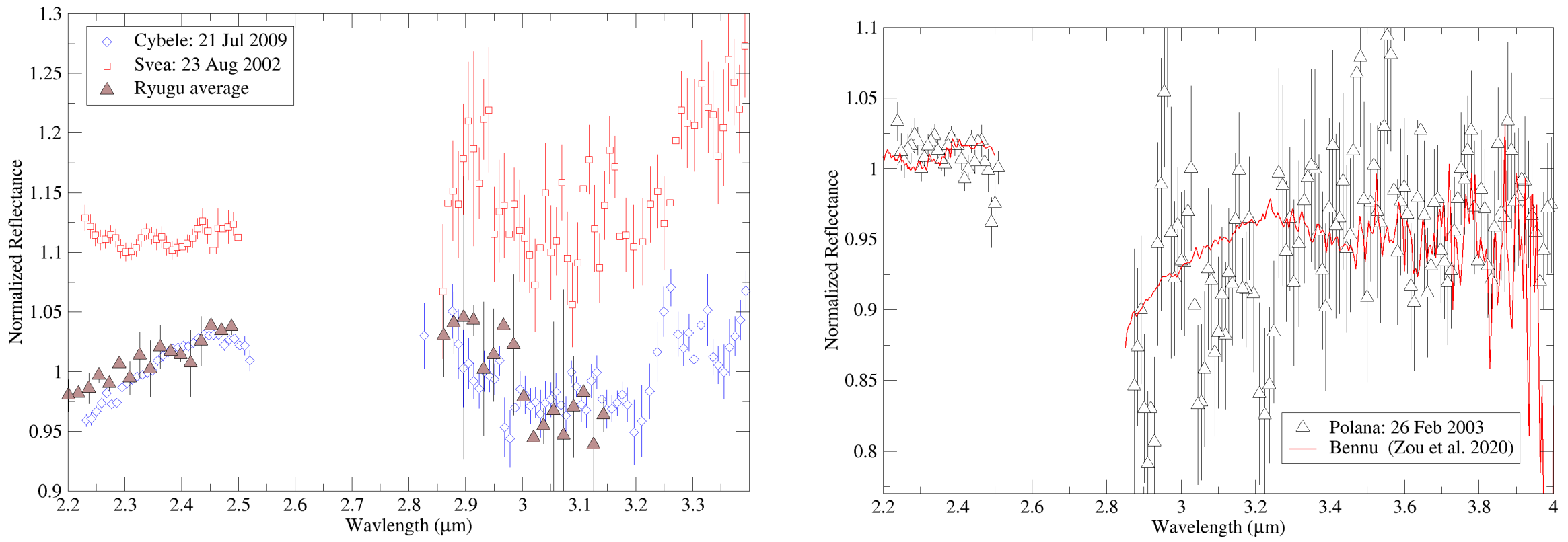

*Figure 6: (left) Two NSTs and Ryugu. The spectrum of Ryugu is from Kitazato et al. (2019), provided by Yata et al. (2021). All three of these objects share similar NST reflectance spectra despite their range in semimajor axis (1.18 AU, 2.47 AU, and 3.43 AU for Ryugu, Svea, and Cybele, respectively), and size (1, 81, and 237 km in the same order). Svea is offset from Cybele and Ryugu for clarity. (right) Polana and Bennu. While the available spectrum of Polana, a putative parent body for low-albedo NEOs, is relatively low-quality, it is consistent with the average spectrum of Bennu from Zou et al. (2021).*

### 4.7. Objects with no band

Fifteen interpretable objects are left unclassified as either NST or ST, because inspection and/or the Gaussian fits described in Section 5 find no absorptions larger than observational uncertainties. We note that in a formal sense, there is a possibility that higher-quality data with smaller uncertainties could detect a weak absorption, but the objects classified as "no band" (NB) have small enough uncertainties that such a hidden absorption would be less than roughly 4-5%.

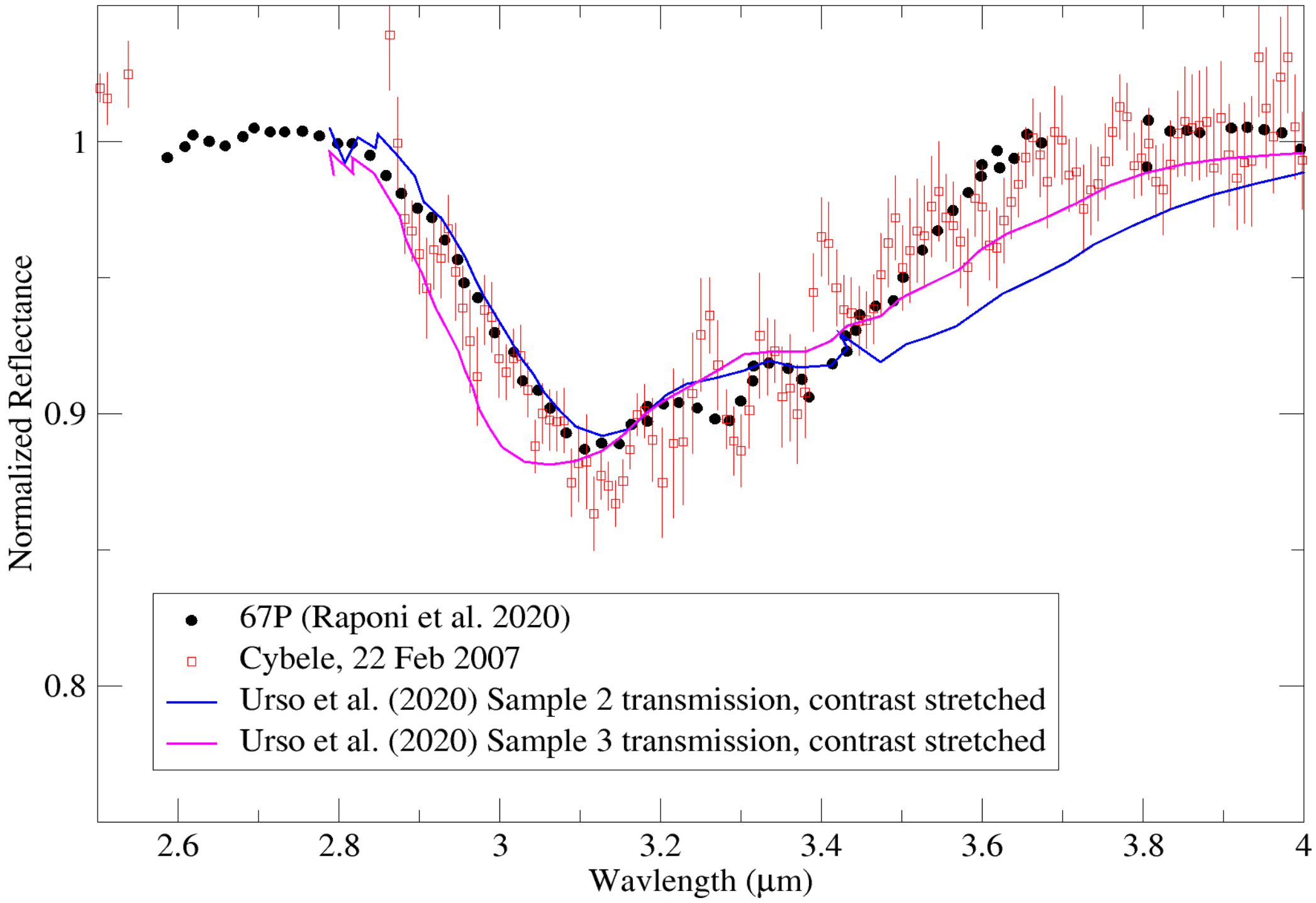

*Figure 7: 67P, Cybele, and two residues of irradiated ice mixtures from Urso et al. (2020). The Urso et al. experiments involved mixtures of different amounts of water ice, methanol, and ammonia, irradiated and then heated to 300 K. The spectra of the residues were presented by Urso et al. in optical depth, they have been converted to transmission and their contrast has been stretched for this plot. Cybele, 67P, and the residues have qualitatively similar spectra in this wavelength region.*

Figure 8 shows five representative objects interpreted as having no band. These five objects range in diameter from 21 km (267 Tirza) to 160 km (596 Scheila) with semi-major axes ranging from 2.40 AU (463 Lola) to 3.12 AU (212 Medea). As with the ST and NST groups, the objects interpreted as showing no band belong to a range of Bus taxonomic classes including those in the C and X complexes, and the D and T classes. However, the majority of them are classified in the X complex. Whether these objects have devolatilized surfaces but are otherwise primitive objects or represent igneous compositions that are coincidentally low-albedo and featureless (like Phobos and Deimos if they are in fact derived from Mars as proposed in recent formation scenarios like Canup and Salmon 2018) is not obvious.

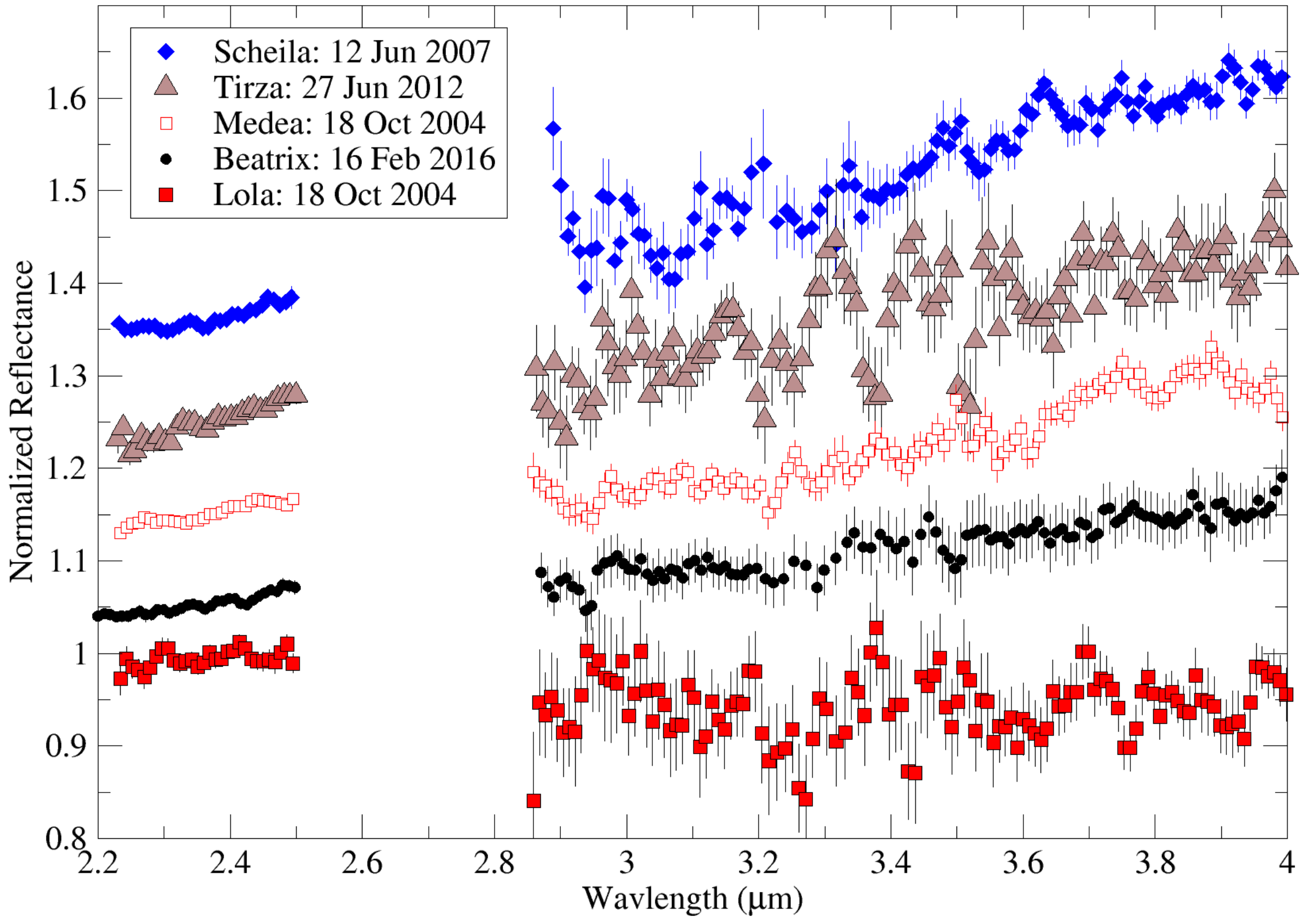

*Figure 8: The four objects presented here are interpreted as having no band (NB). They cover a range of sizes, solar distances, and spectral types, but NB spectra are rare in the spectral sample. These spectra are offset from one another for clarity.*

## 5. Quantitative fitting of band shapes

Following Rivkin et al. (2019), we fit the reflectance in the 3-µm region of the asteroids in the sample (R(*λ*)) with a linear continuum and a single Gaussian with absorption band center (BC), band depth (BD) and band width (BW):

*(Eq 1)* $R(\lambda) = BD \times \exp(-\ln(2)\times((\lambda-BC)/BW)^2)$

We do not propose that either ST or NST spectra are necessarily well fit by a single Gaussian for detailed compositional estimates, but this approach is sufficiently precise for our purposes to determine the absorption band center and band depth. The BC is used to classify objects in the sample as ST or NST below, but detailed compositional analysis would require a more sophisticated approach that is beyond the scope of this paper. Rivkin et al. (2019) found BC estimates for Gaussian fits to NST spectra generally agreed with 6th order polynomial fits to roughly ±0.02 µm, which we adopt as the uncertainty unless a larger uncertainty is returned by the fit itself. The results of the fits for each newly-presented spectrum, along with their formal uncertainties from the fits, are included in Table A3. We note for comparison with values in Table A3 that the Raponi et al. (2020) spectrum of 67P discussed several times in this paper has a BD of 0.115 and a BC of 3.217 µm if data from 2.5–2.85 µm is removed and the band fit as a single Gaussian as the asteroids are.

We also note again that the atmosphere precludes observations between roughly 2.5—2.85 µm, so those wavelengths, which contain the band center and minimum, were removed from the fit. As a result, the fitted results for ST objects are biased to have shallower band depths than would be found if all wavelengths are available, since the fits tended to set the Gaussian amplitude from the actual data rather than extrapolate it into the unobserved region. For the same reason, band centers tended to be fit at wavelengths longer than what we would expect if all wavelengths were available. Despite these biases, which would tend to blur the differences between the ST and NST groups, the following sections show that they are still statistically distinct in many important ways.

Of the 191 spectra, 154 have consistent classifications from both the visual and Gaussian BC classifications, including 78 ST, 61 NST, and 15 NB spectra. Of the remaining spectra, 7 were unclassified visually and 10 spectra had Gaussian BC fits with uncertainties that crossed 3 µm. The remaining 20 measurements include some where the bias toward longer wavelength in the fits pulled the BC to just longward of 3.0 µm, those where lower-quality data reduced the confidence in the visual or Gaussian classifications (or both), those where possibly-discrepant data points near the atmospheric cutoff led to different interpretations in the visual vs. Gaussian classifications, or some combination of these factors.

We note that Bates et al. (2021) present a C chondrite spectrum with a reported BC at 3.02 ± 0.04 µm, which technically counts as NST. However, the measurement uncertainty causes the BC to straddle the NST/ST line. Furthermore, that meteorite (PCA 02010) is also reported to have suffered terrestrial weathering, which is thought to have affected its 3-µm absorption. Finally, inspection of the Bates et al. (2021) PCA 02010 spectrum shows a band minimum near 2.96 µm, with a secondary minimum at longer wavelengths likely responsible for the reported 3.02-µm BC.

Three of the 36 Ch asteroid spectra in Rivkin et al. (2015b) have fitted Gaussian band centers that would place them in the NST category: (207) Hedda (3.028 ± 0.04 µm), (576) Emanuela (3.045 ± 0.022 µm), and (602) Marianna (3.125 ± 0.023 µm). In the case of Hedda, the uncertainty on BC straddles the ST/NST line. In the case of all three objects, inspection shows a spectral shape consistent with ST spectra, with the BC > 3 µm result influenced by lower-quality data points. Nevertheless, in the spirit of being as objective as possible and not wanting to introduce bias based on our expectations, we use the Gaussian fit BC in the statistical discussions below or remove them from the sample, depending on the specific group being considered.

Figure 9 shows binned and uncertainty-weighted average spectra for the ST and NST observations identified from the Gaussian fits, and the average Gaussians calculated for those objects. As noted, Gaussians are imperfect models for these band shapes, particularly the ST spectra, and the true band centers for ST objects are expected to fall in the 2.5—2.85-µm data gap. The BC of the fits are at longer wavelengths than what is indicated by inspection of the average spectra by 0.03–0.05 µm, and the fits underestimate band depth by roughly 2-3%. Both the BC and BD mismatches are comparable to or smaller than the spread in those parameters as fit in the ST and NST populations (Table 10), and so are not surprising. Even keeping the mismatches in mind, Figure 9 demonstrates that the fits reflect the different characters of the ST and NST groups: In general, NST spectra are centered near 3.1 µm and have roughly 10% band depths, while ST spectra have bands that are deeper by roughly a factor of two that are centered at (by definition of the groups) shorter wavelengths.

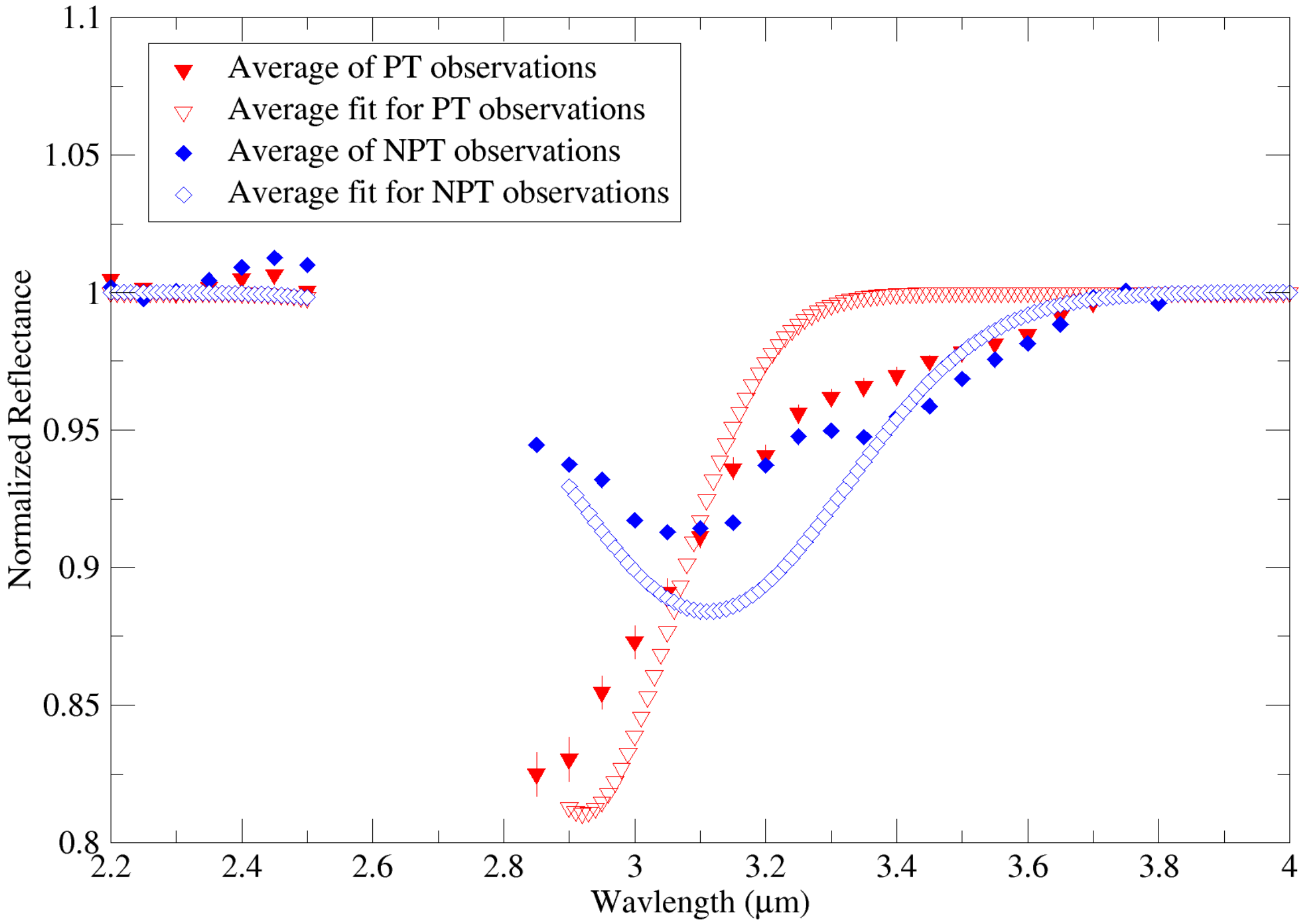

*Figure 9: Average NST and ST spectra, both as calculated from the average Gaussian parameters for those groups and as calculated from the actual spectra. While the Gaussian fits are not adequate for detailed compositional studies, they do allow the basic differences between groups to be studied.*

## 5.1. Variation on Objects

There are both scientific and pragmatic reasons for considering variation in the 3-µm region. Understanding whether a single object can exhibit both ST and NST spectra has important implications for the origin of the minerals that are responsible and the interpretation of how representative observed spectra may be. From a pragmatic point of view, averaging multiple spectra to improve data quality and/or reporting average values for spectral parameters may be misleading if NST and ST spectra are both included in the average, while a consistent spectral appearance increases confidence that such averages are valid. Using average spectral properties for objects rather than treating each observation separately also minimizes the undue numerical influence of objects with repeat visits (Section 3).

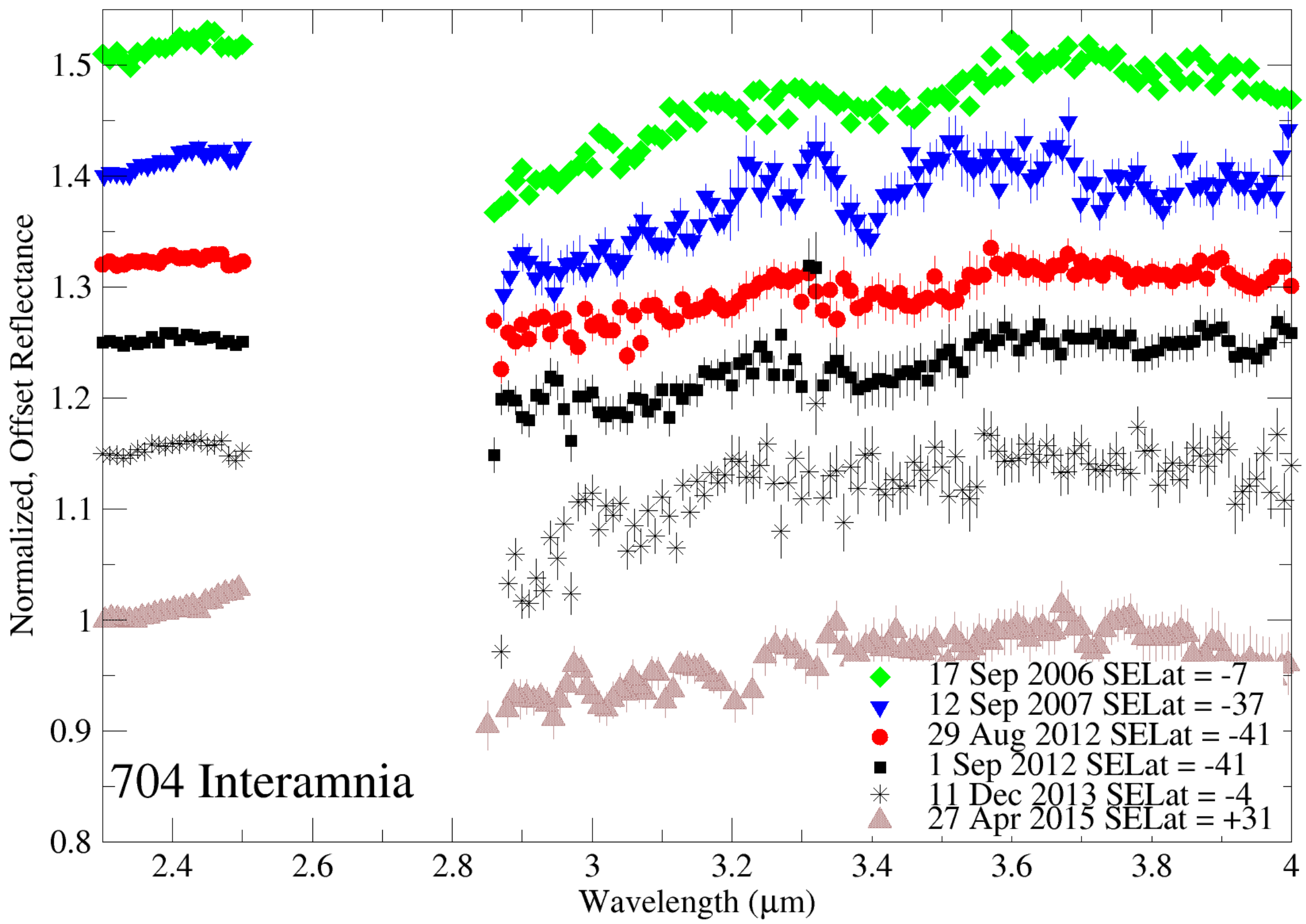

*Figure 10: Adapted from Rivkin et al. (2019). Observations of Interamnia, including a re-reduced 2007 spectrum and the new 2015 spectrum, show relatively consistent behavior at a global scale, though details differ including behavior near 3.3 µm and the specific location of the BC.*

A total of 45 objects were observed multiple times in this work, Rivkin et al. (2015b), and Rivkin et al. (2019). A total of five objects with new data exhibit both ST and NST spectra in Gaussian fits, but inspection shows all of the relevant spectra to be consistent with one another within observational uncertainties. Two objects from Rivkin et al. (2019), Hygiea and Interamnia, have some spectra classified as ST and some classified as NST. Hygiea has a spectrum very much like Ceres, with a band minimum near 3.06 µm and a local reflectance maximum near 2.90—2.92 µm, with reflectance decreasing with decreasing wavelength into the atmospheric opaque region. Depending on data quality, some fits converge such that the decrease into the opaque region is identified as the minimum rather than the feature near 3.06 µm. There may be variation on the surface of Hygiea, but it does not appear to include both ST and NST spectra at global scales.

Interamnia (Figure 10) is potentially a more interesting case. The work of Rivkin et al. (2019) presented the strongest evidence we have that an object can exhibit both ST and NST spectra. However, upon inspection the differences in its spectra appear subtle, especially when updating the 2007 spectrum in Rivkin et al. (2019) for the newly-reduced one (Section 3). While the Gaussian fits do show that some of these spectra have BC > 3 µm, and there are hints of structure near 3.3 µm in some of the spectra in Figure 10, the current evidence for large-scale variation on Interamnia's surface is considerably weaker than was previously thought.

## 5.2. Distribution of NST and ST objects

Figure 11 shows the cumulative distributions of the fitted BC for the spectra in the sample, grouped by location: objects in the inner asteroid belt (semi-major axis between 2.0—2.5 AU), the mid asteroid belt (2.5—2.82 AU), outer asteroid belt (2.82—3.3 AU), and Cybele and Hilda regions (3.3—4.5 AU). The different distributions of band centers are obvious from the plot—with higher concentrations of ST objects in the inner and middle belt and more NST than ST objects in the outer belt and Cybele-Hilda regions. The distribution of ST vs. NST vs. NB objects for different parts of the main belt and beyond is tabulated in Table 7. Using Kuiper's variant of the Kolmogorov-Smirnov test, we find probabilities < 0.0004 that the samples were drawn from the same parent distribution except for inner being drawn from mid (0.01).

*Table 7: ST vs. NST distribution vs. location, based on band centers from Gaussian fits. Note: 4 objects from Rivkin et al. (2015b), two in the middle belt and two in the outer belt, could not be fit successfully with a Gaussian and are omitted from the numbers below.*

| Population | ST | NST | NB | Total |
|---|---|---|---|---|
| Inner belt | 20 | 3 | 3 | 26 |
| Middle belt | 40 | 16 | 3 | 59 |
| Outer belt | 18 | 32 | 3 | 53 |
| Cybele-Hilda | 4 | 11 | 2 | 17 |
| Full sample | 82 | 62 | 11 | 155 |

Figure 12 shows the distributions split by taxonomic class, with mean values presented in Table 8. As expected, nearly all of the Ch/Cgh observations have BC < 3.0 µm, with BC fits for only a few stray objects at longer wavelengths. Among the other groupings, the X complex has a significant fraction of objects with BC shortward of 3 µm but a long tail at longer wavelengths, the B/Cb group is dominated instead by by a rise in the number of objects with BC near 3.1 µm, and there is a relatively consistent fraction across the BC wavelength range for the C/Cg grouping (which also includes objects classified as C or F in the Tholen taxonomy for which no Bus classification is available). Again, it is clear that while some weak correlations do exist, the taxonomic class of a low-albedo asteroid as defined from visible-near IR data does not strongly correlate to a BC unless it is in the Ch or Cgh class. The Kolmogorov-Smirnov tests return probabilities of <0.0005 that any of these distributions were drawn from the same parent distribution as one of the other taxonomic groups.

## 5.3. Largest Objects:

The number of asteroids quickly rises as one looks to smaller and smaller sizes. There are 19 low-albedo, main-belt asteroids with diameters > 200 km. That number doubles by dropping the diameter limit to roughly 160 km, doubles again by 130 km, and there are roughly 150 low-albedo, main-belt asteroids with diameters > 100 km. As the number of objects increases, it becomes more difficult to determine whether a representative sample is being studied.

*Table 8: Average spectral properties of objects in the sample defined by Bus class. NB objects are included in the 2.9- and 3.2-µm BD averages, but omitted from the BD and BC averages.*

| Group | N | Mean 2.9 BD | Mean 3.2 BD | Mean BD | Mean BC |
|---|---|---|---|---|---|
| B/Cb | 25 | 0.103 ± 0.012 | 0.077 ± 0.005 | 0.139 ± 0.011 | 3.054 ± 0.009 |
| C/Cg | 43 | 0.157 ± 0.012 | 0.087 ± 0.005 | 0.145 ± 0.005 | 3.015 ± 0.009 |
| Ch/Cgh | 40 | 0.209 ± 0.01 | 0.079 ± 0.004 | 0.234 ± 0.012 | 2.934 ± 0.006 |
| D/T | 10 | 0.066 ± 0.009 | 0.038 ± 0.005 | 0.134 ± 0.006 | 2.946 ± 0.006 |
| X/Xk | 31 | 0.113 ± 0.010 | 0.073 ± 0.004 | 0.143 ± 0.008 | 3.015 ± 0.012 |
| Xc Class | 8 | 0.134 ± 0.018 | 0.071 ± 0.004 | 0.111 ± 0.003 | 3.078 ± 0.015 |

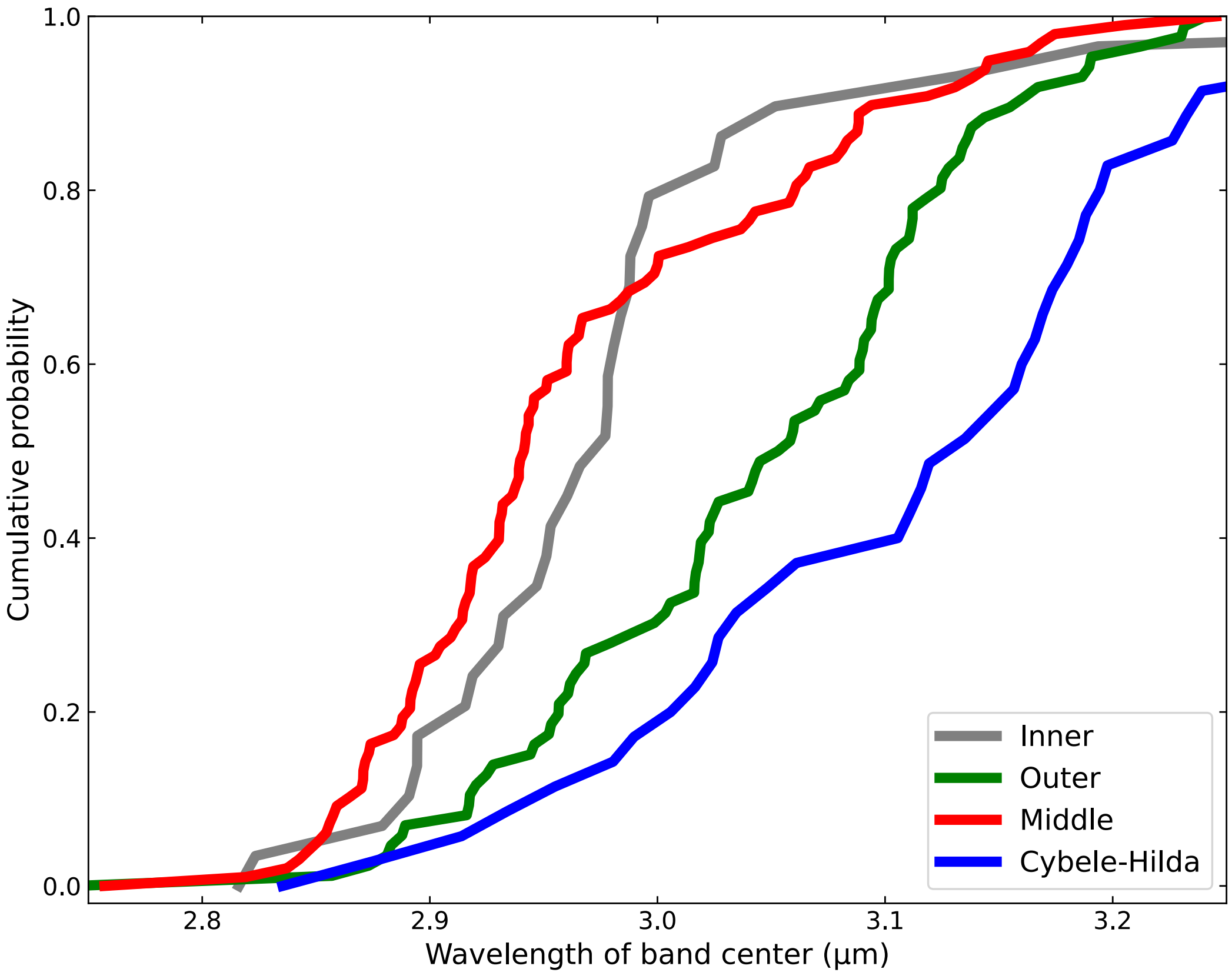

*Figure 11: Cumulative distributions of BC for different asteroidal locations. Steep increases shortward of 3 µm in the inner and middle belt show ST objects dominate there, though NST objects are also present. The bulk of objects have BC > 3 µm in the outer belt and Cybele/Hilda regions, indicating that NST objects are much more common.*

While our sample covers a broader region than the main asteroid belt, and includes several objects with diameters 50 km or smaller, we focus in this section on the 50 and 100 largest objects that orbit in the main asteroid belt (Table 9), with albedos equal to or lower than 11%, as listed by the JPL Small-body Search Engine. The objects in this sample are large enough (diameters of roughly 120 km or larger) that we can expect them to be independently-formed parent bodies, rather than suffering from biases caused by including many family members (Morbidelli et al. 2009). In addition, this minimizes the issue of representativity by allowing a definite closed set of objects to be defined and studied, albeit at the potential cost of an arbitrary size cutoff.

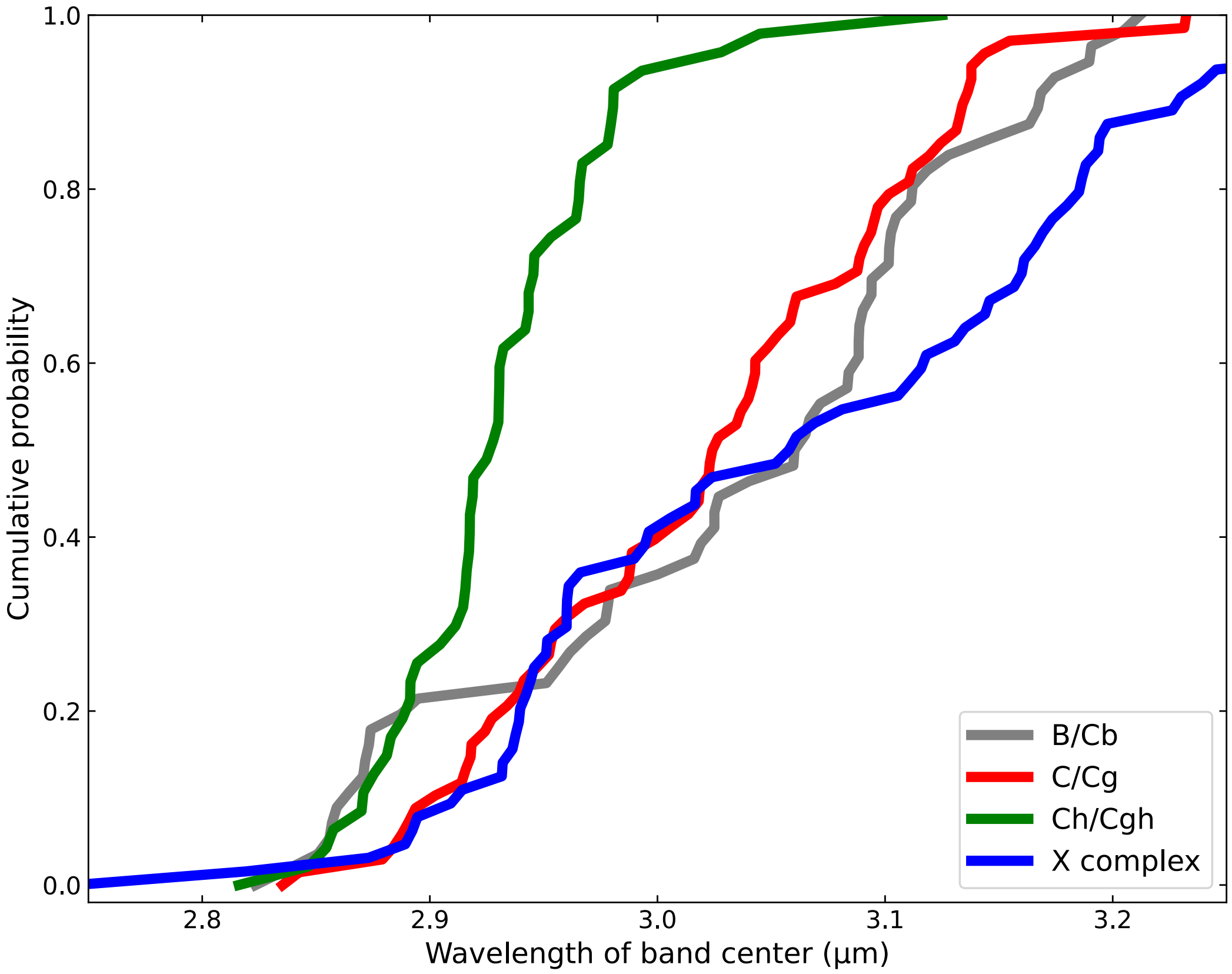

*Figure 12: Cumulative distributions of BC for different Bus taxonomic classes. As shown by Rivkin et al. (2015b), the Ch/Cgh class is dominated by ST objects. However, all of the other groupings have significant NST populations, with the majority of the objects being > 3 µm.*

Table 9 shows the fraction of the 50 and 100-largest low-albedo main-belt asteroids classified as ST and NST based on their average BC, along with a category for objects not seen to have a band. We also include a category for objects that were observed at 3 µm and published in other datasets (Rivkin et al. 2003, Takir et al. 2015, Usui et al. 2019) that can be classified as ST by inspection, and a column for objects that do not have 3-µm measurements but are in the Ch or Cgh spectral classes and are assumed to be ST. Finally, the number of non-Ch/Cgh objects without 3-µm data is listed. No NST objects have been identified in the literature that do not also appear in Rivkin et al. (2019) and/or this work.

*Table 9: Distribution of 3-µm types among largest low-albedo asteroids*

| Group | Minimum diameter | ST | NST | NB | ST from other sources | Ch/Cgh objects, no 3-µm data | No 3-µm data, not Ch/Cgh asteroid |
|---|---|---|---|---|---|---|---|
| 50 largest Main belt | 152 km | 23 | 22 | 1 | 3 | 1 | 0 |
| 100 largest Main belt | 122 km | 36 | 34 | 3 | 4 | 8 | 15 |

The set of the 50-largest low-albedo, main-belt asteroids is nearly complete in terms of 3-µm observations. The only asteroid among the 50 largest without 3-µm observations, (146) Lucina, is a Ch-class asteroid and presumed to be a ST on that basis. As can be seen, these two groups of large objects are roughly evenly split between the NST and ST groups. We note that most of the large ST objects are in the Ch or Cgh class: of the 40 ST objects among the 100 largest main-belt low-albedo asteroids, 22 are Ch or Cgh, and 12 of the 25 ST objects in the group of 50 largest are Ch or Cgh. Recasting these numbers, if a low-albedo object is *lacking* a 0.7-µm band (i.e. it is *not* a Ch or Cgh asteroid), it is much more likely to be an NST than a ST (22 vs. 13 in group of the largest 50, 34 vs. 18 in largest 100).

It is clear that the compositions associated with the NST objects are very common ones in the main asteroid belt, assuming their surfaces are representative of their volumes. This assumption can be tested in detail via spectroscopic measurements of small members of NST families (like Themis, Hygiea, Euphrosyne, etc.). Section 6.3 discusses further the benefit of observing members of asteroid families. Given the large fraction of NST objects among the largest asteroids, we would expect them to contribute to the meteorite collection roughly on par with ST parent bodies unless the physical properties of NSTs and STs differ to the point of affecting collisional or dynamical outcomes. Furthermore, as discussed in Section 4.4 and Section 6, not only are NST compositions common in the main asteroid belt, but they are also found beyond the main asteroid belt in the Cybele region and are consistent with at least one comet, suggesting that they may be common on primitive body surfaces throughout the outer solar system as well. It is also worth noting that NB objects are rare: only (596) Scheila is NB and in the largest 50, a particularly interesting fact given it is included among the list of activated asteroids (Jewitt 2012), though its activity is typically classified as of impact origin rather than sublimation-driven (Jewitt 2012). We also note that Hasegawa et al. (2022) showed that the 3-µm spectrum of Scheila did not change as a result of its activity, though its shorter-wavelength spectral slope did change.

Taking the split of 3-µm types in the population of largest objects at face value, we find that a low-albedo object that is not a small member of a family and that lacks a 0.7-µm band (so, is not classified as a Ch or Cgh asteroid) has a roughly ⅔ chance of being an NST. At sizes where the asteroid population becomes dominated by family members these fractional estimates will change, of course, but it suggests that the original population of planetesimals were roughly split between those whose spectra are dominated by a single absorption band centered < 2.9 µm and those that have additional prominent absorption bands at wavelengths > 3 µm. In addition, it suggests that a sizeable fraction of the latter group have spectra (and presumably compositions) qualitatively consistent with cometary nuclei, and that the former group is itself split into roughly equal populations that show evidence for more- and less-iron-rich compositions (the Ch/Cgh and non-Ch/Cgh ST populations, respectively).

### 5.4. Statistical Tests:

To determine whether the properties of the various subpopulations in the sample of 3-µm measurements differ statistically, we use hypothesis testing for which the null hypothesis is that the mean values of the properties are the same. Because there are relatively few objects in each subpopulation, we use a two-tailed t-test. We assume that the samples are independent and have unknown but equal variances, which we calculate using the pooled variance. The resulting p-value, as listed in Table 11, is the significance level at which the null hypothesis can be rejected. P-values of 0.1, 0.05, or 0.01 indicate that the null hypothesis is rejected and the subpopulations are not statistically similar at confidence levels of 90%, 95%, and 99%, respectively.

We use average values for BC, BD, etc. to prevent objects observed multiple times from being weighted more than objects measured only once. Twenty-two "inconsistent" measurements (discussed in Section 5) were removed from the sample before the t tests were run. Because some of the objects with inconsistent measurements also had consistent measurements from different dates, only seventeen objects were removed from the sample as a result of the twenty-two inconsistent measurements. The few NB objects were also removed. Values for orbital and physical properties were taken from the JPL Horizons database, except for densities, which are taken from the SiMDA compilation database (Kretlow, 2020). Table 10 shows the average values for the parameters of interest for the various subpopulations that were considered.

We first consider the objects that are in the 100 largest, low-albedo, main-belt asteroids discussed in the previous section. The sample has 34 ST objects and 27 NSTs. By definition alone, we expect these groups to differ in band center (BC), and indeed the average BC for the ST group is near the point where the data cuts off due to the atmosphere, while the BC for the NST group is at a significantly longer wavelength rather than near the 3.0-µm dividing point between the groups (see Figure 9). The band depths are also different between the groups at a ≥99% confidence level, with ST objects on average having a larger band depth than NST objects. The varying quality of density measurements makes interpretations somewhat fraught, but there is no evidence of a density difference between them. No significant difference was found between the groups in terms of albedo.

The average orbits for the ST and NST groups also differ at a high confidence level: the ST group has a smaller average semi-major axis and perihelion at a 99% confidence level. However this may be driven by the Ch asteroid population specifically: the set of large Ch/Cgh asteroids also has a different semi-major axis and perihelion than large low-albedo asteroids in other taxonomic classes at a 99% confidence level, even when the latter group is a mixture of ST and NST objects.

*Table 10: Average properties (including Gaussian fits) of various subgroups.*

| Group | N | Avg a | Avg q | Avg density | Avg albedo | Avg period (h) | Avg Gauss BD | Avg Gauss BC | Avg Gauss width |
|---|---|---|---|---|---|---|---|---|---|
| MB ST > 120 km | 34 | 2.817 | 2.378 | 2.249 | 0.056 | 15.89 ± 14.15 | 0.207 ± 0.095 | 2.934 ± 0.033 | 0.204 ± 0.050 |
| MB NST > 120 km | 27 | 2.975 | 2.564 | 2.043 | 0.059 | 12.89 ± 10.28 | 0.115 ± 0.042 | 3.106 ± 0.060 | 0.242 ± 0.058 |
| Big MB ST, Ch/Cgh | 18 | 2.789 | 2.313 | 2.123 | 0.052 | 14.70 ± 9.28 | 0.251 ± 0.108 | 2.938 ± 0.028 | 0.201 ± 0.056 |
| Big MB ST, not Ch/Cgh | 16 | 2.847 | 2.452 | 2.289 | 0.062 | 17.30 ± 18.56 | 0.158 ± 0.041 | 2.931 ± 0.038 | 0.208 ± 0.042 |
| Small MB ST | 38 | 2.575 | 2.127 | 1.777 (N=14) | 0.055 | 14.32 ± 8.93 | 0.213 ± 0.140 | 2.913 ± 0.057 | 0.228 ± 0.063 |
| Small MB NST | 12 | 2.854 | 2.409 | 1.54 (N=4) | 0.057 | 17.18 ± 8.59 | 0.119 ± 0.032 | 3.078 ± 0.045 | 0.230 ± 0.078 |
| Cyb/Hil NST | 9 | 3.540 | 3.098 | 1.753 (N=4) | 0.053 | 7.691 ± 2.55 | 0.125 ± 0.032 | 3.119 ± 0.069 | 0.259 ± 0.048 |
| MB NST | 39 | 2.938 | 2.516 | 1.978 (N=31) | 0.058 | 14.21 ± 9.89 | 0.116 ± 0.040 | 3.097 ± 0.057 | 0.238 ± 0.065 |
| Cyb/Hil ST | 4 | 3.392 | 3.037 | 1.02 (N=3) | 0.173* | 17.79 ± 19.57 | 0.270 ± 0.132 | 2.941 ± 0.042 | 0.206 ± 0.064 |
| MB ST | 72 | 2.689 | 2.246 | 2.101 (N=49) | 0.056 | 15.13 ± 11.85 | 0.211 ± 0.119 | 2.922 ± 0.047 | 0.216 ± 0.058 |
| All ST | 76 | 2.726 | 2.288 | 2.039 (N=52) | 0.062 | 15.27 ± 12.19 | 0.214 ± 0.121 | 2.924 ± 0.047 | 0.215 ± 0.058 |
| All NST | 48 | 3.051 | 2.625 | 1.978 (N=35) | 0.057 | 12.99 ± 9.32 | 0.118 ± 0.038 | 3.101 ± 0.060 | 0.242 ± 0.063 |
| MB Ch/Cgh ST | 34 | 2.690 | 2.210 | 2.038 (N=24) | 0.052 | 14.51 ± 8.50 | 0.243 ± 0.124 | 2.921 ± 0.036 | 0.206 ± 0.049 |
| MB non-Ch/Cgh ST | 38 | 2.688 | 2.278 | 2.162 (N=25) | 0.059 | 15.69 ± 14.29 | 0.181 ± 0.107 | 2.924 ± 0.056 | 0.224 ± 0.064 |

While the NST and ST asteroids seem to have distinct orbital properties and band depth properties, we can also consider the ST asteroids specifically. The large ST objects are split roughly evenly between Ch/Cgh-class objects (18) and other classes (16). In comparison to the ST vs. NST averages, these two subgroups of ST objects do not have statistically significant differences in any of their parameters save BD, which is larger for the Ch/Cgh subgroup at the ≥99% confidence level. This band depth difference could reflect either different hydrated mineral compositions (whether in terms of different compositions leading to different band centers and thus different band depths at the atmospheric cutoff wavelength, different compositions leading to the presence/absence of the 0.7-µm band, or both), or different abundances of similar hydrated minerals resulting in different 0.7-µm band depths, with some 0.7-µm band depths too shallow to detect in available data.

***Table 11: Probability that subgroups are drawn from the same population. Parameters with >90% confidence are bolded***

| | Probability of null result (drawn from same population) | | | | | |
|---|---|---|---|---|---|---|
| Comparison groups | a | q | ρ | pv | BD | BC |
| Big MB ST vs. big MB NST | **0.007** | **0.026** | 0.416 | 0.632 | **0** | **0** |
| Big MB ST: Ch/Cgh vs. other classes | 0.502 | 0.183 | 0.786 | 0.148 | **0.004** | 0.535 |
| Sm MB ST vs. Sm MB NST | **0.001** | **0.001** | — | 0.755 | **0.024** | **0** |
| Cybele/Hilda NST vs. All MB NST | — | — | — | 0.434 | 0.573 | 0.340 |
| Cybele/Hilda NST vs. Big MB NST | — | — | — | 0.292 | 0.556 | 0.603 |
| Big MB ST vs. Small MB ST | **0.0001** | **0.003** | **0.085** | 0.771 | 0.607 | **0.082** |
| Big MB NST vs. Small MB NST | 0.142 | 0.147 | 0.191 | 0.849 | 0.781 | 0.160 |
| All MB ST: Ch/Cgh vs. other classes | 0.972 | 0.334 | 0.601 | 0.217 | **0.028** | 0.811 |
| All MB ST vs. All MB NST | **0** | **0** | 0.544 | 0.521 | **0** | **0** |

We also look at the smaller objects to see how they compare. There are 50 low-albedo, main-belt asteroids in the sample that fall outside the 100 largest, including 17 published in Rivkin et al. (2015b). The differences found between the large ST and NST asteroids are also seen when comparing smaller members of these groups. We also note that the average semimajor axis of the large and small objects differs by 0.1-0.2 AU, as does their average perihelion distance (Table 10). This distance difference is an observational bias, reflecting the relative difficulty of observing the smaller asteroids, which must be observed at closer distances to offset their smaller sizes.

## 5.5. Cybele and Hilda Asteroids:

The sample includes 38 measurements of 15 objects that orbit beyond the main belt, with 31 measurements of 11 objects in the Cybele region and 7 measurements of 4 objects in the Hilda region. Though fainter and generally with larger observational uncertainties than the main belt objects, these asteroids are also seen to include both ST and NST objects (Hargrove et al. 2012, Takir et al. 2012), although they are dominated by the latter. When applying the statistical tests from the previous section to the Cybeles and Hildas, we see that the NSTs in that region do not have statistically significant differences from main-belt NSTs in BC, BD, or albedo, and are consistent with being drawn from the same original population. We do note, however, that the numbers of Hilda and Cybele asteroids in the sample are rather small, especially when looking at subsets like ST or NST objects, and additional measurements are necessary to reliably compare the Cybele and Hilda asteroids to the main belt population.

# 6. Discussion

Here we pause to reiterate some of the major conclusions of the previous sections: 1) In a sample of 158 low-albedo asteroids, the vast majority have 3-µm spectral shapes qualitatively like either Pallas ("ST") or like Themis or Ceres ("NST", with most NST objects more similar to Themis rather than Ceres), with less than 10% showing no band. 2) Some of the NST spectra are very good matches to an average spectrum of comet 67P. 3) The ST and NST groups are similar in albedo, and VIS-NIR taxonomic classifications do not predict whether an object will have a ST or NST band at 3 µm unless that object is in the Ch or Cgh classes. 4) The NST objects appear to orbit further from the Sun than the ST objects on average, though there is significant overlap in their spatial distributions. 6) No compelling evidence exists that objects host large regions with both ST and NST spectra on their surfaces.

What does all of this mean, and what scenarios emerge from explaining it all?

## 6.1. NST Asteroids Beyond the Hilda Region

We see that NST objects share several common characteristics and are widespread in the main belt, Cybele, and Hilda regions, though more commonly found in the outer belt than the inner belt. We can also find objects with those same characteristics that exist outside of these regions: Brown and Rhoden (2014) presented a spectrum of J6 Himalia, the largest of Jupiter's irregular satellites, finding it a very close visual match to the Takir and Emery (2012) spectrum of the asteroid (52) Europa. Takir and Emery report that spectrum as having a BC of 3.15 ± 0.01 µm, placing it well within the NST category. Studies of Jupiter Trojan asteroids have been limited by their faintness. Emery et al. (2011) found the Trojans can be separated into redder- and less-red groups based on their 0.5—2.5-µm spectra, and Brown (2016) found that an average spectrum created from the redder group showed no band while the average spectrum of the less-red group is similar to Themis. We performed an analysis of a digitized version of this latter spectrum, finding a BC of 3.08 ± 0.02 µm and a BD of 12-15% depending on continuum choice, again well inside the range seen for NST objects. We would expect from this that the Lucy mission may encounter one or more NST objects on its tour, which via the L'Ralph instrument (Olkin et al. 2021) would provide valuable comparisons to Rosetta measurements of 67P and important context for main-belt NST objects.

Further from the Sun than the Jupiter Trojans, the difficulty of obtaining 3-µm data for 100–200-km scale transneptunian objects (TNOs) has precluded direct comparison to asteroids, and the stability of ices at those distances vs. asteroidal distances would also complicate such comparisons. Nevertheless, there are intriguing relevant observations: the TNO (120216) 2004 EW95 was observed to have a 0.7-µm band by Seccull et al. (2018), which would suggest it should also have a ST spectrum in the 3-µm region. At 291 km and with an albedo of 4.4%, 2004 EW95 is similar in size to (511) Davida, though larger than any Ch asteroid in the main asteroid belt. Fernandez-Valenzuela (2021) compiled TNO observations at wavelengths > 2.2 µm, finding 6 objects (including 2004 EW95) with diameters < 400 km to have colors consistent with silicate-dominated surfaces, along with the Ceres-sized (120347) Salacia. With the launch of JWST it will be possible to investigate whether these objects have 3-µm spectra like the ST and NST spectra seen in the main-belt, Cybele, and Hilda regions.

## 6.2. Do objects have both ST and NST spectra on their surfaces?

As noted in Section 5.1, there is scant but non-zero evidence for low-albedo objects to show both ST and NST spectra on their surfaces, at least at the large scales that are detectable from Earth. The best case for such an object is Interamnia, a large body that has been interpreted as similar to Ceres and Hygiea in terms of density and lack of surface craters (Hanuš et al. 2020). However, as shown in Figure 10, the case for variation on Interamnia is far from obvious.

Some spatially-resolved data exists from Dawn's visit to Ceres and the OSIRIS-REx encounter with Bennu. Despite the very large size differences between these objects, they were both found to be rather homogeneous in spectral properties. Variation was seen on the NST object Ceres, notably involving the famous faculae, but no areas with ST spectra have been reported (De Sanctis et al. 2019). Similarly, Bennu, which would be classified as a ST object, does not appear to host any areas with NST spectra according to the literature (Barucci et al. 2020).

We do note, however, that measurements by Tatsumi et al. (2021) point to the possibility that Ryugu has both ST and NST material: the boulder Otohime Saxum is reported to have a deeper 0.7-µm band on one facet than on other facets or on Ryugu as a whole, and interestingly, NIRS3 spectra of that same facet show a more ST-like band shape near 3 µm than is seen in other Ryugu spectra. However, interpretations are confounded by the relatively large absolute uncertainties on absolute 0.7-µm band depth in the Hayabusa2 data. Because both Itokawa and Bennu have been observed to have large boulders interpreted as exogenic material (Saito et al. 2006, Le Corre et al. 2021) it is also possible that Otohime does not represent material native to Ryugu, though Tatsumi et al. argue that space weathering and heating variations could cause the differences seen on Ryugu.

## 6.3. Families and 3-µm Groups

Asteroid families are defined both by their dynamical similarities and spectral similarities. Interlopers with coincidentally-similar orbits are identified by albedos or spectral types that are inconsistent with the main group. While it is not known whether 3-µm band shapes can similarly serve as markers of family membership, we can still begin the investigation based on associated data. First, we note that while 3-µm data is not available for more than a handful of dynamical family members, there have been numerous studies at shorter wavelengths. The connection made between Ch asteroids and ST band shapes (Rivkin et al. 2015) allows us to gain some (admittedly imperfect) insight into the distribution of ST and NST objects within families.

De Prá et al. (2021) and Morate et al. (2019) compile results from studies of outer-belt and inner-belt families, respectively, including the relative homogeneity of spectral types and the fraction with 0.7-µm bands (which can serve as a lower limit on fraction of ST objects). The parent bodies of 7 of the families discussed in De Prá et al. (2021) or Morate et al. (2019) are in our sample: Polana (uncertain: see Section 4.5), Erigone (ST), Chaldaea (ST), Svea (NST), Ursula (NST), Hygiea (NST), and Themis (NST). Arredondo et al. report that members of the Erigone and Chaldaea families commonly are found with 0.7-µm bands, consistent with what might be expected for objects derived from ST parent bodies. Morate et al. also report that the 0.7-µm band is absent from observed members of the Polana and Svea families, and De Prá et al.

report its absence from observed members of the Ursula family, again consistent with what would be expected for objects derived from NST parents. Euphrosyne is seen to be an NST object, but there is no visible-wavelength study of its family members analogous to the De Prá et al. and Morate et al. studies that would allow detection of the 0.7-µm band. Whether asteroid (895) Helio is a member of the Euphrosyne family (Masiero et al. 2013) or an interloper (Yang et al. 2020) is a matter of disagreement, but we observe it with an NST spectrum in the 3-µm region, and consistent with either status.

Returning to the Polana family, Section 4.5 discusses its possible relationship with Bennu, which has a ST spectrum. While Polana's uncertain status and the lack of Ch asteroids in its family is consistent with them all being NSTs, it is not inconsistent with it being a ST family. Additional measurements of Polana and its family members in the 3-µm region will be necessary to make more definitive statements.

The two remaining families, Themis and Hygiea, present somewhat more complicated pictures. Both parent bodies are NSTs, and both families are reported to have members that are Ch asteroids, which we would expect to be STs. This would suggest that NSTs and STs can be related, and that we might expect both types of spectra to be found on a single object. However, we note that both of these families might be unusual. The asteroid (159) Aemilia is a Ch asteroid and ST object generally included in lists of the Hygiea family, but Carruba et al. (2014) explicitly identified it as a dynamical interloper. It is thus uncertain whether Aemilia should be included or excluded from Hygiea family membership. None of the other Hygiea family objects are reported by De Prá et al. to have a Ch or Cgh classification, and thus they are consistent with potentially having NST spectra. Until Aemilia's relationship with the Hygiea family is determined, or until a ST object is confidently known to be in the Hygiea family, we are not certain whether this family has both ST and NST members or only NST members.

The Themis family is one of the largest in the main belt, and is thought to be the result of a catastrophic disruption of an object of roughly 350—450 km diameter (Tanga et al. 1999, Durda et al. 2007), about the size of Hygiea. Fonasier et al. (2016) identified several objects within the Themis family (including, interestingly, Themis itself) with a 0.7-µm band, but Kaluna (2015) reported spectra of 45 small objects belonging to the Themis family and Beagle sub-family, with only one of those showing a 0.7-µm band, and only during one of two measurements. The complicated dynamics involving debris from an ancient catastrophic collision and a detailed consideration of interlopers is out of the scope of this paper, but objects made of reaccumulated material could have present-day surfaces dominated by compositions that were formed in parent body interiors (Castillo-Rogez and Schmidt 2010), and an object originally the size of Hygiea could have hydrated mineral compositions in its interior that would show a ST-like spectrum if it were found on an asteroid surface. Like the Hygiea family, however, until more confident identification of interlopers vs. Themis family members can be made, it is uncertain whether both NST and ST objects are found in the Themis family.

If water ice frost were responsible for some of the NST spectral features, we might expect the elevated temperatures involved in collisions to destroy any frost and diminish or remove the absorption. Nevertheless, there is a line of evidence that material with NST spectra can survive large collisions: several of the spectra of Vesta in DeSanctis et al. (2012) show clear evidence of absorptions or shoulders with BC > 3.0 µm, near the average BC seen in the NST spectra. Some measurements of (433) Eros and (1036) Ganymed by Rivkin et al. (2017) also show evidence of NST spectra. This is consistent with contamination of both of these NEOs and Vesta with NST material, as one might expect if NST material was widespread, stable at asteroidal temperatures, and could survive delivery to another object. If so, this would point to a relatively devolatilized composition for NST material, or at least one type of NST material.

Taken as a whole, the indirect data that exist are consistent with both the idea that NST spectra and ST spectra are or are not found in the same asteroid family or on the same surface. We simply do not have the direct data to say whether NST and ST material co-form or co-exist inside or on the surfaces of objects. Measurements in the 3-µm spectral region of small members of families with NST parent bodies will be necessary to answer this question.

### 6.4. How many NST groups are there?

While most of the NST group have spectra that are at least qualitatively similar to Themis, it is not a homogeneous group. A distinction between spectra more similar to Ceres vs. more similar to Themis has been made in both the Takir and Emery (2012). and Rivkin et al. (2019) informal taxonomies (Section 2.2). Takir et al. argued that (52) Europa also displays a qualitatively different spectrum, with a band center near 3.15 µm and a more rapid rise beyond that wavelength to the continuum values. We find some evidence in the new observations of an additional NST group of roughly 3-4 objects with an additional distinctive band shape in the 3-µm region.

Figure 13 shows (747) Winchester as a representative of this group, along with Ceres and Themis for comparison. Winchester shows a shallow but distinctly non-zero band depth of 4% and a calculated BC of 3.04 µm. Figure 14 shows the best examples of objects with similar band shapes, though the relative shallowness of the bands compared to the observational uncertainties means that the BC fit in these objects varies from 3.00–3.07 µm. The members of this tentatively-identified group are found, as with the other groups, in several parts of the asteroid belt and in multiple Bus or Bus-DeMeo taxonomic classes. It is also possible that some additional members of this group exist but have spectra classified as ST or NB if their uncertainties at some wavelengths are particularly high–the shallow 3-µm band compared to typical ST and NST objects increases the risk of a misidentification. While additional work will be necessary to determine whether this grouping is robust, we provisionally suggest “Bamberga-type” as a name since that is the largest object we have found with this spectrum.

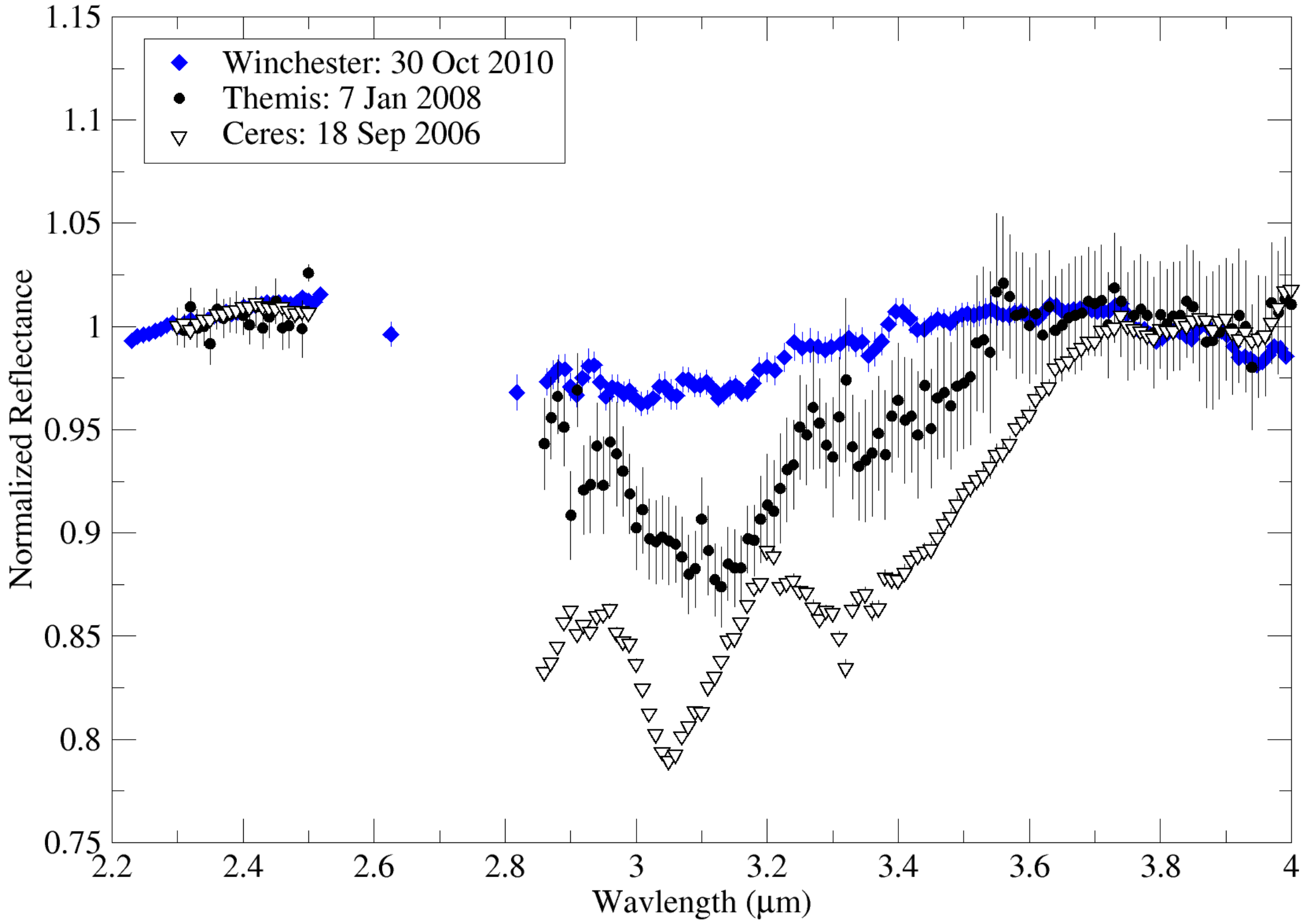

*Figure 13: NST groups. The distinction between Ceres-types and Themis-types is difficult to make in most of the new measurements, though most seem more akin to Themis in band depth and band center. Also presented is Winchester, which has a shallow but definite band that seems to represent a distinct group separate from either Themis- or Ceres- types.*

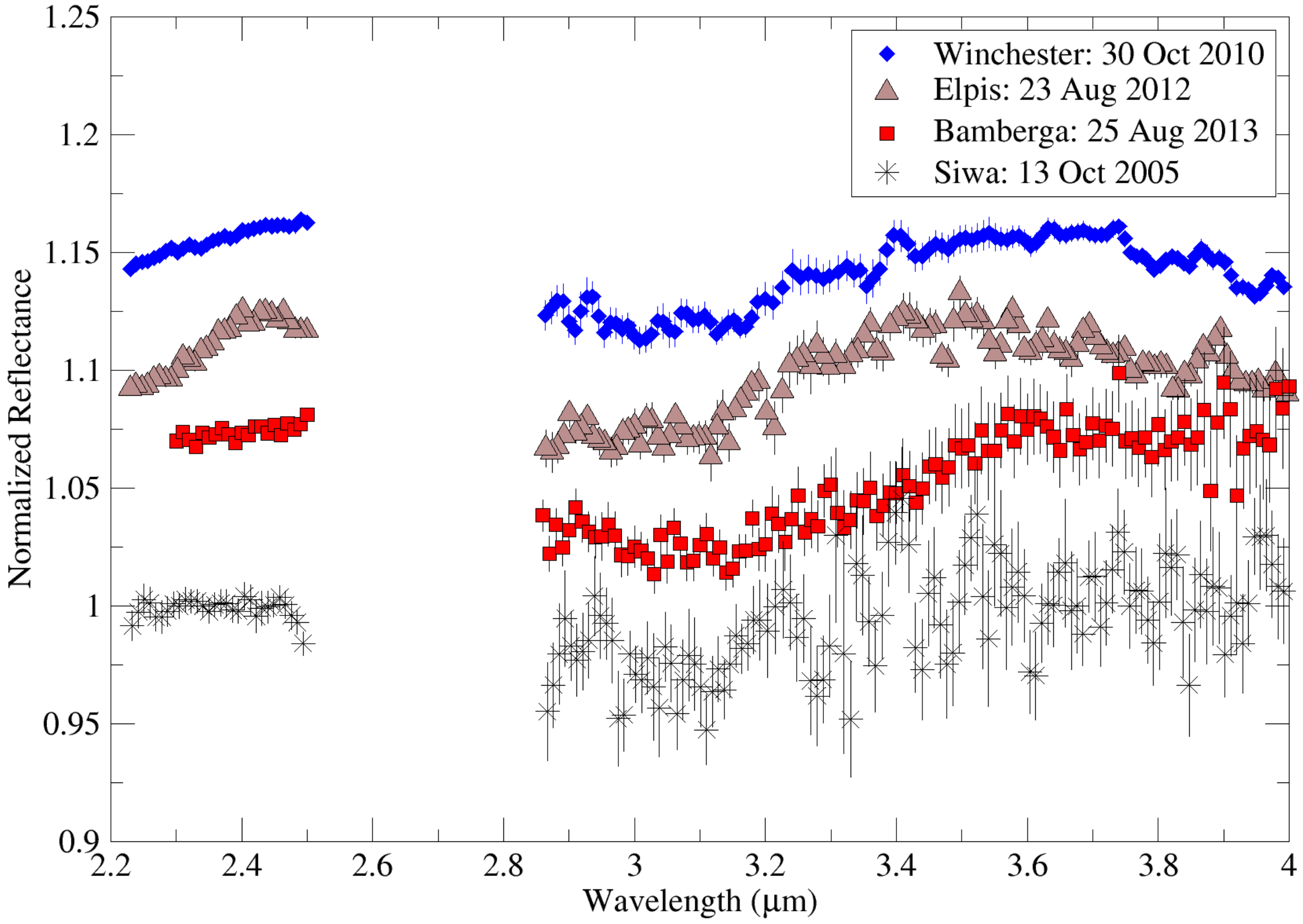

*Figure 14: Though data quality varies, additional objects with bands like Winchester (Figure 13) are present in the sample. Because Bamberga is the largest such object, we refer to these as Bamberga-types. Further study is necessary to determine how robust this grouping is.*

### 6.5. What is the relationship between Themis- and Ceres-types, and what are their compositions?

As discussed in Section 2 and the previous section, there are several spectral subgroups that are included in the NST group in this work. As noted in those sections, most previous discussions separate Ceres and Themis from one another, with different names for the groups they are in. Only a few Ceres types have been identified, while the vast majority of NSTs have spectra more consistent with Themis types. While inspection of their spectra shows distinctions between the groups in BC and the width of their main absorption, as well as a local minimum near 2.9 µm in Ceres types that is absent from Themis types, Rivkin et al. (2019) found it difficult to separate the two groups based on a simple metric or definition.

To address the relationship these groups have with one another, we must revisit interpretations of their compositions. Current interpretations favor hydrated ammoniated minerals and carbonates as the dominant spectral species on Ceres compared to ice frost and organic materials on Themis, though as noted in Section 4.4, the interpretation of 67P's spectrum as due to ammoniated salts, organic materials and submicron ice grains suggests that the surface of Themis-type objects may also have an ammoniated component in addition to an icy component. Using these interpretations, and assuming that other objects with similar spectral features have similar compositions, we would conclude that some NST material is volatile and would be easily destroyed upon heating, and other NST material, for instance the material responsible for the 3.05-µm absorption found by Pilorget et al (2021) in the spectra of returned Ryugu samples (Section 4.5), is stable and would not.

The presence of ice on the surface of Themis is supported indirectly but independently of spectral evidence through its association with the main-belt comets. We note again that some small members of the Themis family exhibit sublimation-driven cometary activity, which is a strong indication that those objects have near-surface ice. Hsieh et al. (2018) performed a systematic search for family associations for active asteroids. In addition to Themis family membership for 133P/Elst-Pizarro and 176P/LINEAR, they found that 324P/La Sagra and P/2016 J1-A could be associated with the Alauda and Theobalda families, respectively, both of which have NST parent bodies. None of the sublimation-driven objects were found to have associations with known ST families. As with Themis, the presence of main-belt comets associated with (702) Alauda and (778) Theobalda is consistent with an icy nature to these parent bodies.

However, the interpretation of ice frost at the surface of Themis is not unanimously held. The short lifetime of ice frost in vacuum at the temperatures expected for Themis has led to alternative proposals for its composition (Beck et al. 2011), but no detailed compositional modeling of alternatives is in the literature. Matters are also confounded by the relatively close band minima of water-ice vs. ammoniated minerals near 3.1 µm, and of carbonates vs. organic materials in the 3.3–3.4-µm region. Berg et al. (2016) and Ehlmann et al. (2018) found a range of band centers for ammoniated minerals, but none had band centers > 3.1 µm. Many of the NST spectra have BC < 3.1 µm, but an appreciable number, including most of the best spectral matches to 67P, have BC > 3.1 µm. It is plausible that all NST material is composed of mixtures of ammoniated, organic, and hydroxylated material that is stable in a vacuum at asteroidal temperatures, but the laboratory spectra available for those materials are not currently consistent with that interpretation.

Taking into account additional non-IRTF data sets suggests that future subdivisions in the NST group could be based on the presence or absence of the hydroxyl absorption near 2.7—2.8 µm. This absorption is seen on Ceres in Dawn data (De Sanctis et al. 2015) and on Hygiea and Themis in AKARI data (Usui et al. 2019), but not seen on 67P in Rosetta data (Raponi et al. 2020) or on Cybele in AKARI data (Usui et al. 2019). Evidence of a 2.7—2.8-µm band can be seen in IRTF data for Ceres and Hygiea as well, based on the decrease shortward of the band minimum near 2.9 µm (Rivkin et al. 2019). The most straightforward interpretation for the difference in 2.7-µm band presence is the concentration of phyllosilicates, which are the product of aqueous alteration. We might also imagine that material like what we see on 67P and Cybele could serve as the precursor material to the aqueously-altered, ammoniated material seen on Ceres and Hygiea, assuming 67P and Cybele are not themselves aqueously altered. Obviously, however, any such conclusions are dependent upon a fuller understanding of NST composition than is currently available.

### 6.6 What is the difference if any between the Ch and non-Ch STs?

While the situation for the ST asteroids seems straightforward, it is not uninteresting. The main composition-related question for this group is why some have a 0.7-µm absorption and some do not. The CM meteorites include specimens with a wide range of 0.7-µm absorption depths: Cloutis et al. (2011) report a range of 0.2–6.3% in their sample of 39 CM chondrites. Absorption bands as shallow as 0.2% would be difficult to convincingly measure in ground-based asteroidal spectra, and an object of CM composition with a band depth that small might easily be classified as something other than a Ch or Cgh asteroid. In principle, then, all ST objects could plausibly share a CM composition. In practice, of course, hydrated non-CM meteorite compositions that generate ST spectra (like those of the CR or CI meteorites) also require parent bodies, and might also be found on ST asteroid surfaces.

Twelve ST objects appear in the AKARI survey (Usui et al. 2019), with reported band minima in the 2.7–2.8-µm region. The 5 Ch/Cgh asteroids have a mean band minimum of 2.766 ± 0.009 in the AKARI data, while the 7 non-Ch/Cgh objects have a mean band center of 2.744 ± 0.013. Takir et al. (2013) found a range of CM chondrite band minima depending on their phyllosilicate composition and alteration index. The AKARI Ch/Cgh asteroids would correspond to the Takir et al. Group 1, associated with less alteration and more iron-rich compositions, while the other AKARI ST asteroids are consistent with somewhat greater alteration and a composition intermediate between iron-rich and more magnesium-rich phyllosilicate endmembers.

While promising, Takir et al. (2013) did not make measurements in the 0.7-µm region, so cannot make direct comparisons to Ch vs. non-Ch asteroids. Cloutis et al. (2011) also discuss the petrologic subtypes developed for CM meteorites by different investigators, finding them to correlate imperfectly with their spectral characteristics: the best correlation they found was between the 0.7-µm band depth and the mineralogical alteration index of Browning et al. (1996) and petrologic subtype criteria developed by Trigo-Rodriguez et al. (2006), Rubin et al. (2007), and de Leuw et al. (2009). However, the 0.7-µm band was seen in samples across the alteration indices and subtypes, making its application to the asteroid spectra not straightforward. Cloutis et al. suggested that differences in opaque abundance between samples counteracted against phyllosilicate abundance, reducing the expected correlations. In addition, there are a lot of possible complicating factors for asteroid spectra, including grain size, the amount of spectrally opaque materials made during the alteration process, and observational effects, among others.

### 6.7 Identifying Outer Solar System Material in the Main Asteroid Belt

Dynamical modelers of early solar system history have sought constraints on the original formation locations of small bodies in order to test their work. Several papers have followed the conventional wisdom that the reddest objects in the small body population probably formed the furthest from the Sun, and the number and distribution of D-class asteroids in the main belt has been used as inputs or constraints to dynamical models of material transport (e.g. Vokrouhlický et al. 2016). It has also been known for some time that D-class spectra can be generated by other types of material: Emery et al. (2004, 2011) showed that spectra of pyroxene can explain the D-class spectra of Trojan asteroids, rather than requiring abundant organic material.

The findings of this work demonstrate that the number of D-class asteroids cannot be used as a proxy for the amount of transported TNO material with any degree of confidence. There is variety in the 3-µm band shapes of all of the non-Ch/Cgh classes, including the D class. The closest matches to 67P include members of both the C and X complexes, and similarly those objects lacking any band at all include members of multiple classes. It seems very likely that the amount of TNO material delivered to the asteroid belt is much larger than previously thought, if objects with NST spectra represent that material.

Recent work by Hasegawa et al. (2021) identified two very red objects in the main asteroid belt, interpreting their spectral slopes as evidence of an origin "in the vicinity of Neptune". Both of these objects (203 Pompeja and 269 Justitia) reach brightnesses that make them viable targets for LXD observations like those presented in this work. Such observations would allow a direct comparison of their compositions to members of the ST and NST groups.

### 6.8 A Possible Scenario:

A simple scenario can be imagined that explains the data, or is at least consistent with the evidence. We can posit that all of the low-albedo objects formed beyond the water ice line, with a fraction of them forming in more distant regions where ammonia ice is also stable and available for accretion. Given sufficient radiogenic heat, the objects that formed in ammonia-free regions would experience melting of interior ice, aqueous alteration, and formation of phyllosilicates. The aqueous alteration reactions may have been exothermic (Brearley 2006), leading to a runaway process and widespread phyllosilicate formation even in objects that only just reach the ice melt threshold. Phyllosilicates are stable on surfaces in asteroidal conditions today, allowing them to be detected. This part of the scenario has been proposed many times before (Brearley 2006), and may serve to create the ST objects, with the difference between the Ch/Cgh ST objects and the other ST objects a function of the original rock/ice ratio, amount of radioactive material, intensity of alteration, or other factors. The aqueous alteration process serves to lithify material, and while carbonaceous chondrites as a class represent a wide range of strengths and durabilities, they are all sufficiently strong to make it to Earth's surface as meteorites.

The objects that accreted ammonia ice as well as water ice will have different alteration paths. Objects that formed further from the Sun may have accreted a smaller concentration of radioisotopes, or accreted them after more half-lives have elapsed than inner-solar-system objects did (Castillo-Rogez and Schmidt 2010). Radiogenic heat would be consumed melting ammonia (and other) ice at lower temperatures before water ice could be melted and aqueous alteration begun, and these reactions may not have been exothermic. Large objects like Ceres and Hygiea in this scenario could have retained and generated heat long enough to successfully melt water ice and begin the aqueous alteration processes that created the ammoniated minerals interpreted to be at their surfaces today. Smaller objects may not have been able to retain heat to the point of allowing aqueous alteration and formation of ammoniated minerals by that route. Because ammonium formation requires the presence of water (Castillo-Rogez et al. 2018), objects with ammonia ice that did not reach the melting point of water would be expected to refreeze their ammonia as temperatures waned. The co-existence of water ice and $NH_4^+$ in minerals in the absence of phyllosilicates, as interpreted to be present on the surface of 67P by Raponi et al. (2020), suggests that the ammoniation reactions may have consumed enough energy to have prevented phyllosilicate production on 67P, though consideration of the geochemistry is well outside the scope of this paper.

Sufficient processing of surface ices via cosmic ray or other irradiation, such as simulated in the Urso et al. (2020) experiments, followed by transport to the inner solar system and sublimation of remaining surface ices, may result in NST spectra like what is seen on Cybele or 67P, inferred in the latter case to be due to ammoniated salts. Additionally, the presence of ammoniated salts in the original asteroid would have facilitated melting, which in turn would facilitate formation of organics and ammoniated phyllosilicates in objects of a smaller size than those without ammoniated salts (Kargel et al. 1992). Since ammoniated salts, ammoniated phyllosilicates, and organic compounds all have absorption features near 3.1 µm, this process does provide at least two ways to shift the band depth and center of the absorption band without invoking changes to Fe/Mg abundance in the mineral matrix (Kebukawa et al. 2020). In this case, all NST objects would have similar initial compositions, and the two sub-groups (Ceres vs Themis/67P) develop from different styles of processing of those original materials. Ceres-like objects form one type of NST material from high temperature hydration and ammoniation, whereas 67P-like objects form another type of NST material from irradiation of the original surface ices. In this scenario, the 67P-like surface may only have a relatively thin surface layer of NST material, with unirradiated (and still volatile-rich?) material at relatively shallow depths. This scenario would also provide a path for smaller objects, such as Trojan asteroids and Himalia, to have NST spectra without undergoing differentiation and intense aqueous alteration.

It is not clear how lithified these NST objects may be if they did not experience widespread aqueous alteration, though they are clearly able to create families as evidenced by the Themis family, Euphrosyne family, and others. Given the similarity of and possible kinship between NSTs and comets, we can look to the latter objects for clues. Yang et al (2009) concluded that the nucleus of comet 17P/Holmes was composed of low-tensile-strength porous aggregates of ice grains and refractory particles held together by ice and/or organic compounds, based on detections of both warm dust and cold ice grains within 17P's coma after an outburst. After ejection from 17P, they suggest sublimation of ice leads to breakup of these aggregates into dust and ice, the latter of which eventually sublimes away completely. Yang et al. also pointed out that similar detections of both warm dust and cold ice grains were made after the Deep Impact experiment at comet 9P/Tempel 1 (Sunshine et al. 2007). It is difficult to see how such objects could remain coherent enough to survive the processes that lead to eventual collection as meteorites.

Finally, the NB objects could represent a variety of paths, including those similar to cometary evolution. Their lack of an absorption band may represent ammonia-free low-albedo material that did not experience aqueous alteration and lost surface ice to sublimation, resulting in a featureless surface mantled with anhydrous material. It could also represent outer solar system material transported to the inner solar system sufficiently quickly after formation that volatile ices sublimed before the irradiation timescale for making the residues responsible for NST spectra in this scenario. Finally, NB objects could represent either ST or NST material heated by internal processes early in their histories to the point of dehydration and devolatilization, representing for instance compositions like the CV or CO chondrites. Or, naturally, the NB population could be a combination of all of the above. It is not clear which of these paths, or perhaps other paths, might make the most sense for the NB Trojan asteroid population or those cometary nuclei that don't show strong absorption features at these wavelengths (Groussin et al. 2013).

This general scenario is consistent with the work of others. Raymond and Izidoro (2017) demonstrate that the distribution of C-type asteroids in the main asteroid belt could be explained as an original population of planetesimals that formed in the 4-9 AU region and that were subsequently scattered by the giant planets into the inner solar system, with the source for implanted asteroids likely being even wider. The ice lines for several volatiles fall in this semi-major axis range, including not only water and ammonia, but methanol and HCN in very early solar system history (Dodson-Robinson et al. 2009), and as stability regions for volatiles evolve with time and the amount of radioactive isotopes available for accretion decreases with time, it is not hard to imagine that these planetesimals might have experienced different levels of aqueous alteration and retained different mixtures of water, ammonia, and organic materials. A study by Sridhar et al. (2021) of the morphology of magnetite in carbonaceous chondrites found evidence for two aqueous alteration pathways, one for CM chondrites and one for CI chondrites, Tagish Lake, and WIS 91600, and they concluded that these two morphologies might represent magnetite formed on parent bodies within the ammonia ice line (the CM chondrites) and beyond it (the remaining samples). As discussed below, it is not yet clear whether the spectral evidence can be reconciled with this conclusion.

A more detailed consideration of the relationship between comets like 67P and the NST objects in asteroidal orbits would need to consider the specifics of cometary activity and the typical pace of mass loss from comets. It is estimated that 67P loses the equivalent of roughly 1 m of material from its surface per orbit and perhaps 400 m has already been lost from its surface due to previous activity (Nesvorný et al. 2018), pointing toward a homogeneous volume. Future work, beyond the scope of this paper, would be needed to investigate the balance between the relative timescales of irradiation and residue formation vs. dust emission and resurfacing in the context of small-body delivery from the TNO region to the main asteroid belt.

Any of these paths are scientifically interesting. In the case that a single component is responsible for the similarity between 67P and the subset of the NST sample, this would imply that the surface compositions of some large asteroids are remarkably similar to an active Jupiter Family Comet (JFC) with a relatively recent injection into the inner solar system, and that the outer belt and some JFCs may share more similarities in their formation region then previously recognized. If the surfaces of these objects are similar to 67P due to the presence of ammoniated salts, this would seem to explain the lack of the absorption seen in meteorites; ammoniated salts are destroyed at relatively low temperatures (Gounelle and Zolensky 2014). In the case where the degeneracies of absorption bands in the 3-µm region from ammoniate salts, phyllosilicates, and organics render the band difficult to interpret, it demands an independent set of tests to constrain the components of the surface composition.

## 7. Open questions and Implications

The dataset and interpretations discussed above address fundamental questions about the distribution of hydrated material in the asteroid belt, and further indicate that the overlap between asteroids and comets may be much larger and general than realized. As is typical with large datasets, however, a new set of mysteries and open questions have emerged as well as implications for the work.

*Are there meteorites from NST objects?* As noted several times, there are no meteorites with published spectra that have band centers near 3.1 µm or appear like most NST spectra. This is in contrast to ST spectra, which are widespread in carbonaceous chondrites. There is no particular *a priori* reason to think that NST objects, which are numerous, should not provide meteorites in proportion to their numbers. So, where are they?

There have been a few papers offering explanations for the apparent non-existence of meteorites from NST objects. Rivkin et al. (2014) considered why Ceres specifically appears unrepresented in the meteorite collection, suggesting that it was a consequence of would-be Ceres family members with high ice fractions disintegrating after sublimation. Rivkin and DeMeo (2019) noted the paucity of C-complex asteroids in the NEO population compared to delivery models, which could point to a structural constraint in at least some C-complex asteroids against forming family members that can be transported to the inner solar system. Vernazza et al. (2017) concluded from mid-infrared measurements that interplanetary dust particles (IDPs) are samples from “low-density icy asteroids”, which could correspond to the NST population. While Nesvorný et al. (2011) found via dynamical studies that micrometeorites were dominated by cometary material, the similarity we find between at least some NST spectra and 67P (section 4.4) shows that both interpretations may be consistent with one another.

Gounelle and Zolensky (2014) note that ammonium salts were found in Orgueil soon after its fall but they have subsequently been lost. The recent work of Sridhar et al. (2021) argues that we do have meteorites from beyond the ammonia ice line in the form of CI chondrites, which would include Orgueil, Tagish Lake, and WIS 91600. Because none of these objects are seen to have NST spectra, this would also suggest that any ammonium salts present pre-arrival have disappeared since the meteorites fell. However, given the spectral properties of ammonium chloride (Fastelli et al. 2020) it is not clear that addition of ammonium salts alone to an otherwise ST spectrum would be able to reproduce the typical NST spectral shapes seen in the dataset. A full explanation may involve a combination of differentiated and undifferentiated bodies and a range of relative amounts of ice and silicate accretion.

*Do band shapes evolve through an asteroid’s history?* The stability of NST compositions at asteroidal temperatures is unknown, if only because the exact compositions in question are still uncertain (Section 6.5). Temperature changes due to impacts, particularly family forming impacts, may have driven volatile loss and compositional changes in addition to any changes due to possible transport from the outer to the inner solar system. As noted in Section 6.3, 3-µm spectral measurements of secure members of families with NST parent bodies will be needed to establish whether these changes occur, which will also allow an assessment of whether 3-µm band shapes can be used as markers of family membership the way that taxonomic classes are used to identify interlopers in families. Band shape evolution would be expected as a natural consequence of the scenario suggested in Section 6.8, though after billions of years in their current orbits we would expect absorption bands to have reached an equilibrium shape.

*What are spectral slopes telling us?* While it is clear that the use of absorption features is more diagnostic than using spectral slopes for determining composition, it is equally clear that there is some information contained in spectral slopes that is not yet understood. The fact that Ch/Cgh asteroids have a relatively limited range of spectral slopes shows that spectral features and slopes are not entirely independent of one another, at least not in all cases. Similarly, the transition from objects with shallow/flat spectral slopes dominating the main belt to steeper slopes dominating the Trojan clouds is presumably not a coincidence. While comparisons between TNO taxonomy and asteroid taxonomy are not straightforward due to the presence of abundant ices on TNO surfaces changing their spectra, this trend continues further, with the D-class-like BR and IR classes more frequently found closer to the Sun and the redder-than-D-class RR class being the most abundant beyond roughly 30 AU (Fulchignoni et al. 2008). Space weathering, which has been shown to have a variety of effects on spectral slopes of low-albedo objects depending on particular circumstances (Kaluna et al. 2017, Lantz et al. 2018), is another complicating factor. Understanding the compositional causes of spectral slopes in one small-body population will likely help us understand it in other small-body populations.

## 8. Conclusions

We compile and present 267 spectra from 159 low-albedo asteroids covering the 2—4-µm region, including 191 newly-published spectra representing 120 asteroids. We confirm the presence of multiple band shapes in this region, and define two broad groups of spectra based on the wavelength of the band center (BC): “Sharp types” (ST) with BC < 3 µm and “Non-Sharp types” (NST) with BC > 3 µm. In almost all cases where objects are observed multiple times, the resulting spectra have consistent ST vs. NST classifications, and as a result we can comfortably consider entire objects to be ST or NST (with the recognition that additional data may cause attendant discomfort). In terms of previous work, the NST objects can be considered a superset of all of the objects not part of the Pallas types or “Sharp types”, depending on which informal taxonomy is being used. Most NSTs appear to be Themis-types (in the Rivkin et al. 2019 terminology), and the Pallas-type vs. Themis-type dichotomy appears to be as fundamental a distinction for low-albedo asteroids as the S-complex vs. C-complex distinction among asteroids in general.

All of the low-albedo taxonomic classes based on 0.5—2.5-µm spectra data have both ST and NST members. In the case of the Ch/Cgh classes, which are defined by the presence of a 0.7-µm band, those few objects with NST classifications are thought to be misclassified due to data with large observational uncertainties. For the other classes, however, it appears that one cannot confidently predict the hydrated mineralogy and 3-µm BC of an asteroid knowing only its visible-near IR taxonomic type.

We found several asteroids with spectra that are good qualitative matches to the average spectrum of 67P in 3-µm spectral shape, extending the findings of Rivkin et al. (2019) to smaller objects. The spectral similarity of a half-dozen or more objects > 100 km in diameter to a cometary nucleus is evidence that these large objects either have comet-like compositions at their surfaces, and that despite a factor of 20-50 difference in size the asteroids did not suffer processes to destroy such compositions, or that some combination of ammoniated salts, organics, and phyllosilicates can reproduce a very similar spectrum.

A series of statistical tests shows that the differences between the main-belt ST and NST populations (and subgroups defined by size) in band depths (BD), semi-major axis, and perihelion distance are statistically significant at the 90% level or greater. These same differences are present and comparable in the smaller populations of Cybele/Hilda ST and NST objects. This is consistent with ST and NST objects being drawn from different original populations and delivered to slightly different places. Statistically significant differences in BD are also seen between the Ch/Cgh ST population and the non-Ch/Cgh ST population. Other differences are not seen to be statistically significant in various subgroups, including albedo and density.

Future work should include investigations of ST vs. NST distributions within asteroid families to test the stability of NST material in different parts of the solar system and whether parent bodies can give rise to both ST and NST objects. It should also include further investigations of the diversity of material within the ST and NST groups, just as unusual 0.5–2.5-µm taxonomic classes like those that Vesta (Fulvio et al. 2018) or Aquitania (Sunshine et al. 2008) belong to provide insight into the processes on silicate asteroids. It would be useful to observe the 3-µm region on more distant objects to determine whether ST and/or NST material is the dominant non-ice material on primitive objects. We also caution against using D-type asteroids, or any taxonomic classifications based only on 0.5—2.5-µm data, as a proxy for outer solar system material. It appears likely that there is more (perhaps vastly more) outer solar system material in the main asteroid belt and Cybele/Hilda regions than previously thought.

***Acknowledgements:*** This work is the result of years of effort and numerous discussions. The authors wish to recognize and acknowledge the very significant cultural role and reverence that the summit of Maunakea has always had within the indigenous Hawaiian community. We are most fortunate to have the opportunity to conduct observations from this mountain, and recognize we are guests. We thank those who work at the NASA Infrared Telescope Facility, without whom this would have been impossible: the telescope operators, the day crew, the instrument scientists, administrators, admin assistants, and others. We similarly thank the staff at Hale Pohaku, even including the kitchen crew that wasn't as good as the other one. Thanks to Eric Volquardsen for allowing us to use the data he collected, and thanks to the online IRTF Legacy Archive and the IRTF Data Archive at IRSA for enabling that use. Consistent support from the NASA Planetary Astronomy program, including grants NNX14AJ39G, NNX09AB45G, NNG05GR60G, and NAG5-10604, enabled these observations and analysis. The SSERVI RESOURCE team and their PI Jen Heldmann provided support for the finishing stages of this project. This research has made use of NASA's Astrophysics Data System.

***Appendix A: Supplementary Data***

Table A1: Observational Circumstances Taxonomic class is reported from Bus and Binzel (2002) with the following exceptions: if the first character is lowercase, the classification is taken from Tholen (1984), and if the first two characters are "S3", the remaining letters are the classification in the Bus system, taken from Lazzaro et al (2004).

| Date | Number | Name | Tax Class | V mag | Earth dist | Solar Dist | Phase Angle |
|---|---|---|---|---|---|---|---|
| 2-Mar-2005 | 2 | Pallas | B | 7.5 | 1.394 | 2.319 | 11.4 |
| 12-Jun-2007 | 2 | Pallas | B | 10.2 | 3.131 | 3.34 | 17.7 |
| 13-Sep-2007 | 2 | Pallas | B | 9 | 2.257 | 3.232 | 5.3 |
| 23-Jun-2011 | 2 | Pallas1 | B | 9.8 | 2.707 | 3.403 | 14 |
| 23-Jun-2011 | 2 | Pallas2 | B | 9.8 | 2.707 | 3.403 | 14 |
| 23-Aug-2012 | 2 | Pallas | B | 9 | 2.171 | 3.022 | 12.2 |
| 29-Aug-2012 | 2 | Pallas | B | 8.9 | 2.11 | 3.009 | 10.5 |
| 6-Dec-2013 | 2 | Pallas1 | B | 8.4 | 1.812 | 2.13 | 27.5 |
| 6-Dec-2013 | 2 | Pallas2 | B | 8.4 | 1.812 | 2.13 | 27.5 |
| 6-Dec-2013 | 2 | Pallas3 | B | 8.4 | 1.812 | 2.13 | 27.5 |
| 2-Dec-2015 | 10 | Hygiea | C | 11.4 | 3.13 | 3.04 | 18.3 |
| 16-Feb-2016 | 10 | Hygiea | C | 10.2 | 2.902 | 2.962 | 10.8 |
| 18-Feb-2016 | 10 | Hygiea | C | 10.1 | 2.076 | 2.96 | 10.2 |
| 12-Oct-2005 | 36 | Atalante | Ch | 13.1 | 1.992 | 2.396 | 24.1 |
| 30-Oct-2010 | 36 | Atalante | Ch | 10.9 | 1.007 | 1.939 | 14 |
| 17-Sep-2006 | 45 | Eugenia | C | 11.4 | 1.811 | 2.812 | 2.1 |
| 22-Jul-2013 | 46 | Hestia | Xc | 13.2 | 2.024 | 2.569 | 21.7 |
| 30-Apr-2007 | 47 | Aglaja | B | 13 | 2.198 | 3.058 | 11.6 |
| 21-Jul-2013 | 47 | Aglaja | B | 11.4 | 1.499 | 2.506 | 3.9 |
| 16-Feb-2016 | 52 | Europa | C | 10.2 | 1.815 | 2.801 | 1.7 |
| 31-Aug-2004 | 53 | Kalypso | xc | 12.6 | 1.803 | 2.808 | 2.1 |
| 31-Oct-2010 | 54 | Alexandra | C | 11.7 | 1.641 | 2.465 | 15.8 |
| 30-Apr-2007 | 56 | Melete | Xk | 11.9 | 1.497 | 2.457 | 9.1 |
| 29-Aug-2012 | 56 | Melete | Xk | 11.4 | 1.225 | 2.159 | 13.6 |
| 23-Aug-2012 | 59 | Elpis | B | 11.8 | 1.582 | 2.417 | 16.8 |
| 21-Feb-2007 | 65 | Cybele | Xc | 12.1 | 2.5 | 3.359 | 9.7 |
| 22-Feb-2007 | 65 | Cybele | Xc | 12.1 | 2.49 | 3.358 | 9.2 |
| 20-Mar-2007 | 65 | Cybele | Xc | 11.5 | 2.339 | 3.333 | 1.3 |
| 29-Apr-2007 | 65 | Cybele | Xc | 12.2 | 2.477 | 3.294 | 11.8 |
| 20-Jun-2008 | 65 | Cybele | Xc | 11.2 | 2.063 | 3.073 | 2.6 |
| 21-Jul-2009 | 65 | Cybele | Xc | 12.4 | 2.229 | 3.077 | 12.3 |
| 4-Jan-2013 | 65 | Cybele | Xc | 12.7 | 2.903 | 3.509 | 13.9 |
| 12-Oct-2005 | 66 | Maja | Ch | 12.6 | 1.337 | 2.289 | 9.8 |
| 21-Sep-2005 | 72 | Feronia | tdg | 12.3 | 1.267 | 2.002 | 24.8 |
| 2-Dec-2015 | 74 | Galatea | C | 12.4 | 1.51 | 2.13 | 24.6 |
| 1-Sep-2012 | 76 | Freia | X | 13.1 | 2.561 | 3.566 | 1.7 |
| 6-Dec-2013 | 76 | Freia | X | 12.3 | 1.963 | 2.923 | 5.3 |
| 18-Feb-2016 | 81 | Terpsichore | Cb | 13.8 | 2.392 | 3.151 | 13.2 |
| 16-Feb-2016 | 83 | Beatrix | X | 12.7 | 1.655 | 2.391 | 19 |
| 2-Apr-2007 | 85 | Io | B | 12.3 | 1.992 | 2.514 | 22 |
| 12-Oct-2005 | 87 | Sylvia | X | 12.3 | 2.578 | 3.376 | 11.6 |
| 10-Dec-2006 | 87 | Sylvia | X | 12.7 | 2.905 | 3.667 | 11 |
| 20-Jun-2008 | 87 | Sylvia | X | 13.4 | 3.762 | 3.734 | 15.6 |
| 5-Jan-2013 | 87 | Sylvia | X | 12.2 | 2.651 | 3.583 | 5.8 |
| 24-Aug-2017 | 87 | Sylvia | X | 11.6 | 2.177 | 3.159 | 5.2 |
| 13-Oct-2005 | 88 | Thisbe | B | 11.5 | 1.814 | 2.686 | 12.6 |
| 28-Aug-2011 | 90 | Antiope | C | 11.7 | 1.674 | 2.683 | 1.5 |
| 27-Feb-2003 | 93 | Minerva | C | 12.3 | 1.913 | 2.715 | 14.6 |
| 22-Feb-2007 | 93 | Minerva | C | 12.2 | 2.029 | 2.998 | 4.6 |
| 14-Aug-2009 | 93 | Minerva | C | 12.7 | 2.133 | 2.693 | 20.3 |
| 21-Jul-2013 | 94 | Aurora | C | 13.1 | 2.585 | 3.041 | 18.6 |
| 29-Apr-2007 | 96 | Aegle | T | 12.1 | 1.889 | 2.677 | 16 |
| 30-Jan-2012 | 96 | Aegle | T | 11.5 | 1.711 | 2.679 | 4.9 |
| 22-Nov-2007 | 102 | Miriam | C | 13.6 | 1.23 | 2.087 | 17.6 |
| 16-Feb-2016 | 106 | Dione | Cgh | 13 | 2.764 | 2.703 | 20.7 |
| 18-Nov-2005 | 107 | Camilla | X | 12 | 2.41 | 3.368 | 5.1 |
| 21-Jun-2008 | 107 | Camilla | X | 12.9 | 3.009 | 3.467 | 16.1 |
| 24-Aug-2017 | 107 | Camilla | X | 12.6 | 2.665 | 3.633 | 5.3 |
| 2-Jan-2013 | 117 | Lomia | X | 12.5 | 2.036 | 2.96 | 7.9 |
| 1-Sep-2012 | 120 | Lachesis | C | 13.2 | 2.561 | 3.286 | 13.9 |
| 17-May-2006 | 121 | Hermione | C | 13.2 | 2.956 | 3.874 | 7.2 |
| 15-Jun-2007 | 121 | Hermione | C | 12.3 | 2.481 | 3.494 | 1.6 |
| 20-Jun-2008 | 121 | Hermione | C | 13.1 | 2.729 | 3.064 | 19.1 |
| 20-Mar-2002 | 128 | Nemesis | C | 12.2 | 2.11 | 3.088 | 4.1 |
| 6-Mar-2015 | 128 | Nemesis | C | 12.3 | 2.389 | 3.098 | 14.6 |
| 27-Apr-2015 | 128 | Nemesis | C | 13.2 | 2.73 | 2.937 | 20 |
| 11-Jun-2007 | 137 | Meliboea | Ch | 12.1 | 1.637 | 2.477 | 16.3 |
| 29-Sep-2013 | 139 | Juewa | X | 12.6 | 2.25 | 3.247 | 1.9 |
| 18-Feb-2016 | 139 | Juewa | X | 12.5 | 1.963 | 2.36 | 24.3 |
| 2-Mar-2003 | 140 | Siwa | Xc | 13.4 | 2.332 | 3.266 | 6.9 |
| 12-Oct-2005 | 140 | Siwa | Xc | | 1.563 | 2.539 | 6 |
| 13-Oct-2005 | 140 | Siwa | Xc | 11.9 | 1.562 | 2.541 | 5.6 |
| 26-Feb-2003 | 142 | Polana | B | 13.5 | 1.371 | 2.315 | 9.6 |
| 22-Jul-2013 | 150 | Nuwa | Cb | 13.2 | 2.094 | 2.92 | 13.8 |
| 18-Feb-2016 | 150 | Nuwa | Cb | 13.6 | 2.34 | 3.132 | 12.5 |
| 15-Jun-2007 | 153 | Hilda | X | 12.4 | 2.418 | 3.429 | 2 |
| 20-Jun-2008 | 153 | Hilda | X | 13.8 | 3.097 | 3.604 | 15.1 |
| 28-Feb-2003 | 154 | Bertha | C | 12.7 | 2.231 | 2.92 | 16 |
| 29-Aug-2004 | 160 | Una | C | 13 | 1.741 | 2.724 | 6.1 |
| 7-Sep-2009 | 165 | Loreley | Cb | 12.2 | 1.955 | 2.918 | 7.2 |
| 31-Aug-2004 | 173 | Ino | Xk | 11.8 | 1.597 | 2.42 | 16.8 |
| 15-Aug-2009 | 173 | Ino | Xk | 11.6 | 1.479 | 2.197 | 22.8 |
| 20-Sep-2010 | 175 | Andromache | Cg | 12.4 | 1.669 | 2.544 | 13.8 |
| 26-Jun-2013 | 181 | Eucharis | Xk | 13.5 | 2.818 | 3.758 | 6.8 |
| 1-Sep-2012 | 185 | Eunike | C | 12.4 | 2.103 | 2.405 | 24.7 |
| 20-Mar-2007 | 190 | Ismene | X | 13 | 2.836 | 3.83 | 0.7 |
| 5-Nov-2020 | 190 | Ismene | X | 12.6 | 2.472 | 3.459 | 1.8 |
| 8-Dec-2017 | 191 | Kolga | Cb | 13.4 | 1.986 | 2.806 | 13.3 |
| 16-May-2006 | 194 | Prokne | C | 12.9 | 2.288 | 2.902 | 17.9 |
| 11-Jun-2007 | 194 | Prokne | C | 11.4 | 1.442 | 2.033 | 27.9 |
| 18-Feb-2016 | 209 | Dido | Xc | 13 | 2.195 | 3.12 | 7.6 |
| 27-Jun-2013 | 210 | Isabella | Cb | 13.9 | 1.906 | 2.784 | 12.8 |
| 18-Oct-2004 | 212 | Medea | dcx: | 12.2 | 1.801 | 2.794 | 2.3 |
| 17-Oct-2004 | 220 | Stephania | xc | 13 | 0.9 | 1.881 | 7.5 |
| 12-Jun-2007 | 225 | Henrietta | f | 13 | 1.741 | 2.502 | 18.5 |
| 15-Jun-2007 | 225 | Henrietta | f | 13 | 1.719 | 2.5 | 18 |
| 14-Sep-2007 | 225 | Henrietta | f | 13.2 | 1.837 | 2.494 | 20.4 |
| 2-Dec-2015 | 240 | Vanadis | C | 11.5 | 1.14 | 2.12 | 4.5 |
| 16-Feb-2016 | 240 | Vanadis | C | 13.4 | 1.835 | 2.185 | 26.7 |
| 18-Sep-2006 | 241 | Germania | B | 12.5 | 2.096 | 2.794 | 17.2 |
| 2-Mar-2003 | 247 | Eukrate | Xc | 12.5 | 1.96 | 2.91 | 6.6 |
| 18-Dec-2005 | 259 | Aletheia | X | 13.1 | 2.584 | 3.52 | 5.8 |
| 25-Jun-2013 | 261 | Prymno | X | 12.9 | 1.397 | 2.323 | 13.4 |
| 27-Jun-2012 | 267 | Tirza | D | 13.5 | 1.491 | 2.498 | 4.1 |
| 27-Jun-2012 | 268 | Adorea | S3X | 13.3 | 2.153 | 2.754 | 19.4 |
| 5-Nov-2020 | 284 | Amalia | Ch | 13.5 | 1.368 | 2.236 | 15.7 |
| 13-Jun-2019 | 304 | Olga | Xc | 13.3 | 1.334 | 1.918 | 30.1 |
| 18-Sep-2006 | 308 | Polyxo | T | 12.8 | 2.07 | 2.651 | 20.1 |
| 8-Jan-2013 | 308 | Polyxo | T | 12.4 | 1.949 | 2.843 | 9.1 |
| 30-Aug-2004 | 313 | Chaldaea | c | 13.7 | 2.013 | 2.799 | 15.3 |
| 18-Feb-2016 | 324 | Bamberga | S3Cb | 13 | 3.02 | 3.589 | 14.1 |
| 3-Mar-2019 | 326 | Tamara | c | 12.7 | 1.387 | 2.244 | 16.3 |
| 23-Aug-2002 | 329 | Svea | S3C | 13.3 | 1.549 | 2.464 | 12.6 |
| 14-Sep-2007 | 334 | Chicago | c | 13.3 | 2.877 | 3.827 | 5.7 |
| 8-Dec-2017 | 334 | Chicago | c | 13.5 | 3.05 | 3.98 | 5.2 |
| 12-Oct-2005 | 335 | Roberta | B | 12.3 | 1.471 | 2.454 | 5.3 |
| 13-Oct-2005 | 335 | Roberta | B | 12.3 | 1.471 | 2.456 | 4.9 |
| 17-Dec-2005 | 335 | Roberta | B | 13.7 | 1.471 | 2.456 | 20.1 |
| 20-Jun-2008 | 335 | Roberta | B | 12.2 | 1.352 | 2.146 | 21.5 |
| 15-May-2006 | 336 | Lacadiera | Xk | 12.8 | 1.208 | 2.108 | 16.6 |
| 30-Jan-2012 | 360 | Carlova | C | 12.4 | 1.698 | 2.632 | 8.5 |
| 25-Nov-2008 | 361 | Bononia | S3D | 13.2 | 2.245 | 3.203 | 5.2 |
| 17-Dec-2005 | 363 | Padua | X | 12.6 | 1.728 | 2.71 | 1.6 |
| 18-Dec-2005 | 363 | Padua | X | 12.6 | 1.73 | 2.711 | 2 |
| 16-Feb-2016 | 363 | Padua | X | 13.4 | 1.995 | 2.93 | 7.5 |
| 23-Aug-2012 | 368 | Haidea | d | 13.2 | 1.479 | 2.471 | 5.9 |
| 22-Sep-2005 | 372 | Palma | B | 11.9 | 2.043 | 2.996 | 7.3 |
| 2-Jan-2013 | 372 | Palma | B | 12.4 | 2.259 | 2.718 | 20.2 |
| 15-Aug-2009 | 375 | Ursula | Xc | 12 | 1.961 | 2.82 | 13.1 |
| 2-Dec-2015 | 375 | Ursula | Xc | 12.5 | 2.25 | 3.06 | 12.1 |
| 29-Aug-2004 | 386 | Siegena | C | 13 | 2.494 | 2.938 | 19.3 |
| 31-Oct-2010 | 409 | Aspasia | Xc | 12.1 | 1.951 | 2.677 | 17.1 |
| 21-Jun-2008 | 420 | Bertholda | p | 13.7 | 2.667 | 3.506 | 10.8 |
| 27-Jun-2012 | 422 | Berolina | dx | 12.9 | 1.031 | 2.022 | 8.9 |
| 28-Jun-2012 | 423 | Diotima | C | 12.5 | 2.395 | 2.977 | 18 |
| 22-Dec-2020 | 423 | Diotima | C | 11.9 | 2.21 | 3.176 | 4 |
| 4-Jan-2013 | 426 | Hippo | S3X | 13.4 | 2.139 | 2.601 | 21.2 |
| 2-Dec-2015 | 429 | Lotis | S3Xk | 12.8 | 1.37 | 2.35 | 3.3 |
| 13-Sep-2007 | 431 | Nephele | B | 12.5 | 1.645 | 2.637 | 4.7 |
| 13-Sep-2007 | 444 | Gyptis | C | 12.6 | 2.012 | 2.435 | 23.8 |
| 8-Dec-2017 | 451 | Patientia | S3Cb | 10.7 | 1.854 | 2.835 | 2.1 |
| 17-Dec-2005 | 455 | Bruchsalia | S3Xk | 12 | 1.462 | 2.44 | 3.4 |
| 18-Dec-2005 | 455 | Bruchsalia | S3Xk | 12 | 1.468 | 2.443 | 3.9 |
| 18-Oct-2004 | 463 | Lola | x | 13.3 | 0.896 | 1.889 | 3.5 |
| 4-Jan-2013 | 466 | Tisiphone | c | 14.1 | 2.767 | 3.051 | 18.6 |
| 11-Oct-2005 | 476 | Hedwig | X | 12.6 | 1.748 | 2.678 | 9.6 |
| 12-Oct-2005 | 476 | Hedwig | X | 12.6 | 1.745 | 2.679 | 9.3 |
| 13-Oct-2005 | 476 | Hedwig | X | 12.6 | 1.741 | 2.68 | 9 |
| 18-Dec-2005 | 476 | Hedwig | X | 13.4 | 2.087 | 2.728 | 18 |
| 25-Jun-2013 | 476 | Hedwig | X | 12.8 | 1.727 | 2.466 | 19.5 |
| 29-Sep-2013 | 479 | Caprera | C | 12.4 | 1.237 | 2.226 | 5.4 |
| 12-Oct-2005 | 510 | Mabella | S3T | 13 | 1.517 | 2.509 | 3.1 |
| 26-Jun-2013 | 510 | Mabella | S3T | 13.1 | 1.325 | 2.112 | 22.2 |
| 2-Mar-2003 | 511 | Davida | C | 11.1 | 2.073 | 2.637 | 20.1 |
| 25-Jul-2006 | 511 | Davida | C | 11.5 | 2.544 | 3.548 | 2.9 |
| 17-Sep-2006 | 511 | Davida | C | 12.2 | 2.789 | 3.478 | 13.5 |
| 18-Dec-2005 | 514 | Armida | S3Xe | 13.5 | 2.005 | 2.956 | 6 |
| 17-Dec-2005 | 524 | Fidelio | S3Ch | 13.1 | 1.382 | 2.329 | 8.6 |
| 18-Dec-2005 | 524 | Fidelio | S3Ch | 13.1 | 1.382 | 2.329 | 8.9 |
| 8-Jan-2013 | 536 | Merapi | S3X | 13.5 | 2.664 | 3.602 | 5.5 |
| 24-Aug-2017 | 536 | Merapi | S3X | 13.3 | 2.371 | 3.2 | 12.1 |
| 2-Dec-2015 | 547 | Praxedis | Xk | 12.8 | 1.29 | 2.22 | 10.7 |
| 27-Jun-2013 | 559 | Nanon | Xk | 13.1 | 1.58 | 2.577 | 5.6 |
| 29-Aug-2006 | 566 | Stereoskopia | c | 12.4 | 2.076 | 3.081 | 2.4 |
| 23-Aug-2012 | 566 | Stereoskopia | c | 12.8 | 2.177 | 3.148 | 6.1 |
| 23-Aug-2002 | 585 | Bilkis | c | 14.2 | 1.74 | 2.74 | 2.8 |
| 12-Jun-2007 | 596 | Scheila | T | 12.1 | 1.443 | 2.443 | 5.3 |
| 23-Aug-2002 | 688 | Melanie | C | 13.8 | 1.34 | 2.34 | 3.9 |
| 17-Oct-2004 | 690 | Wratislavia | S3B | 11.6 | 1.598 | 2.576 | 5.3 |
| 30-Oct-2010 | 690 | Wratislavia | S3B | 12.6 | 1.982 | 2.774 | 14.7 |
| 7-Sep-2009 | 694 | Ekard | S3Ch | 11.7 | 0.98 | 1.82 | 24.5 |
| 25-Jun-2013 | 702 | Alauda | B | 12.9 | 2.669 | 3.147 | 17.8 |
| 27-Apr-2015 | 704 | Interamnia | B | 11.7 | 2.604 | 3.522 | 7.8 |
| 11-Oct-2005 | 705 | Erminia | C | 12.7 | 1.917 | 2.867 | 7.5 |
| 12-Oct-2005 | 705 | Erminia | C | 12.7 | 1.914 | 2.867 | 7.3 |
| 31-Oct-2010 | 705 | Erminia | C | 12.7 | 1.914 | 2.865 | 7 |
| 26-Jun-2013 | 712 | Boliviana | X | 13.4 | 2.188 | 2.337 | 25.7 |
| 30-Oct-2010 | 747 | Winchester | C | 11.9 | 1.74 | 2.122 | 27.6 |
| 18-Feb-2016 | 762 | Pulcova | S3Cb | 13.4 | 2.604 | 2.922 | 19.5 |
| 1-Sep-2004 | 773 | Irmintraud | T | 13.3 | 1.777 | 2.72 | 9.4 |
| 2-Dec-2015 | 778 | Theobalda | S3C | 13.2 | 1.515 | 2.456 | 8.7 |
| 30-Aug-2004 | 783 | Nora | C | 14.5 | 1.34 | 1.81 | 33.3 |
| 21-Jun-2008 | 790 | Pretoria | S3X | 12.2 | 1.923 | 2.93 | 3.4 |
| 10-Aug-2008 | 790 | Pretoria | S3X | 13.1 | 2.254 | 2.906 | 17.4 |
| 12-Oct-2005 | 838 | Seraphina | p | 13.9 | 1.616 | 2.558 | 9.3 |
| 13-Oct-2005 | 838 | Seraphina | p | 13.9 | 1.612 | 2.559 | 8.9 |
| 10-Dec-2006 | 895 | Helio | B | 12.1 | 1.752 | 2.733 | 2.3 |
| 13-Jun-2019 | 914 | Palisana | S3Ch | 11.4 | 1.011 | 1.984 | 11.9 |
| 1-Sep-2004 | 924 | Toni | X | 13.9 | 1.944 | 2.689 | 17.1 |
| 23-Aug-2002 | 1343 | Nicole | C | 14.4 | 1.44 | 2.42 | 8.6 |
| 2-Dec-2015 | 1963 | Bezovec | c | 13.5 | 1.05 | 1.97 | 14.3 |

Table A2: Physical properties of objects discussed in this work. A value of -1 for densities (ρ) and associated uncertainty represents data was unavailable or not used due to large uncertainties. Densities are taken from the SiMDA compilation database (Kretlow, 2020). Taxonomic classes are as in Table A1.

| number | name | H | tax class | D (km) | alb | a (AU) | q (AU) | ρ (g/cm3) | err |
|---|---|---|---|---|---|---|---|---|---|
| 1 | Ceres | 3.4 | C | 939 | 0.09 | 2.765 | 2.55 | 2.16 | 0.01 |
| 2 | Pallas | 4.2 | B | 545 | 0.101 | 2.772 | 2.137 | 2.77 | 0.21 |
| 10 | Hygiea | 5.5 | C | 407 | 0.0717 | 3.140 | 2.791 | 2.07 | 0.26 |
| 13 | Egeria | 6.8 | Ch | 202 | 0.045 | 2.577 | 2.357 | 1.73 | 0.84 |
| 19 | Fortuna | 7.3 | Ch | 200 | 0.037 | 2.441 | 2.059 | 1.86 | 0.29 |
| 24 | Themis | 7.3 | B | 198 | 0.067 | 3.130 | 2.746 | 1.63 | 0.4 |
| 31 | Euphrosyne | 6.8 | Cb | 267 | 0.053 | 3.1544 | 2.467 | 1.39 | 0.55 |
| 34 | Circe | 8.6 | Ch | 133 | 0.023 | 2.686 | 2.4 | 4.63 | 1.21 |
| 36 | Atalante | 8.5 | Ch | 133 | 0.029 | 2.747 | 1.909 | 2.92 | 1.32 |
| 38 | Leda | 8.3 | Cgh | 92 | 0.0617 | 2.741 | 2.318 | 1.18 | 1.46 |
| 41 | Daphne | 7.4 | Ch | 205 | 0.0395 | 2.76 | 2.003 | 1.97 | 0.29 |
| 48 | Doris | 7.1 | Ch | 216 | 0.05 | 3.11 | 2.886 | 1.24 | 0.38 |
| 51 | Nemausa | 7.6 | Ch | 138 | 0.1 | 2.366 | 2.205 | 1.71 | 0.46 |
| 52 | Europa | 6.4 | C | 304 | 0.057 | 3.011 | 2.753 | 1.5 | 0.31 |
| 53 | Kalypso | 8.8 | xc | 97 | 0.028 | 2.618 | 2.09 | 1.49 | 0.46 |
| 54 | Alexandra | 7.8 | C | 160 | 0.059 | 2.712 | 2.177 | 2.23 | 1.29 |
| 56 | Melete | 8.3 | Xk | 121 | 0.057 | 2.597 | 1.998 | -1 | -1 |
| 59 | Elpis | 8.0 | B | 165 | 0.044 | 2.714 | 2.394 | 1.34 | 0.35 |
| 65 | Cybele | 6.6 | Xc | 237 | 0.071 | 3.427 | 3.041 | 1.99 | 0.5 |
| 66 | Maja | 9.5 | Ch | 72 | 0.016 | 2.645 | 2.187 | -1 | -1 |
| 70 | Panopaea | 8.1 | Ch | 128 | 0.038 | 2.615 | 2.144 | 2.33 | 0.84 |
| 72 | Feronia | 8.9 | tdg | 75 | 0.083 | 2.267 | 1.993 | -1 | -1 |
| 74 | Galatea | 8.7 | C | 119 | 0.0431 | 2.781 | 2.125 | 0.99 | 0.8 |
| 76 | Freia | 7.9 | X | 145 | 0.058 | 3.414 | 2.847 | -1 | -1 |
| 78 | Diana | 8.1 | Ch | 121 | 0.064 | 2.619 | 2.078 | 1.62 | 0.7 |
| 81 | Terpsichore | 8.6 | Cb | 118 | 0.045 | 2.853 | 2.251 | 2.92 | 0.86 |
| 85 | Io | 7.7 | B | 155 | 0.0666 | 2.655 | 2.137 | 1.39 | 0.77 |
| 87 | Sylvia | 6.9 | X | 253 | 0.046 | 3.488 | 3.154 | 1.34 | 0.19 |
| 88 | Thisbe | 7.2 | B | 232 | 0.0671 | 2.767 | 2.322 | 2.14 | 0.34 |
| 90 | Antiope | 8.3 | C | 116 | 0.058 | 3.154 | 2.622 | 1.15 | 0.16 |
| 93 | Minerva | 7.9 | C | 154 | 0.056 | 2.754 | 2.373 | 2 | 0.39 |
| 94 | Aurora | 7.6 | C | 205 | 0.049 | 3.158 | 2.86 | 1.08 | 1.08 |
| 96 | Aegle | 7.7 | T | 178 | 0.048 | 3.051 | 2.617 | 2.67 | 1.97 |
| 102 | Miriam | 9.2 | C | 83 | 0.051 | 2.661 | 1.994 | -1 | -1 |
| 105 | Artemis | 8.6 | Ch | 95 | 0.0466 | 2.374 | 1.953 | 1.9 | 0.53 |
| 106 | Dione | 7.5 | Cgh | 208 | 0.059 | 3.175 | 2.673 | 1.96 | 0.86 |
| 107 | Camilla | 7.1 | X | 210 | 0.059 | 3.487 | 3.262 | 1.77 | 0.39 |
| 109 | Felicitas | 8.75 | Ch | 83 | 0.0705 | 2.696 | 1.888 | -1 | -1 |
| 117 | Lomia | 8 | X | 209 | 0.037 | 2.991 | 2.917 | 3.77 | 1.31 |
| 120 | Lachesis | 7.8 | C | 155 | 0.058 | 3.115 | 2.956 | 1.64 | 0.44 |
| 121 | Hermione | 7.4 | Ch | 209 | 0.482 | 3.458 | 3.018 | 1.48 | 0.14 |
| 127 | Johanna | 8.6 | Ch | 122 | 0.042 | 2.755 | 2.571 | 1.66 | 1.23 |
| 128 | Nemesis | 7.8 | C | 163 | 0.067 | 2.751 | 2.397 | 1.63 | 0.66 |
| 130 | Elektra | 7.1 | Ch | 181 | 0.083 | 3.124 | 2.472 | 1.76 | 0.23 |
| 137 | Meliboea | 8.1 | Ch | 129 | 0.065 | 3.119 | 2.463 | 2.52 | 1.25 |
| 139 | Juewa | 8.0 | X | 151 | 0.052 | 2.782 | 2.296 | 2.82 | 0.63 |
| 142 | Polana | 10.3 | B | 55 | 0.047 | 2.418 | 2.092 | -1 | -1 |
| 144 | Vibilia | 8.1 | Ch | 142 | 0.0597 | 2.653 | 2.031 | 2.28 | 0.36 |
| 150 | Nuwa | 8.5 | Cb | 119 | 0.052 | 2.984 | 2.616 | -1 | -1 |
| 153 | Hilda | 7.5 | X | 171 | 0.062 | 3.971 | 3.42 | -1 | -1 |
| 154 | Bertha | 7.7 | C | 193 | 0.044 | 3.192 | 2.95 | 2.12 | 1.22 |
| 156 | Xanthippe | 8.7 | Ch | 143 | 0.03 | 2.729 | 2.11 | 3.35 | 0.69 |
| 159 | Aemilia | 8.3 | Ch | 125 | 0.064 | 3.103 | 2.773 | 2.67 | 1.01 |
| 163 | Erigone | 9.5 | Ch | 82 | 0.033 | 2.367 | 1.912 | -1 | -1 |
| 165 | Loreley | 7.8 | Cb | 180 | 0.086 | 3.124 | 2.859 | 3.07 | 3.16 |
| 168 | Sibylla | 7.9 | Ch | 144 | 0.0568 | 3.377 | 3.124 | -1 | -1 |
| 173 | Ino | 8.0 | Xk | 126 | 0.096 | 2.744 | 2.169 | 1.04 | 0.33 |
| 176 | Iduna | 7.9 | Ch | 107 | 0.065 | 3.189 | 2.638 | -1 | -1 |
| 181 | Eucharis | 7.9 | Xk | 115 | 0.101 | 3.125 | 2.478 | 5.46 | 2.43 |
| 194 | Prokne | 7.8 | C | 162 | 0.057 | 2.618 | 1.993 | 1.34 | 0.32 |
| 200 | Dynamene | 8.3 | Ch | 128 | 0.053 | 2.737 | 2.375 | 2.66 | 0.94 |
| 207 | Hedda | 9.9 | Ch | 61 | 0.0339 | 2.284 | 2.216 | -1 | -1 |
| 210 | Isabella | 9.5 | Cb | 87 | 0.044 | 2.722 | 2.383 | 0.92 | 0.46 |
| 211 | Isolda | 8.0 | Ch | 141 | 0.062 | 3.039 | 2.559 | 1.63 | 0.45 |
| 220 | Stephania | 11.3 | xc | 32 | 0.055 | 2.348 | 1.743 | -1 | -1 |
| 225 | Henrietta | 8.7 | f | 96 | 0.062 | 3.390 | 2.498 | -1 | -1 |
| 240 | Vanadis | 9.1 | C | 88 | 0.031 | 2.664 | 2.11 | 1.72 | 0.56 |
| 241 | Germania | 7.8 | B | 169 | 0.0575 | 3.051 | 2.733 | 2.19 | 0.81 |
| 247 | Eukrate | 8.2 | Xc | 131 | 0.064 | 2.741 | 2.069 | 0.41 | 0.21 |
| 259 | Aletheia | 7.8 | X | 174 | 0.067 | 3.134 | 2.727 | 2.52 | 0.9 |
| 261 | Prymno | 9.4 | X | 50 | 0.137 | 2.332 | 2.125 | -1 | -1 |
| 268 | Adorea | 8.4 | S3X | 145 | 0.041 | 3.093 | 2.669 | 1.49 | 0.95 |
| 284 | Amalia | 10.1 | Ch | 53 | 0.06 | 2.359 | 1.835 | -1 | -1 |
| 304 | Olga | 9.9 | Xc | 66 | 0.047 | 2.404 | 1.871 | 0.53 | 0.27 |
| 308 | Polyxo | 8.1 | T | 129 | 0.047 | 2.748 | 2.639 | 3.06 | 1.76 |
| 313 | Chaldaea | 8.9 | c | 71 | 0.044 | 2.375 | 1.942 | 0.95 | 0.4 |
| 324 | Bamberga | 7.0 | S3Cb | 221 | 0.0617 | 2.686 | 1.766 | 1.59 | 0.19 |
| 326 | Tamara | 9.4 | c | 93 | 0.0368 | 2.317 | 1.877 | 0.95 | 0.54 |
| 329 | Svea | 9.6 | S3C | 81 | 0.039 | 2.478 | 2.416 | -1 | -1 |
| 334 | Chicago | 7.7 | c | 199 | 0.041 | 3.889 | 3.795 | -1 | -1 |
| 335 | Roberta | 9.0 | B | 97 | 0.049 | 2.474 | 2.047 | -1 | -1 |
| 336 | Lacadiera | 9.8 | Xk | 63 | 0.054 | 2.253 | 2.036 | -1 | -1 |
| 345 | Tercidina | 8.7 | Ch | 90 | 0.052 | 2.326 | 2.181 | -1 | -1 |
| 360 | Carlova | 8.5 | C | 129 | 0.043 | 3.003 | 2.483 | 6.62 | 4.51 |
| 363 | Padua | 9.1 | X | 97 | 0.059 | 2.746 | 2.55 | 4.1 | 2.25 |
| 372 | Palma | 7.5 | B | 174 | 0.059 | 3.148 | 2.343 | 1.65 | 0.29 |
| 375 | Ursula | 7.4 | Xc | 216 | 0.065 | 3.122 | 2.8 | 2.06 | 1.13 |
| 386 | Siegena | 7.7 | C | 165 | 0.0692 | 2.895 | 2.409 | 2.66 | 1.15 |
| 404 | Arsinoe | 9.0 | Ch | 95 | 0.0451 | 2.593 | 2.077 | -1 | -1 |
| 405 | Thia | 8.5 | Ch | 109 | 0.0467 | 2.584 | 1.956 | 1.31 | 0.18 |
| 407 | Arachne | 8.9 | S3Ch | 95 | 0.052 | 2.624 | 2.44 | -1 | -1 |
| 409 | Aspasia | 7.6 | Xc | 171 | 0.054 | 2.578 | 2.39 | 3.62 | 1.07 |
| 423 | Diotima | 7.3 | C | 176 | 0.058 | 3.067 | 2.959 | 1.33 | 0.46 |
| 431 | Nephele | 8.7 | B | 102 | 0.055 | 3.138 | 2.605 | -1 | -1 |
| 451 | Patientia | 6.7 | S3Cb | 254 | 0.085 | 3.061 | 2.838 | 1.79 | 0.36 |
| 466 | Tisiphone | 8.4 | c | 95 | 0.092 | 3.354 | 3.053 | 0.65 | 0.35 |
| 476 | Hedwig | 8.6 | X | 138 | 0.035 | 2.65 | 2.457 | 1.6 | 0.41 |
| 479 | Caprera | 9.8 | C | 60 | 0.071 | 2.719 | 2.124 | -1 | -1 |
| 510 | Mabella | 9.8 | S3T | 60 | 0.062 | 2.611 | 2.112 | -1 | -1 |
| 511 | Davida | 6.3 | C | 270 | 0.076 | 3.166 | 2.569 | 2.15 | 0.78 |
| 524 | Fidelio | 9.9 | S3Ch | 65 | 0.048 | 2.635 | 2.297 | -1 | -1 |
| 554 | Peraga | 9 | Ch | 96 | 0.049 | 2.375 | 2.015 | 1.58 | 0.23 |
| 559 | Nanon | 9.5 | Xk | 80 | 0.05 | 2.711 | 2.534 | -1 | -1 |
| 566 | Stereoskopia | 8 | C | 167 | 0.04 | 3.383 | 2.973 | 1.91 | 0.49 |
| 576 | Emanuela | 9.4 | S3Cgh | 77 | 0.052 | 2.986 | 2.416 | -1 | -1 |
| 585 | Bilkis | 10.5 | c | 50 | 0.014 | 2.433 | 2.122 | -1 | -1 |
| 602 | Marianna | 8.3 | S3Ch | 110 | 0.052 | 3.09 | 2.31 | -1 | -1 |
| 654 | Zelinda | 8.5 | Ch | 160 | 0.027 | 2.296 | 1.768 | 1.31 | 0.24 |
| 688 | Melanie | 10.9 | C | 42 | 0.068 | 2.698 | 2.322 | -1 | -1 |
| 690 | Wratislavia | 8 | S3B | 135 | 0.0604 | 3.142 | 2.591 | 1.3 | 0.44 |
| 694 | Ekard | 9.1 | S3Ch | 122 | 0.045 | 2.671 | 1.808 | -1 | -1 |
| 702 | Alauda | 7.3 | B | 191 | 0.061 | 3.193 | 3.14 | 2.7 | 1.94 |
| 704 | Interamnia | 6.4 | B | 306 | 0.078 | 3.060 | 2.582 | 2.1 | 0.38 |
| 705 | Erminia | 8.4 | C | 132 | 0.031 | 2.922 | 2.775 | 4.02 | 2.39 |
| 712 | Boliviana | 8.5 | X | 124 | 0.039 | 2.575 | 2.097 | 1.83 | 0.62 |
| 747 | Winchester | 7.8 | C | 172 | 0.0503 | 2.998 | 1.983 | 1.65 | 0.57 |
| 754 | Malabar | 9.2 | Ch | 95 | 0.042 | 2.987 | 2.836 | -1 | -1 |
| 773 | Irmintraud | 9.1 | T | 92 | 0.048 | 2.859 | 2.634 | 1.31 | 0.72 |
| 776 | Berbericia | 7.7 | Cgh | 152 | 0.065 | 2.934 | 2.445 | 1.3 | 1.2 |
| 778 | Theobalda | 9.7 | S3C | 55 | 0.079 | 3.178 | 2.366 | 1.1 | 0.65 |
| 790 | Pretoria | 8.1 | S3X | 170 | 0.038 | 3.408 | 2.888 | -1 | -1 |
| 914 | Palisana | 9.2 | S3Ch | 76 | 0.095 | 2.458 | 1.931 | 1.48 | 0.74 |
| 1343 | Nicole | 11.1 | C | 24 | 0.136 | 2.569 | 2.278 | -1 | -1 |
| 1467 | Mashona | 8.57 | S3Ch | 104 | 0.061 | 3.38 | 2.951 | 0.93 | 0.49 |

***Table A3: Gaussian fits to continuum-removed spectra. BD is presented as a negative number so that using these numbers in Equation 1 will produce an absorption rather than an emission. Observations that could not be successfully fit with a Gaussian are included for completeness, with all entries equal to -1.***

| Date | Numb | Name | BD | err | BC | err | Width | err |
|---|---|---|---|---|---|---|---|---|
| 2-Mar-05 | 2 | Pallas | -0.1662 | 0.0048 | 2.8559 | 0.0096 | 0.1405 | 0.0067 |
| 12-Jun-07 | 2 | Pallas | -0.1606 | 0.0054 | 2.8647 | 0.0083 | 0.1649 | 0.0063 |
| 13-Sep-07 | 2 | Pallas | -0.1718 | 0.0066 | 2.8509 | 0.0116 | 0.1904 | 0.0084 |
| 23-Jun-11 | 2 | Pallas1 | -0.1853 | 0.0058 | 2.8707 | 0.0086 | 0.1979 | 0.0067 |
| 23-Jun-11 | 2 | Pallas2 | -0.1714 | 0.0073 | 2.8947 | 0.0117 | 0.1941 | 0.0089 |
| 23-Aug-12 | 2 | Pallas | -0.1781 | 0.0081 | 2.8732 | 0.0153 | 0.1945 | 0.0121 |
| 29-Aug-12 | 2 | Pallas1 | -0.162 | 0.0062 | 2.8873 | 0.0128 | 0.1941 | 0.0111 |
| 6-Dec-13 | 2 | Pallas1 | -0.1255 | 0.0111 | 2.8716 | 0.0267 | 0.1226 | 0.0191 |
| 6-Dec-13 | 2 | Pallas2 | -0.1546 | 0.0076 | 2.8591 | 0.0154 | 0.143 | 0.0102 |
| 6-Dec-13 | 2 | Pallas3 | -0.1691 | 0.0181 | 2.874 | 0.035 | 0.1699 | 0.0243 |
| 2-Dec-15 | 10 | Hygiea | -0.1597 | 0.006 | 3.1436 | 0.0165 | 0.3408 | 0.0137 |
| 16-Feb-16 | 10 | Hygiea | -0.0953 | 0.0031 | 3.0968 | 0.0135 | 0.3052 | 0.0194 |
| 18-Feb-16 | 10 | Hygiea | -0.1207 | 0.004 | 3.1121 | 0.0127 | 0.03367 | 0.0208 |
| 12-Oct-05 | 36 | Atalante | -0.15 | 0.011 | 2.8913 | 0.0166 | 0.2865 | 0.019 |
| 30-Oct-10 | 36 | Atalante | -0.1948 | 0.0047 | 2.9306 | 0.0087 | 0.2471 | 0.0086 |
| 17-Sep-06 | 45 | Eugenia | -0.146 | 0.0043 | 3.0136 | 0.0103 | 0.2866 | 0.0079 |
| 22-Jul-13 | 46 | Hestia | -0.1155 | 0.0215 | 2.9516 | 0.0678 | 0.1831 | 0.0527 |
| 30-Apr-07 | 47 | Aglaja | -0.1107 | 0.006 | 2.9791 | 0.0181 | 0.2177 | 0.0236 |
| 21-Jul-13 | 47 | Aglaja | -0.0776 | 0.0099 | 2.9565 | 0.0352 | 0.1632 | 0.0418 |
| 16-Feb-16 | 52 | Europa | -0.1156 | 0.0026 | 3.1244 | 0.0077 | 0.2776 | 0.0119 |
| 31-Aug-04 | 53 | Kalypso | -0.2182 | 0.0056 | 2.961 | 0.0083 | 0.2828 | 0.0078 |
| 31-Oct-10 | 54 | Alexandra | -0.2249 | 0.0032 | 2.9183 | 0.005 | 0.1817 | 0.0045 |
| 30-Apr-07 | 56 | Melete | -0.106 | 0.0034 | 2.9422 | 0.0111 | 0.2414 | 0.0117 |
| 29-Aug-12 | 56 | Melete | -0.1492 | 0.006 | 2.9944 | 0.0144 | 0.2978 | 0.0115 |
| 23-Aug-12 | 59 | Elpis | -0.0512 | 0.0013 | 3.0001 | 0.0095 | 0.2241 | 0.0119 |
| 21-Feb-07 | 65 | Cybele | -0.124 | 0.0034 | 3.0611 | 0.0113 | 0.2711 | 0.0122 |
| 22-Feb-07 | 65 | Cybele | -0.1353 | 0.0037 | 3.1692 | 0.0095 | 0.283 | 0.0124 |
| 20-Mar-07 | 65 | Cybele | -0.1514 | 0.0041 | 3.16 | 0.0089 | 0.2961 | 0.0103 |
| 29-Apr-07 | 65 | Cybele | -0.1884 | 0.0038 | 3.2392 | 0.0061 | 0.3659 | 0.0116 |
| 20-Jun-08 | 65 | Cybele | -0.1322 | 0.0063 | 3.1882 | 0.0144 | 0.2854 | 0.0223 |
| 21-Jul-09 | 65 | Cybele | -0.1332 | 0.0069 | 3.1977 | 0.0158 | 0.2979 | 0.0271 |
| 4-Jan-13 | 65 | Cybele | -0.1176 | 0.0063 | 3.1735 | 0.0159 | 0.2916 | 0.0259 |
| 12-Oct-05 | 66 | Maja | -0.2462 | 0.0092 | 2.8545 | 0.0075 | 0.1981 | 0.0066 |
| 21-Sep-05 | 72 | Feronia | -0.0488 | 0.0031 | 2.9877 | 0.0223 | 0.1855 | 0.026 |
| 2-Dec-15 | 74 | Galatea | -0.0851 | 0.0058 | 3.0429 | 0.0235 | 0.2308 | 0.0299 |
| 1-Sep-12 | 76 | Freia | -0.1088 | 0.0116 | 3.1459 | 0.0214 | 0.1773 | 0.0288 |
| 6-Dec-13 | 76 | Freia | -0.0692 | 0.0082 | 2.9897 | 0.0385 | 0.2202 | 0.042 |
| 18-Feb-16 | 81 | Terpsichore | -0.1481 | 0.0223 | 3.0269 | 0.0486 | 0.2549 | 0.0392 |
| 16-Feb-16 | 83 | Beatrix | -0.0258 | 0.0039 | 3.5985 | 0.0713 | 0.4628 | 0.0885 |
| 2-Apr-07 | 85 | Io | -0.1456 | 0.0029 | 3.1293 | 0.0084 | 0.3244 | 0.0102 |
| 12-Oct-05 | 87 | Sylvia | -0.0715 | 0.0028 | 3.1658 | 0.0111 | 0.269 | 0.0176 |
| 10-Dec-06 | 87 | Sylvia | -0.1341 | 0.0051 | 3.1943 | 0.0116 | 0.3608 | 0.0193 |
| 20-Jun-08 | 87 | Sylvia | -0.0934 | 0.0199 | 3.2262 | 0.0189 | 0.0988 | 0.0265 |
| 5-Jan-13 | 87 | Sylvia | -0.2076 | 0.0197 | 3.3087 | 0.023 | 0.3432 | 0.0438 |
| 24-Aug-17 | 87 | Sylvia | -0.0738 | 0.0056 | 3.1566 | 0.0204 | 0.239 | 0.0268 |
| 13-Oct-05 | 88 | Thisbe | -0.1666 | 0.0042 | 3.1745 | 0.0092 | 0.3918 | 0.0167 |
| 28-Aug-11 | 90 | Antiope | -0.1704 | 0.0033 | 3.1379 | 0.007 | 0.3084 | 0.0112 |
| 27-Feb-03 | 93 | Minerva | -0.0809 | 0.0033 | 3.0878 | 0.0173 | 0.2965 | 0.0165 |
| 22-Feb-07 | 93 | Minerva | -0.1316 | 0.0062 | 3.1379 | 0.0173 | 0.2795 | 0.0199 |
| 14-Aug-09 | 93 | Minerva | -0.0827 | 0.007 | 2.9179 | 0.0312 | 0.2068 | 0.0275 |
| 21-Jul-13 | 94 | Aurora | -0.123 | 0.007 | 3.0529 | 0.0197 | 0.2457 | 0.0297 |
| 29-Apr-07 | 96 | Aegle | -0.1579 | 0.0046 | 2.9203 | 0.0106 | 0.2356 | 0.0105 |
| 30-Jan-12 | 96 | Aegle | -0.1477 | 0.0053 | 2.9567 | 0.0135 | 0.2149 | 0.0149 |
| 22-Nov-07 | 102 | Miriam | -0.2112 | 0.0096 | 2.8937 | 0.0096 | 0.276 | 0.0103 |
| 16-Feb-16 | 106 | Dione | -0.2755 | 0.0102 | 2.9642 | 0.0123 | 0.1928 | 0.0154 |
| 18-Nov-05 | 107 | Camilla | -0.1022 | 0.0029 | 3.0165 | 0.0114 | 0.2452 | 0.0162 |
| 21-Jun-08 | 107 | Camilla | -0.2002 | 0.0301 | 3.3024 | 0.0387 | 0.4442 | 0.0746 |
| 24-Aug-17 | 107 | Camilla | -0.1083 | 0.0062 | 3.1351 | 0.0149 | 0.2686 | 0.0293 |
| 2-Jan-13 | 117 | Lomia | -0.1594 | 0.048 | 2.8892 | 0.0577 | 0.2083 | 0.0389 |
| 1-Sep-12 | 120 | Lachesis | -0.1293 | 0.0115 | 2.9523 | 0.0283 | 0.1962 | 0.0315 |
| 17-May-06 | 121 | Hermione | -0.2242 | 0.0127 | 2.955 | 0.017 | -0.1692 | 0.0178 |
| 15-Jun-07 | 121 | Hermione | -0.2007 | 0.0118 | 3.0268 | 0.0206 | 0.2533 | 0.0209 |
| 20-Jun-08 | 121 | Hermione | -0.2367 | 0.0091 | 2.9339 | 0.0107 | 0.238 | 0.0085 |
| 20-Mar-02 | 128 | Nemesis | -0.1381 | 0.0088 | 2.9874 | 0.0208 | 0.1917 | 0.0285 |
| 6-Mar-15 | 128 | Nemesis | -0.1485 | 0.0163 | 2.984 | 0.0387 | 0.2021 | 0.0434 |
| 27-Apr-15 | 128 | Nemesis | -0.1135 | 0.0081 | 2.9986 | 0.0235 | 0.2105 | 0.0321 |
| 11-Jun-07 | 137 | Meliboea | -0.1688 | 0.0052 | 2.9278 | 0.0091 | 0.2285 | 0.0086 |
| 29-Sep-13 | 139 | Juewa | -0.1298 | 0.0052 | 3.1439 | 0.0158 | 0.2874 | 0.0177 |
| 18-Feb-16 | 139 | Juewa | -0.1322 | 0.0029 | 3.1306 | 0.0094 | 0.3283 | 0.011 |
| 2-Mar-03 | 140 | Siwa | -1 | -1 | -1 | -1 | -1 | -1 |
| 12-Oct-05 | 140 | Siwa | -1 | -1 | -1 | -1 | -1 | -1 |
| 13-Oct-05 | 140 | Siwa | -0.0327 | 0.0036 | 3.0578 | 0.0192 | 0.1653 | 0.0309 |
| 26-Feb-03 | 142 | Polana | -0.0918 | 0.012 | 2.9771 | 0.0405 | 0.2736 | 0.0377 |
| 22-Jul-13 | 150 | Nuwa | -0.1664 | 0.0161 | 3.0902 | 0.0353 | 0.2636 | 0.0527 |
| 18-Feb-16 | 150 | Nuwa | -0.1226 | 0.0224 | 3.1906 | 0.0477 | 0.2301 | 0.0603 |
| 15-Jun-07 | 153 | Hilda | -0.0874 | 0.0067 | 3.1852 | 0.0277 | 0.3177 | 0.0457 |
| 20-Jun-08 | 153 | Hilda | -0.0619 | 0.0074 | 3.1056 | 0.033 | 0.2267 | 0.0512 |
| 28-Feb-03 | 154 | Bertha | -0.1538 | 0.0035 | 3.1327 | 0.0074 | 0.2509 | 0.0113 |
| 29-Aug-04 | 160 | Una | -0.1242 | 0.0059 | 3.0781 | 0.0174 | 0.2553 | 0.0242 |
| 7-Sep-09 | 165 | Loreley | -0.0623 | 0.0128 | 3.0035 | 0.0629 | 0.1768 | 0.0832 |
| 31-Aug-04 | 173 | Ino | -0.0918 | 0.0035 | 2.9601 | 0.0139 | 0.187 | 0.016 |
| 15-Aug-09 | 173 | Ino | -0.1464 | 0.0056 | 2.9511 | 0.0132 | 0.2355 | 0.0139 |
| 20-Sep-10 | 175 | Andromache | -0.1577 | 0.0061 | 3.0186 | 0.0146 | 0.3161 | 0.0137 |
| 26-Jun-13 | 181 | Eucharis | -0.3105 | 0.0558 | 2.7443 | 0.0084 | -0.1509 | 0.0106 |
| 1-Sep-12 | 185 | Eunike | -0.1081 | 0.0087 | 3.0366 | 0.0267 | 0.2173 | 0.0319 |
| 20-Mar-07 | 190 | Ismene | -0.0507 | 0.0064 | 3.1108 | 0.021 | 0.1362 | 0.0269 |
| 5-Nov-20 | 190 | Ismene | -1 | -1 | -1 | -1 | -1 | -1 |
| 8-Dec-17 | 191 | Kolga | -0.1651 | 0.0201 | 3.0191 | 0.0136 | 0.1036 | 0.0172 |
| 16-May-06 | 194 | Prokne | -0.1989 | 0.0084 | 2.9272 | 0.0151 | 0.1766 | 0.0157 |
| 11-Jun-07 | 194 | Prokne | -0.21 | 0.0039 | 2.9024 | 0.0053 | 0.1978 | 0.0046 |
| 18-Feb-16 | 209 | Dido | -0.0667 | 0.0093 | 3.0691 | 0.0425 | 0.216 | 0.0564 |
| 27-Jun-13 | 210 | Isabella | -0.6203 | 0.0527 | 2.8373 | 0.0105 | 0.2042 | 0.0093 |
| 18-Oct-04 | 212 | Medea | -0.0231 | 0.0031 | 3.0821 | 0.0506 | 0.2532 | 0.0622 |
| 17-Oct-04 | 220 | Stephania | -0.0826 | 0.012 | 2.9962 | 0.0495 | 0.2364 | 0.051 |
| 12-Jun-07 | 225 | Henrietta | -0.169 | 0.0222 | 2.8353 | 0.0166 | 0.1918 | 0.0157 |
| 15-Jun-07 | 225 | Henrietta | -0.2068 | 0.0083 | 3.2323 | 0.0143 | 0.3816 | 0.0268 |
| 14-Sep-07 | 225 | Henrietta | -0.1307 | 0.0075 | 3.1193 | 0.0226 | 0.3167 | 0.0263 |
| 2-Dec-15 | 240 | Vanadis | -0.1677 | 0.0035 | 2.8841 | 0.0045 | 0.2396 | 0.0039 |
| 16-Feb-16 | 240 | Vanadis | -0.1697 | 0.0078 | 2.9157 | 0.0132 | 0.191 | 0.0115 |
| 18-Sep-06 | 241 | Germania | -0.1293 | 0.0068 | 3.1671 | 0.0194 | 0.2948 | 0.0294 |
| 2-Mar-03 | 247 | Eukrate | -0.081 | 0.0122 | 3.2455 | 0.0404 | 0.2538 | 0.0523 |
| 18-Dec-05 | 259 | Aletheia | -0.0932 | 0.0155 | 2.8731 | 0.0299 | 0.2355 | 0.028 |
| 25-Jun-13 | 261 | Prymno | -0.077 | 0.0109 | 3.0518 | 0.0422 | 0.2171 | 0.065 |
| 27-Jun-12 | 267 | Tirza | -1 | -1 | | -1 | -1 | -1 |
| 27-Jun-12 | 268 | Adorea | -0.0468 | 0.0088 | 3.1181 | 0.0435 | 0.1966 | 0.0601 |
| 5-Nov-20 | 284 | Amalia | -0.6841 | 0.1751 | 2.8161 | 0.016 | 0.1377 | 0.01 |
| 18-Sep-06 | 308 | Polyxo | -0.2142 | 0.0133 | 3.0006 | 0.0211 | 0.2884 | 0.0197 |
| 8-Jan-13 | 308 | Polyxo | -0.1731 | 0.0082 | 2.8956 | 0.0192 | 0.1687 | 0.0178 |
| 30-Aug-04 | 313 | Chaldaea | -0.3601 | 0.0145 | 2.8909 | 0.0089 | 0.2112 | 0.0074 |
| 18-Feb-16 | 324 | Bamberga | -0.148 | 0.0137 | 3.1634 | 0.0305 | 0.262 | 0.0384 |
| 3-Mar-19 | 326 | Tamara | -0.1813 | 0.0086 | 2.947 | 0.0127 | 0.2255 | 0.0112 |
| 23-Aug-02 | 329 | Svea | -0.1314 | 0.0086 | 3.1313 | 0.0165 | 0.2156 | 0.0219 |
| 14-Sep-07 | 334 | Chicago | -0.1791 | 0.0159 | 3.0483 | 0.0102 | 0.106 | 0.014 |
| 8-Dec-17 | 334 | Chicago | -0.1485 | 0.015 | 3.0347 | 0.0283 | 0.2073 | 0.0399 |
| 12-Oct-05 | 335 | Roberta | -0.0603 | 0.005 | 2.9782 | 0.00306 | 0.2047 | 0.0364 |
| 13-Oct-05 | 335 | Roberta | -0.0666 | 0.0042 | 3.025 | 0.0221 | 0.2327 | 0.0312 |
| 17-Dec-05 | 335 | Roberta | -0.441 | 0.0448 | 2.8234 | 0.0109 | 0.2066 | 0.0113 |
| 20-Jun-08 | 335 | Roberta | -0.0611 | 0.0056 | 2.9512 | 0.0281 | 0.3395 | 0.0228 |
| 15-May-06 | 336 | Lacadiera | -0.0486 | 0.0098 | 2.9603 | 0.0495 | 0.1413 | 0.0546 |
| 30-Jan-12 | 360 | Carlova | -0.0817 | 0.0036 | 3.1105 | 0.013 | 0.2377 | 0.0196 |
| 25-Nov-08 | 361 | Bononia | -1 | -1 | -1 | -1 | -1 | -1 |
| 17-Dec-05 | 363 | Padua | -0.2209 | 0.0115 | 2.9092 | 0.0124 | 0.2109 | 0.0108 |
| 18-Dec-05 | 363 | Padua | -0.1753 | 0.009 | 2.9391 | 0.0162 | 0.2222 | 0.0148 |
| 16-Feb-16 | 363 | Padua | -0.1958 | 0.0082 | 2.9601 | 0.0128 | 0.2228 | 0.0144 |
| 23-Aug-12 | 368 | Haidea | -0.2036 | 0.005 | 3.0163 | 0.0098 | 0.3504 | 0.0071 |
| 22-Sep-05 | 372 | Palma | -0.1356 | 0.0055 | 3.1048 | 0.0106 | 0.2124 | 0.0145 |
| 2-Jan-13 | 372 | Palma | -0.1647 | 0.0091 | 3.212 | 0.0178 | 0.2871 | 0.0221 |
| 15-Aug-09 | 375 | Ursula | -0.1371 | 0.0052 | 3.2301 | 0.0138 | 0.3375 | 0.0242 |
| 2-Dec-15 | 375 | Ursula | -0.129 | 0.0085 | 3.1611 | 0.0208 | 0.2667 | 0.0259 |
| 29-Aug-04 | 386 | Siegena | -0.238 | 0.0089 | 2.8878 | 0.008 | -0.2038 | 0.0065 |
| 31-Oct-10 | 409 | Aspasia | -0.1493 | 0.0045 | 2.8923 | 0.0078 | 0.2206 | 0.0069 |
| 21-Jun-08 | 420 | Bertholda | -1 | -1 | -1 | -1 | -1 | -1 |
| 27-Jun-12 | 422 | Berolina | -0.1824 | 0.0163 | 2.9838 | 0.0274 | 0.1839 | 0.0361 |
| 28-Jun-12 | 423 | Diotima | -0.1171 | 0.0089 | 3.2314 | 0.0232 | 0.2858 | 0.0293 |
| 22-Dec-20 | 423 | Diotima | -0.1511 | 0.0136 | 3.1547 | 0.0304 | 0.2707 | 0.0367 |
| 4-Jan-13 | 426 | Hippo | -0.1338 | 0.0085 | 3.1865 | 0.0244 | 0.3379 | 0.0351 |
| 2-Dec-15 | 429 | Lotis | -0.1252 | 0.0053 | 3.0239 | 0.0142 | 0.2749 | 0.0133 |
| 13-Sep-07 | 431 | Nephele | -0.1612 | 0.0033 | 3.1114 | 0.0074 | 0.2888 | 0.011 |
| 13-Sep-07 | 444 | Gyptis | -0.0864 | 0.0063 | 2.9242 | 0.0218 | 0.2601 | 0.0203 |
| 8-Dec-17 | 451 | Patientia | -0.1237 | 0.0149 | 3.0884 | 0.0479 | 0.2666 | 0.0526 |
| 17-Dec-05 | 455 | Bruchsalia | -0.126 | 0.0054 | 3.0812 | 0.018 | 0.3191 | 0.0204 |
| 18-Dec-05 | 455 | Bruchsalia | -0.1134 | 0.0077 | 2.9316 | 0.0163 | 0.2313 | 0.0139 |
| 18-Oct-04 | 463 | Lola | -0.0356 | 0.0073 | 3.1936 | 0.0654 | 0.2769 | 0.0693 |
| 4-Jan-13 | 466 | Tisiphone | -0.1918 | 0.0271 | 2.9139 | 0.0379 | 0.2792 | 0.0436 |
| 11-Oct-05 | 476 | Hedwig | -0.1537 | 0.0048 | 2.932 | 0.0093 | 0.2539 | 0.009 |
| 12-Oct-05 | 476 | Hedwig | -0.1385 | 0.0058 | 2.9458 | 0.0144 | 0.2186 | 0.0155 |
| 13-Oct-05 | 476 | Hedwig | -0.1481 | 0.0067 | 2.9363 | 0.0152 | 0.2202 | 0.0152 |
| 18-Dec-05 | 476 | Hedwig | -0.1841 | 0.0195 | 2.9397 | 0.0386 | 0.1857 | 0.0351 |
| 25-Jun-13 | 476 | Hedwig | -0.191 | 0.0063 | 2.9376 | 0.0107 | 0.2373 | 0.0099 |
| 29-Sep-13 | 479 | Caprera | -0.1149 | 0.0046 | 2.9413 | 0.0145 | 0.3357 | 0.0165 |
| 12-Oct-05 | 510 | Mabella | -0.1314 | 0.0408 | 2.7571 | 0.0188 | 0.1714 | 0.024 |
| 26-Jun-13 | 510 | Mabella | -0.0604 | 0.0035 | 2.9604 | 0.0209 | 0.2622 | 0.0213 |
| 2-Mar-03 | 511 | Davida | -0.1157 | 0.0047 | 2.9607 | 0.0159 | 0.2613 | 0.0165 |
| 25-Jul-06 | 511 | Davida | -0.1233 | 0.0059 | 2.9888 | 0.0166 | 0.244 | 0.0138 |
| 17-Sep-06 | 511 | Davida | -0.1586 | 0.0068 | 2.9679 | 0.0139 | 0.2206 | 0.0142 |
| 18-Dec-05 | 514 | Armida | -0.1621 | 0.0134 | 2.9442 | 0.0251 | 0.3048 | 0.0262 |
| 17-Dec-05 | 524 | Fidelio | -0.1354 | 0.011 | 2.967 | 0.0229 | 0.2528 | 0.021 |
| 18-Dec-05 | 524 | Fidelio | -0.2221 | 0.0207 | 2.8467 | 0.0124 | 0.2248 | 0.0124 |
| 8-Jan-13 | 536 | Merapi | -1 | -1 | -1 | -1 | -1 | -1 |
| 24-Aug-17 | 536 | Merapi | -0.1256 | 0.0079 | 3.1799 | 0.0229 | 0.3181 | 0.032 |
| 2-Dec-15 | 547 | Praxedis | -0.0755 | 0.0091 | 2.9142 | 0.0314 | 0.3009 | 0.0249 |
| 27-Jun-13 | 559 | Nanon | -0.4497 | 0.0311 | 2.8192 | 0.0072 | 0.2111 | 0.0076 |
| 29-Aug-06 | 566 | Stereoskopia | -0.084 | 0.0037 | 3.0058 | 0.0147 | 0.2303 | 0.0164 |
| 23-Aug-12 | 566 | Stereoskopia | -0.101 | 0.0065 | 3.0241 | 0.0262 | 0.3095 | 0.0276 |
| 23-Aug-02 | 585 | Bilkis | -0.1643 | 0.0143 | 2.9881 | 0.0371 | 0.4082 | 0.0501 |
| 12-Jun-07 | 596 | Scheila | -0.0361 | 0.0067 | 3.241 | 0.0304 | 0.1854 | 0.0376 |
| 23-Aug-02 | 688 | Melanie | -0.259 | 0.0142 | 2.9391 | 0.0139 | 0.3009 | 0.0145 |
| 17-Oct-04 | 690 | Wratislavia | -0.1218 | 0.0036 | 3.1026 | 0.0108 | 0.2691 | 0.0153 |
| 30-Oct-10 | 690 | Wratislavia | -0.1018 | 0.0036 | 3.0887 | 0.0116 | 0.3545 | 0.0178 |
| 7-Sep-09 | 694 | Ekard | -0.1608 | 0.0091 | 2.888 | 0.0161 | 0.2366 | 0.013 |
| 25-Jun-13 | 702 | Alauda | -0.1995 | 0.007 | 3.1897 | 0.0151 | 0.3515 | 0.0145 |
| 27-Apr-15 | 704 | Interamnia | -0.0703 | 0.0079 | 3.0161 | 0.0329 | 0.2015 | 0.048 |
| 11-Oct-05 | 705 | Erminia | -0.2457 | 0.0125 | 3.1363 | 0.0214 | 0.3714 | 0.0301 |
| 12-Oct-05 | 705 | Erminia | -0.046 | 0.0046 | 3.1016 | 0.0201 | 0.19 | 0.0301 |
| 31-Oct-10 | 705 | Erminia | -0.0748 | 0.0041 | 3.0225 | 0.0165 | 0.2121 | 0.0259 |
| 26-Jun-13 | 712 | Boliviana | -0.1842 | 0.0129 | 2.9661 | 0.0193 | 0.2598 | 0.0145 |
| 30-Oct-10 | 747 | Winchester | -0.0409 | 0.0031 | 3.0416 | 0.0173 | 0.1864 | 0.0253 |
| 18-Feb-16 | 762 | Pulcova | -0.1346 | 0.0115 | 3.1279 | 0.031 | 0.2628 | 0.0351 |
| 1-Sep-04 | 773 | Irmintraud | -0.1791 | 0.0268 | 2.9986 | 0.0272 | 0.14 | 0.0363 |
| 2-Dec-15 | 778 | Theobalda | -0.0739 | 0.0079 | 3.023 | 0.0368 | 0.2753 | 0.0263 |
| 30-Aug-04 | 783 | Nora | -1 | -1 | -1 | -1 | -1 | -1 |
| 21-Jun-08 | 790 | Pretoria | -0.1579 | 0.0188 | 3.3009 | 0.0246 | 0.2337 | 0.0327 |
| 10-Aug-08 | 790 | Pretoria | -0.1214 | 0.0059 | 3.1158 | 0.0174 | 0.2765 | 0.0254 |
| 12-Oct-05 | 838 | Seraphina | -1 | -1 | | -1 | -1 | -1 |
| 13-Oct-05 | 838 | Seraphina | -0.1398 | 0.011 | 3.0057 | 0.0284 | 0.33 | 0.0283 |
| 10-Dec-06 | 895 | Helio | -0.091 | 0.0035 | 3.0602 | 0.01 | 0.2095 | 0.0157 |
| 1-Sep-04 | 924 | Toni | -0.2371 | 0.0198 | 3.0169 | 0.0097 | 0.1211 | 0.0129 |
| 23-Aug-02 | 1343 | Nicole | -0.1822 | 0.2498 | 2.8426 | 0.2598 | 0.1221 | 0.1357 |
| 2-Dec-15 | 1963 | Bezovec | -0.1878 | 0.0079 | 2.8791 | 0.0101 | 0.3634 | 0.012 |

***Figure A1: Continuum-removed reflectance spectra for new objects. Asteroid number and date of observation is included in each panel, and each panel shares the same x and y axis bounds and tick marks.***

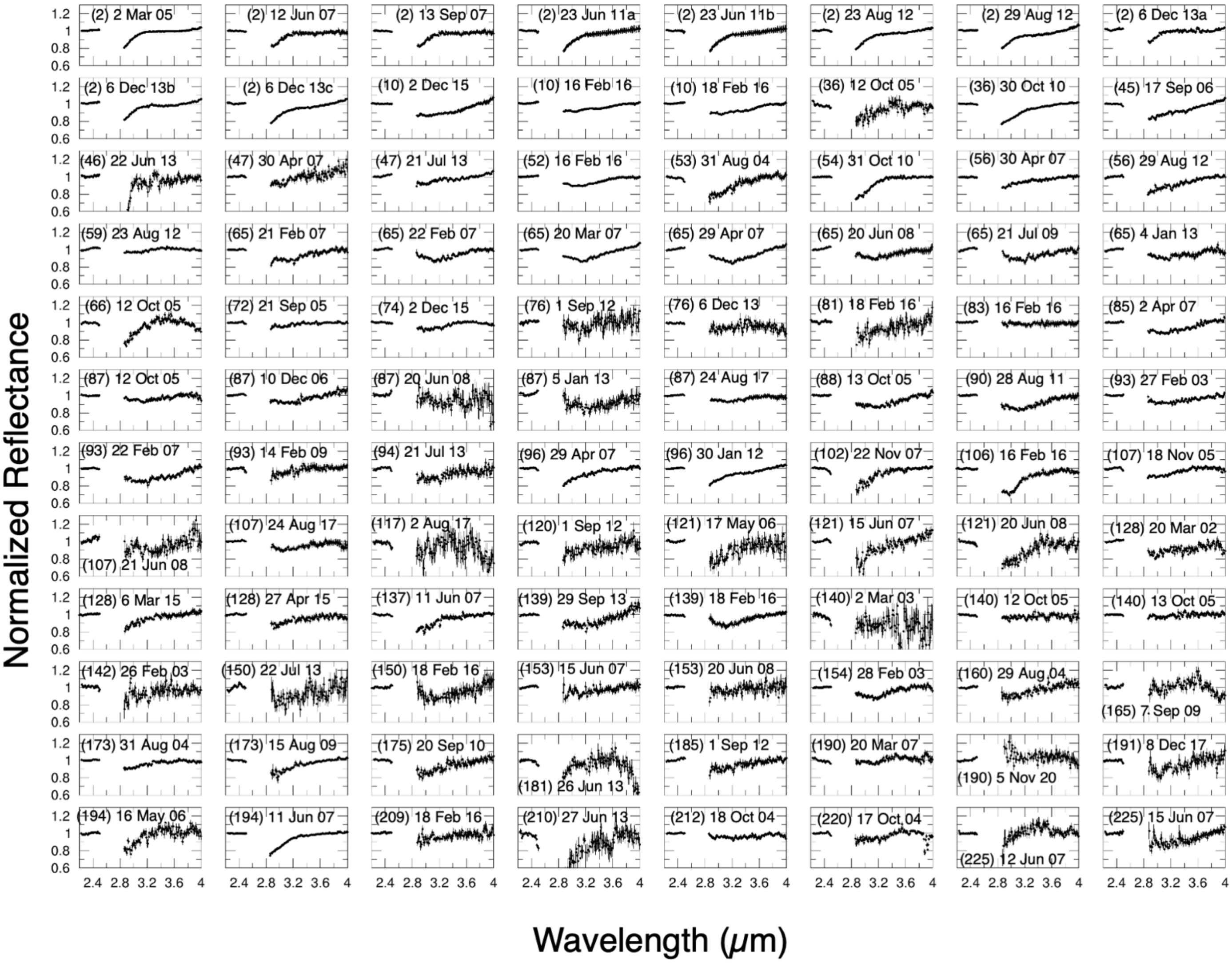

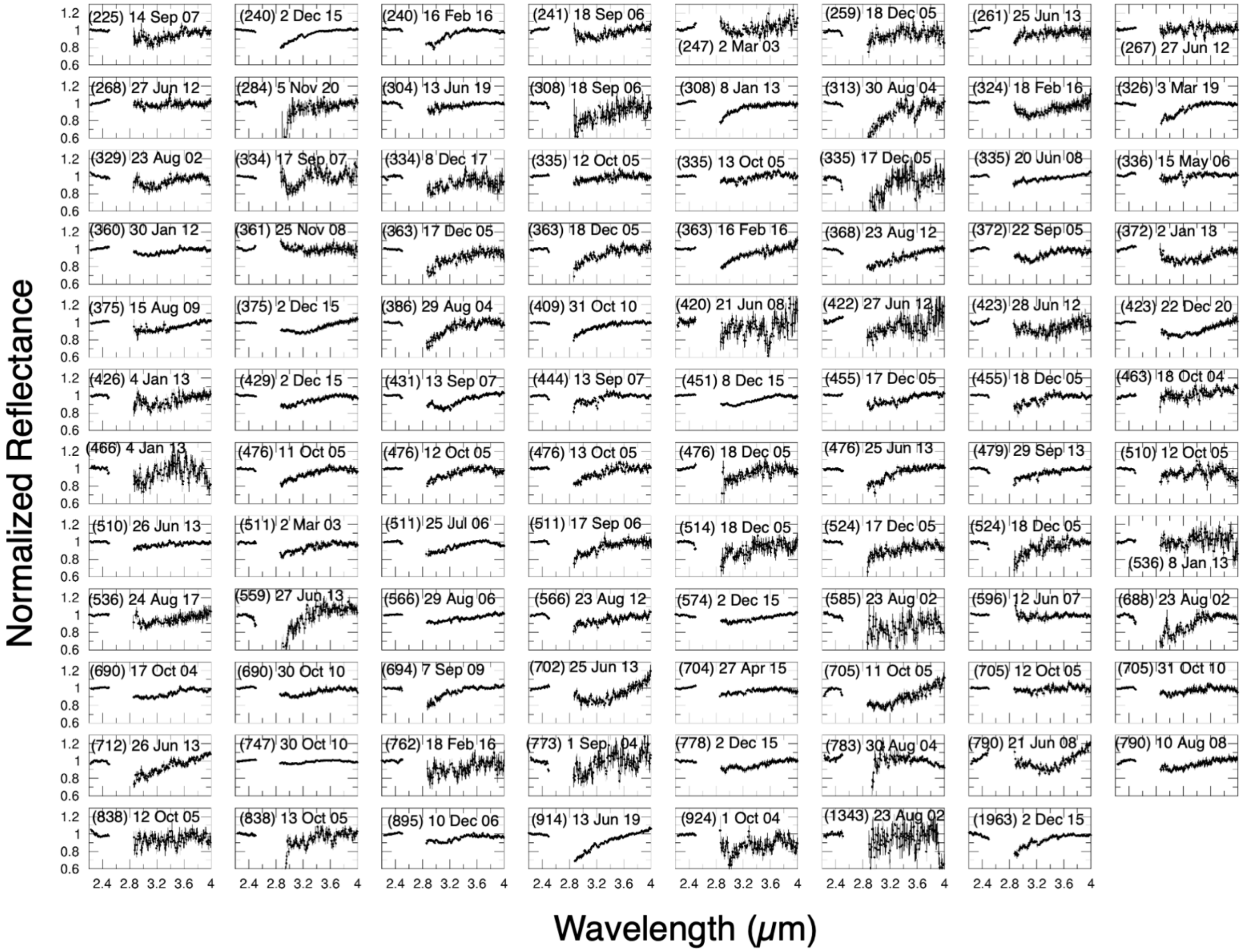
Normalized Reflectance
1.2
1
0.8
0.6
(225) 14 Sep 07
(240) 2 Dec 15
(240) 16 Feb 16
(241) 18 Sep 06
(247) 2 Mar 03
(259) 18 Dec 05
(261) 25 Jun 13
(267) 27 Jun 12
(268) 27 Jun 12
(284) 5 Nov 20
(304) 13 Jun 19
(308) 18 Sep 06
(308) 8 Jan 13
(313) 30 Aug 04
(324) 18 Feb 16
(326) 3 Mar 19
(329) 23 Aug 02
(334) 17 Sep 07
(334) 8 Dec 17
(335) 12 Oct 05
(335) 13 Oct 05
(335) 17 Dec 05
(335) 20 Jun 08
(336) 15 May 06
(360) 30 Jan 12
(361) 25 Nov 08
(363) 17 Dec 05
(363) 18 Dec 05
(363) 16 Feb 16
(368) 23 Aug 12
(372) 22 Sep 05
(372) 2 Jan 13
(375) 15 Aug 09
(375) 2 Dec 15
(386) 29 Aug 04
(409) 31 Oct 10
(420) 21 Jun 08
(422) 27 Jun 12
(423) 28 Jun 12
(423) 22 Dec 20
(426) 4 Jan 13
(429) 2 Dec 15
(431) 13 Sep 07
(444) 13 Sep 07
(451) 8 Dec 15
(455) 17 Dec 05
(455) 18 Dec 05
(463) 18 Oct 04
(466) 4 Jan 13
(476) 11 Oct 05
(476) 12 Oct 05
(476) 13 Oct 05
(476) 18 Dec 05
(476) 25 Jun 13
(479) 29 Sep 13
(510) 12 Oct 05
(510) 26 Jun 13
(511) 2 Mar 03
(511) 25 Jul 06
(511) 17 Sep 06
(514) 18 Dec 05
(524) 17 Dec 05
(524) 18 Dec 05
(536) 8 Jan 13
(536) 24 Aug 17
(559) 27 Jun 13
(566) 29 Aug 06
(566) 23 Aug 12
(574) 2 Dec 15
(585) 23 Aug 02
(596) 12 Jun 07
(688) 23 Aug 02
(690) 17 Oct 04
(690) 30 Oct 10
(694) 7 Sep 09
(702) 25 Jun 13
(704) 27 Apr 15
(705) 11 Oct 05
(705) 12 Oct 05
(705) 31 Oct 10
(712) 26 Jun 13
(747) 30 Oct 10
(762) 18 Feb 16
(773) 1 Sep 04
(778) 2 Dec 15
(783) 30 Aug 04
(790) 21 Jun 08
(790) 10 Aug 08
(838) 12 Oct 05
(838) 13 Oct 05
(895) 10 Dec 06
(914) 13 Jun 19
(924) 1 Oct 04
(1343) 23 Aug 02
(1963) 2 Dec 15
2.4
2.8
3.2
3.6
4
Wavelength (μm)